\documentclass{aa}
\usepackage{graphicx}
\usepackage{supertabular}
\usepackage{epsfig}
\usepackage{ifthen}
\usepackage[figuresright]{rotating}
\usepackage{natbib}
\usepackage{longtable}
\usepackage{lscape}

\bibpunct{(}{)}{;}{a}{}{,} 

\newcommand{\Day}{$^{d}$}

\newcommand{\Mag}{$^{m}$}

\begin{document}


   \title{Near-IR variability properties of a selected sample of AGB
 stars\thanks{Tables A is only available in electronic form
 at the CDS via anonymous ftp to cdsarc.u-strasbg.fr (130.79.128.5) or
 via http://cdsweb.u-strasbg.fr/cgi-bin/qcat?J/A+A/}}

   \subtitle{}

   \author{F.~M. Jim\'enez-Esteban   \inst{1,2,3}
        \and P. Garc\'\i a-Lario  \inst{4}
        \and D. Engels            \inst{3}
        \and A. Manchado          \inst{5,6}
   } 

   \offprints{F.M. Jim\'enez-Esteban, \email{f.jimenez-esteban@oan.es}}

\institute{Observatorio Astron\'omico Nacional, Apartado 112, E-28803 Alcal\'a de Henares, Spain.
        \and FRACTAL SLNE, Avda San Sebasti\'an 5 3-D, E-38003 Santa Cruz de Tenerife, Spain. 
	\and Hamburger Sternwarte, Gojenbergsweg 112, D-21029 Hamburg, Germany. 
        \and Research and Scientific Support Department of ESA, European Space Astronomy Centre, Villafranca del Castillo, Apartado de Correos 50727, E-28080 Madrid, Spain.
        \and Instituto de Astrof\'\i sica de Canarias, Via Lactea s/n, E-38200 La Laguna, Tenerife, Spain.
	\and Consejo Superior de Investigaciones Cient\'\i ficas, Spain.
}

   \date{Received 4 August 2005 / Accepted 8 June 2006}

   \abstract{

     We present the results of a near-infrared monitoring programme of
a selected sample of stars, initially suspected to be Mira variables
and OH/IR stars, covering more than a decade of observations. The
objects monitored cover the typical range of IRAS colours shown by
O-rich stars on the asymptotic giant branch and show a surprisingly
large diversity of variability properties. Sixteen objects are
confirmed as large-amplitude variables. Periods between 360 and
1\,800\Day\, and typical amplitudes
1\Mag\,$\la$\,$\Delta$K\,$\la$\,2\Mag could be determined for nine of
them. In three light curves, we find a systematic decrease in the mean
brightness, and two light curves show pronounced asymmetry. One
source, IRAS\,07222--2005, shows infrared colours typical of Mira
variables, but it pulsates with a much longer period
($\approx$1200\Day) than a normal Mira. Two objects are either close
to (IRAS\,03293+6010) or probably in (IRAS\,18299--1705) the post-AGB
phase. In IRAS\,16029--3041 we found a systematic increase in the H--K
colour of $\approx$1\Mag, which we interpret as evidence of a recent
episode of enhanced mass loss. IRAS\,18576+0341, a heavily obscured
luminous blue variable was also monitored. The star showed a continued
decrease in brightness over a period of 7 years (1995 -- 2002).

   
   }

   \maketitle

   \section{\label{intro}Introduction}

During the evolution on the asymptotic giant branch (AGB) stars begin
to pulsate and appear as large-amplitude long-period variables (LALPV)
during the final phase. Well-known objects are the Mira variables with
optical amplitudes $>$\,2.5\Mag\, and typical periods
$<$\,400\Day. Less known are the AGB variables that are heavily
obscured in the optical. The periods of these ``OH/IR stars'' are
considerably longer than those of optical Mira variables. Observing in
the infrared, \cite{Engels83} derived the periods of 17 objects they
monitored, finding values between 500\Day\, and 1\,800\Day\, and very
large amplitudes (up to $\approx$\,4\Mag) in the K
band. \cite{Herman85a} found similar periods from the monitoring of
the OH maser emission. The upper limits of the period distribution on
the AGB could not be determined, however, as the monitor projects were
not continued long enough.

We started a new open-end infrared monitoring programme in 1991 making
use of the service time option at the Observatorio del Teide. The aim
was to determine variability properties of a sample of AGB stars with
a wide range of mid-infrared IRAS colours, including some stars
suspected of having extremely long periods. Although in the meantime
various authors
\citep{Nakashima00,Blommaert98,Wood98b,Wood98a,Jones94,vanLangevelde93,vanLangevelde90,LeBertre93}
have provided periods for additional OH/IR stars, the number of OH/IR
stars studied for variability is still low compared to the number of
Mira variables monitored.

In this paper we present the observations of 24 sources suspected of
being AGB stars and obtained between 1991 and 2002. Due to the service time
character of the programme, the observations were made irregularly so
their number is sometimes very limited.  Nevertheless, we are able to
provide a handful of new period and amplitude determinations of Miras
and OH/IR stars.

   \section{\label{sample}Sample selection}


\begin{sidewaystable*}
\begin{minipage}[t][180mm]{\textwidth} 
\caption{IRAS sources included in the monitoring programme.}
\label{table1}

\begin{tabular}[t]{lrcrccccllc}
\hline\hline\noalign{\smallskip}
IRAS           & [12]$-$[25] & 2MASS Coordinates    & N$_{o/b}$ & Ref. (1) & Period     & Amplitude   & $\overline{K}$ (mag) & $f$                        & Classification & Ref. (2)\\ 
               &             &   (J2000)            &           &          & (days)     &  K (mag)    & $K_{o}$ (mag)        & $\dot{K}$ (mag\,yr$^{-1}$) &          &       \\ 
\noalign{\smallskip}\hline\noalign{\smallskip}
\object{02095$-$2355}   &$-$0.75 & 02 11 48.02 $-$23 41 43.8 & 20/2  & 1,2       &                 & 0.17 $\pm$ 0.04 & 2.08 $\pm$ 0.04 &                            & M-giant  & 2 \\
\object{03293+6010}     &   0.68 & 03 33 30.59 +60 20 09.4   & 12/2  & 2,3       & 1800 $\pm$  400 & 1.07 $\pm$ 0.07 & 7.42 $\pm$ 0.01 &                            & LALPV-1  & 1 \\ 
\object{04130+3918}     &$-$1.27 & 04 16 24.21 +39 25 45.1   & 19/1  & 2         &  470 $\pm$   30 & 0.93 $\pm$ 0.02 & 3.55 $\pm$ 0.01 &                            & LALPV-1  & 1 \\ 
\object{05131+4530}     &   0.61 & 05 16 47.47 +45 34 04.1   & 27/5  & 2-6       & 1100 $\pm$  100 & 2.48 $\pm$ 0.02 & 6.68 $\pm$ 0.01 & 0.29                       & LALPV-3  & 1 \\ 
\object{06297+4045}     &$-$0.09 & 06 33 15.79 +40 42 51.4   & 25/3  & 2,5,7     &  520 $\pm$   10 & 1.08 $\pm$ 0.01 & 2.97 $\pm$ 0.01 &                            & LALPV-1  & 1 \\ 
\object{07222$-$2005}   &$-$0.74 & 07 24 24.43 $-$20 11 56.0 & 17/3  & 2,8       & 1200 $\pm$  200 & 1.12 $\pm$ 0.02 & 4.73 $\pm$ 0.01 & (28 $\pm$ 3)$\cdot10^{-3}$ & LALPV-2  & 1 \\ 
\object{16029$-$3041}   &   0.69 & 16 06 08.36 $-$30 49 34.0 & 14/3  & 2,8-10    &                 & 2.75 $\pm$ 0.04 & 7.03 $\pm$ 0.04 &                            & Peculiar & 1 \\ 
\object{16037+4218}     &$-$0.55 & 16 05 28.91 +42 10 29.4   & 37/1  & 2         &  360 $\pm$   20 & 0.73 $\pm$ 0.01 & 3.80 $\pm$ 0.01 & (52 $\pm$ 1)$\cdot10^{-3}$ & LALPV-2  & 1 \\ 
\object{16437$-$3140}   &   0.61 & 16 46 58.83 $-$31 46 13.0 &  8/4  & 2,8,11,12 &                 & 3.21 $\pm$ 0.02 & 8.13 $\pm$ 0.02 &                            & Variable & 1 \\ 
\object{17103$-$0559}   &$-$0.54 & 17 12 58.13 $-$06 02 57.1 & 20/1  & 2         &  420 $\pm$   30 & 1.19 $\pm$ 0.01 & 4.16 $\pm$ 0.01 & (265 $\pm$ 2)$\cdot10^{-3}$& LALPV-2  & 1 \\ 
\object{17105$-$2804}   &$-$0.61 & 17 13 42.01 $-$28 07 49.9 & 14/1  & 2         &  420 $\pm$   20 & 1.11 $\pm$ 0.01 & 4.64 $\pm$ 0.01 &                            & LALPV-1  & 1 \\ 
\object{17313$-$1531}   &$-$0.12 & 17 34 11.30 $-$15 33 01.5 & 12/1  & 2         &                 & 1.51 $\pm$ 0.05 & 4.96 $\pm$ 0.05 &                            & Variable & 1 \\ 
\object{17347$-$3139}   &   1.81 & 17 38 00.61 $-$31 40 55.2 &  3/1  & 2         &                 & 0.18 $\pm$ 0.08 &10.21 $\pm$ 0.08 &                            & Bipolar PN & 3 \\ 
\object{17411$-$3154}*  &   0.83 & 17 44 23.89 $-$31 55 39.5 &  1/-- &           & $\approx$1440   &                 &                 &                            & LALPV    & 4 \\  
\object{18025$-$2113}   &$-$0.17 & 18 05 35.50 $-$21 13 42.3 & 14/1  & 2,8       &                 & 0.6  $\pm$ 0.2  & 2.6  $\pm$ 0.2  &                            & Supergiant  & 5 \\ 
\object{18299$-$1705}   &   0.96 & 18 32 50.75 $-$17 02 48.5 &  9/2  & 2,11      &                 & 0.33 $\pm$ 0.04 & 5.17 $\pm$ 0.04 &                            & PPN      & 1 \\ 
\object{18314$-$1131}   &$-$0.53 & 18 34 16.40 $-$11 29 29.2 & 10/1  & 2         &                 & 1.70 $\pm$ 0.07 & 3.71 $\pm$ 0.07 &                            & Variable & 1 \\ 
\object{18327$-$0715}   &   0.66 & 18 35 29.22 $-$07 13 11.0 &  2/6  & 2,13,14   &                 & 0.84 $\pm$ 0.06 & 8.43 $\pm$ 0.06 &                            & Variable & 1 \\ 
\object{18429$-$1721}   &$-$0.53 & 18 45 51.33 $-$17 17 59.4 &  7/1  & 2         &                 & 1.24 $\pm$ 0.04 & 2.31 $\pm$ 0.04 &                            & Variable & 1 \\ 
\object{18454$-$1226}   &$-$0.69 & 18 48 17.53 $-$12 22 43.2 &  8/1  & 2         &                 & 0.2  $\pm$ 0.3  & 2.0  $\pm$ 0.3  &                            & M-giant  & 1 \\ 
\object{18576+0341}     &   2.15 & 19 00 10.89 +03 45 47.1   & 15/14 & 2,15-19   &                 & 2.1  $\pm$ 0.1  & 7.5  $\pm$ 0.1  &                            & LBV      & 6 \\ 
\object{19129+2803}     &$-$0.52 & 19 14 59.10 +28 09 14.1   & 23/1  & 2         &  420 $\pm$   20 & 1.14 $\pm$ 0.01 & 5.19 $\pm$ 0.01 & 0.41                       & LALPV-3  & 1 \\ 
\object{19147+5004}     &$-$0.66 & 19 16 04.03 +50 09 36.6   & 19/2  & 2,20      &                 & 0.4  $\pm$ 0.1  & 2.4  $\pm$ 0.1  &                            & Lb-Variable & 7 \\ 
\object{23492+0846}     &$-$0.71 & 23 51 47.89 +09 03 23.4   &  6/2  & 1,2       &                 & 0.1  $\pm$ 0.2  & 1.4  $\pm$ 0.2  &                            & M-giant  & 2 \\ 
\noalign{\smallskip}\hline&  
\end{tabular} 

{\em Notes:} \\
$N_{o/b}$: The number of K-band observations, own and taken from the bibliography; Ref. (1): references for the photometric data taken from the bibliography; $K_{o}$ and $\dot{K}$: parameters of the light curves with linear variation of the mean magnitude. $f$: asymmetric factor of the light curves; Ref. (2): references for the classification assigned. \\
~* Coordinates derived relative to nearby 2MASS sources on the MAGIC image. The period was derived by \citealp{Olivier01} from measurements in the L' band.\\
LALPV-1: symmetric light curve with constant mean magnitude; LALPV-2: symmetric light curve with linear variation of the mean magnitude; LALPV-3: asymmetric light curve with constant mean magnitude.\\
References (1): 1- \citealp{Whitelock95}; 2- \citealp{Cutri03}; 3- \citealp{Noguchi93}; 4- \citealp{Blommaert93}; 5- \citealp{Xiong94}; 6- \citealp{Jiang96}; 7- \citealp{Hyland72}; 8- \citealp{Epchtein97}; 9- \citealp{Persi90}; 10- \citealp{Fouque92}; 11- \citealp{Manchado89}; 12- \citealp{vanderVeen90}; 13- \citealp{Gehrz85}; 14- \citealp{vanderVeen89}; 15- \citealp{Garcia-Lario97}; 16- \citealp{Meixner99}; 17- \citealp{Ueta01}; 18- \citealp{Clark03}; 19- \citealp{Pasquali02}; 20- \citealp{Kerschbaum96}. \\
References (2): 1- This work; 2- \citealp{Whitelock95}; 3- \citealp{PereaCalderon06}; 4- \citealp{Olivier01}; 5- \citealp{Josselin98}; 6- \citealp{Clark03}; 7- \citealp{Kholopov98}.\\

\end{minipage}
\end{sidewaystable*}



\begin{figure*}
\begin{center}
\includegraphics[width=13.0cm]{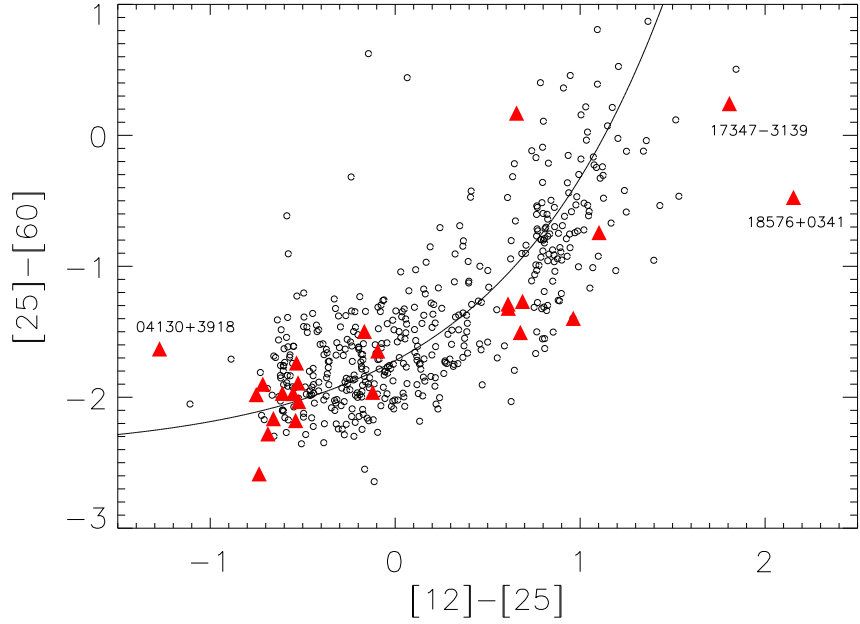}
    \caption{\label{fig_monit:IRAS_2cd} IRAS colour-colour diagram
    showing the location of the stars included in the monitoring
    programme (filled triangles), together with oxygen-rich AGB stars
    from the `Arecibo sample' (JE05) and from the `GLMP sample' (JE06)
    (open circles).  The continuous line is the `O-rich AGB sequence',
    and the IRAS colours are defined as:
    [12]$-$[25]\,=\,$-$2.5\,log$\frac{F_{\nu}(12)}{F_{\nu}(25)}$ and
    [25]$-$[60]\,=\,$-$2.5\,log$\frac{F_{\nu}(25)}{F_{\nu}(60)}$.}
\end{center}
\end{figure*}


The sample contains 24 stars selected from the IRAS Point Source
Catalogue with well-measured fluxes (FQUAL\,$\ge$\,2,
\citealp{Beichman88}) in the IRAS photometric bands centred at 12, 25,
and 60\,$\mu$m and with colours that resemble those typically shown by
O-rich AGB stars. The selection was made in 1990 and none of the stars
had been studied before. All objects monitored are listed in Table
\ref{table1}. The table gives the IRAS name, the [12]$-$[25] IRAS
colour, the J2000 equatorial coordinates, and several columns with the
results of the variability analysis, which are described in Sect.
\ref{resul-lc}. All sources but IRAS\,02095--2355, IRAS\,04130+3918,
IRAS\,18576+0341, and IRAS\,23492+0846 were previously detected as OH
masers at 1612 MHz
\citep{teLintel-Hekkert89,teLintel-Hekkert91a,Zijlstra89,David93},
confirming their oxygen-rich chemistry. IRAS\,02095--2355 and
IRAS\,23492+0846 turned out to be M-giants \citep{Whitelock95}, while
IRAS\,04130+3918 has recently been identified as a carbon star
\citep{Alksnis01} and IRAS\,18576+0341 as a luminous blue variable
star \citep{Clark03}.

In Fig. \ref{fig_monit:IRAS_2cd} we show the location of the sample in
an IRAS colour-colour diagram. The sources were intentionally selected
to be located at different positions along the `O-rich AGB sequence',
as is defined by the oxygen-rich AGB star samples of Jim\'enez-Esteban
et al. (\citeyear{Jimenez-Esteban05}, \citeyear{Jimenez-Esteban06};
hearafter JE05, JE06). Two exceptions are IRAS\,17347--3139 and
IRAS\,18576+0341, which are placed in the region corresponding to
sources with detached shells (like post-AGB stars or planetary
nebulae). The Mira variables populate the blue and the OH/IR stars the
red part. It was therefore expected to find Mira-like variability with
increasing periods and amplitudes along the `O-rich AGB sequence'.

   \section{\label{obs}Observations}

The observations presented here were carried out over a period of
10\,years, from 1991 March to 2002 August. During the first years we
performed aperture photometry, mainly using service time at the
infrared 1.5\,m Carlos S\'anchez Telescope (CST) of the Observatorio
del Teide (Canary Islands, Spain). Later, we complemented the
monitoring programme with infrared images taken at the 1.23\,m
telescope at Calar Alto Observatory (Almer\'\i a, Spain).
  
\subsection{Aperture photometry}

Aperture photometry at the 1.5\,m CST was performed with a Circular
Variable Filter (CVF) spectrophotometer. The CVF is equipped with an
InSb photometric detector, which works at the temperature of liquid
nitrogen. A photometric aperture of 15\arcsec\, with a chopper throw
of 30\arcsec\, in right ascension was used to subtract the
contribution from the background sky. We used the infrared filters J
(1.25\,$\mu$m), H (1.65\,$\mu$m), K (2.20\,$\mu$m), and occasionally
L' (3.8\,$\mu$m). The Teide photometric system and its transformation
to other standard photometric systems are described in
\cite{Arribas87}.  All observations were taken under photometric
weather conditions. For the photometric calibration, we used the list
of standard stars in \cite{Koornneef83b}, observing at least two of
these several times every night at different air masses to obtain the
extinction coefficients. The service observations were solicited
imposing a minimum S/N ratio of 50 in the K band. All magnitudes given
in this paper are in the Teide photometric system.

\subsection{Imaging photometry}

In the final years of the monitoring programme we also used the
1.23\,m telescope of the Calar Alto Observatory in three different
runs, equipped with the infrared camera MAGIC
\citep{Herbst93}. Observations were made in the broad-band filters J,
H, and K. The camera has a 256\,$\times$\,256 NICMOS3 HgCdTe detector
array with a pixel scale of 1.2\arcsec, which provides an approximate
field of view of 5\arcmin\,$\times$\,5\arcmin. Several standard stars
from the list of \cite{Elias82} were used for the calibration of
near-infrared images taken under photometric weather
conditions. Images taken during non-photometric nights were calibrated
using 2MASS Point Source Catalogue \citep{Cutri03}. The observational
technique and the data reduction process are described in detail in
JE05.

We found no differences in magnitudes and colours between the Calar
Alto photometric system and the ``CIT'' system used by
\cite{Elias82}. To transform the Calar Alto photometry (CA) to the
Teide photometric system (OT), we used the transformations given by
\cite{Arribas87}, which relate the Teide to the ``CIT''
system. Actually, only the J-band photometry is affected:
K$_{OT}$\,=\,K$_{CIT}$\,=\,K$_{CA}$,
H$_{OT}$\,=\,H$_{CIT}$\,=\,H$_{CA}$, and
(J$-$H)$_{OT}$\,=\,1.11\,(J$-$H)$_{CIT}$\,=\,1.11\,(J$-$H)$_{CA}$.

   \section{\label{resul-photo}Near-infrared photometry}     


\begin{figure*}
\begin{center}
\epsfxsize=6.5cm
\epsfbox{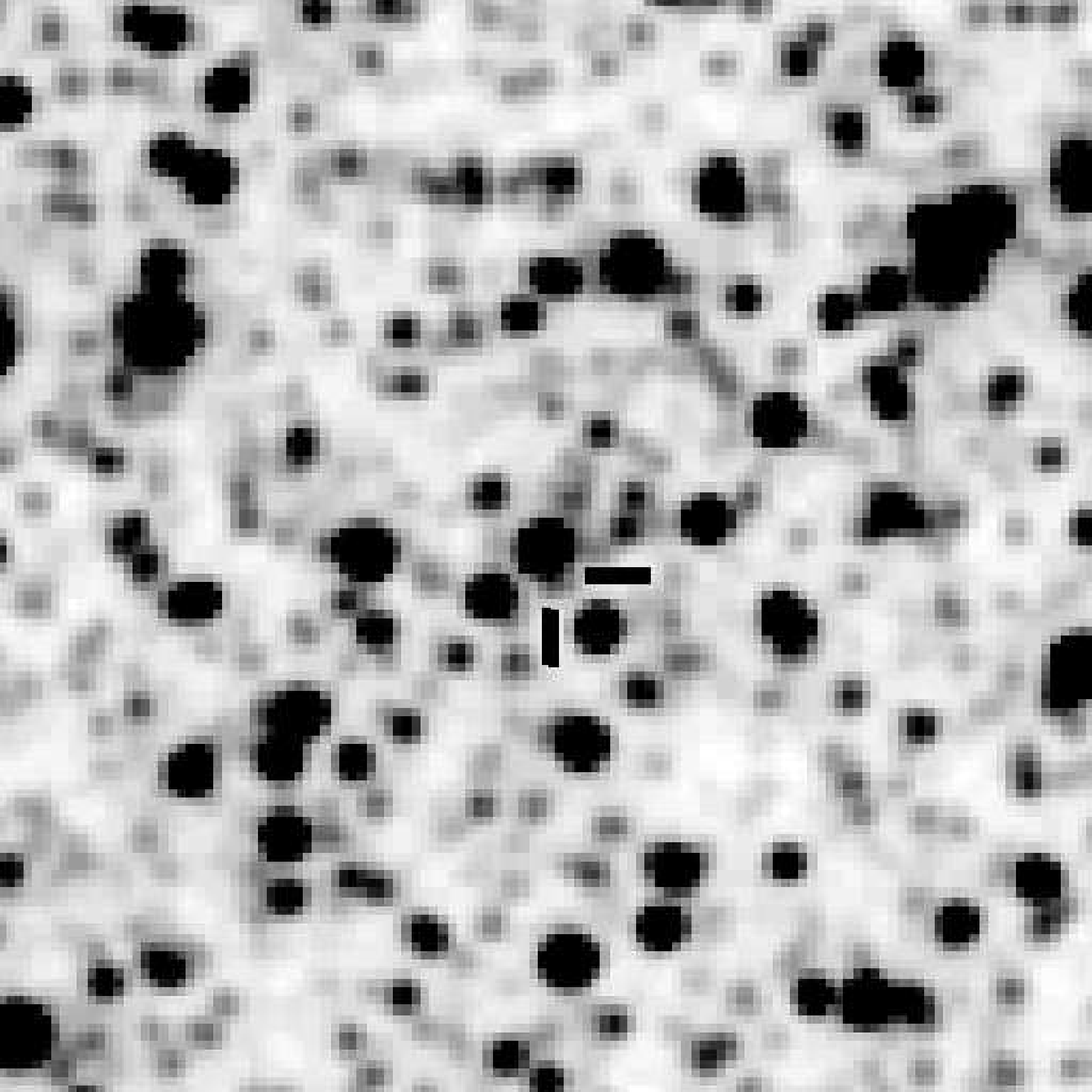}
\hspace{1cm}
\epsfxsize=6.5cm
\epsfbox{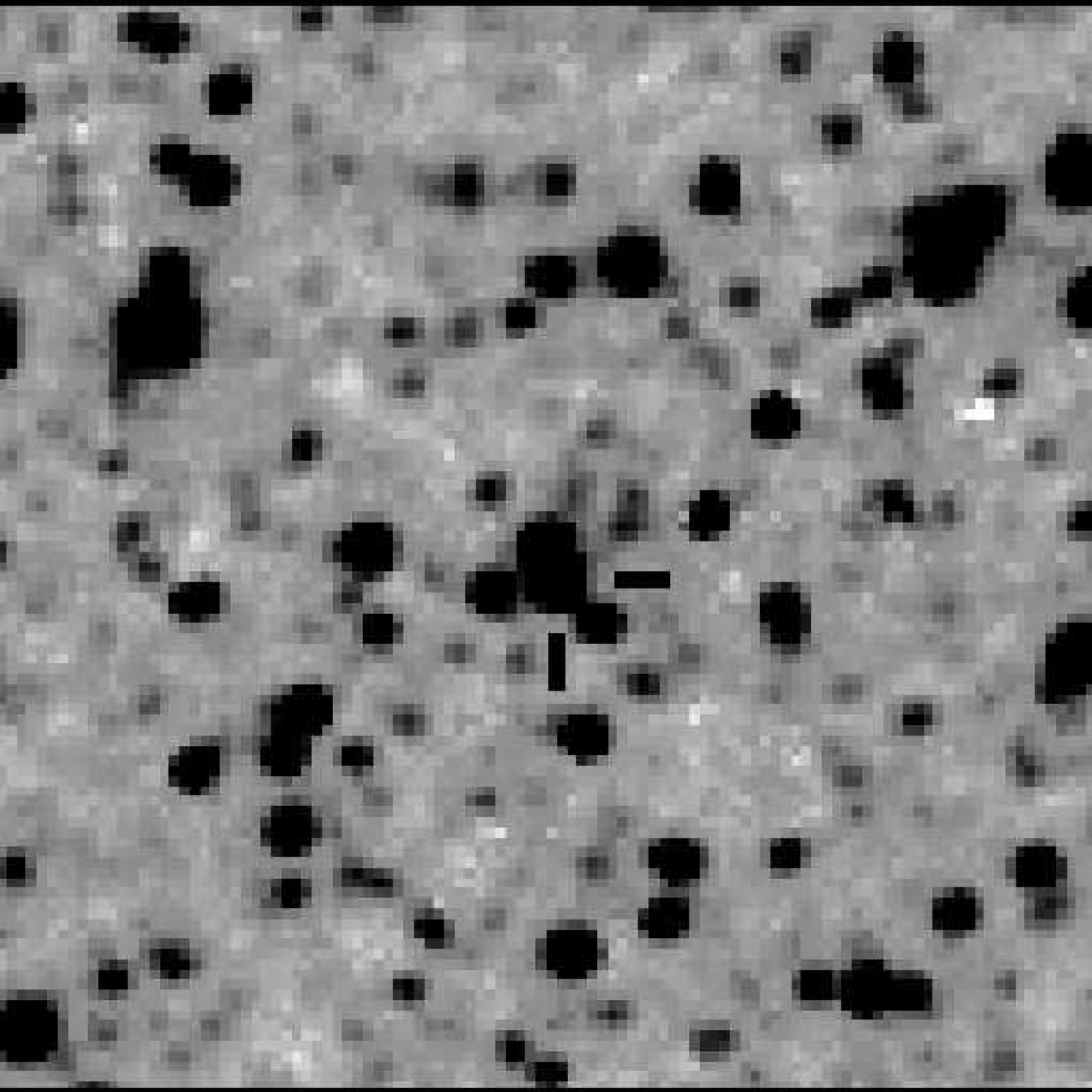}

\end{center}

\caption{\label{fig_monit:2epoch17411}K-band images of
IRAS\,17411--3154 in two different epochs: 1998 (2MASS; left panel)
and 2001 (MAGIC; right panel).}

\end{figure*}


The near-infrared counterparts of the IRAS sources were initially
searched for with the CVF spectrophotometer by scanning in the
vicinity of the IRAS coordinates, which have typical errors
20\arcsec\,--\,30\arcsec. In most cases only one source showing
colours typical of AGB stars was found. This source actually was also
the reddest in the field and, as such, easily recognizable as the right
counterpart.

The counterparts were later verified by using the MAGIC images from
the Calar Alto observations and by checking the spectral energy
distributions with the help of the MSX Point Source Catalogue
\citep{Egan03a} and the 2MASS Point Source Catalogue. Except for
IRAS\,17411--3154, the original counterparts could be confirmed. The
best available positions as given by the 2MASS are listed in Table
\ref{table1}.

In the case of IRAS\,17411--3154, we found the counterpart to be
strongly blended by a foreground star and to be detectable only during
the bright phases of the light variation cycle and only in the K-band
(see Fig. \ref{fig_monit:2epoch17411}). The object was monitored in
parallel by \cite{Olivier01} at 3.45$\mu$m, and turned out to be a
LALPV with a period of P\,=\,1\,440\Day. The coordinates given in
Table \ref{table1} for this source were derived relative to the 2MASS
positions of nearby sources in the field.


The near-infrared JHKL' photometric data, together with their
associated errors, are listed in Table A of the Appendix. Although the
data has been collected during 121 observing nights, the number of
observations per object was much smaller and rather inhomogeneous.  A
minimum of six and a maximum of 37 observations were accumulated for
individual objects. The errors obtained for the aperture photometry
are in the range 0.01\,--\,0.1\Mag\, for the majority of cases, while
those obtained for imaging photometry are typically
$\approx$\,0.03\Mag.

Judging from the MAGIC infrared images, we found a number of cases where
the infrared fluxes obtained with the CVF spectrophotometer must have
been contaminated by nearby stars contained in the aperture.
IRAS\,16437--3140, for instance, was strongly blended with an object
having similar brightness in the J-band, but much fainter in the
H and K bands. We therefore removed the photometry in the J filter for
this source from Table A.  Other sources affected were
IRAS\,17347--3139 and IRAS\,17411--3154, where none of the aperture
photometry data turned out to be useful.

As the light variations were rather similar in all three bands, we
present graphically here only light curves from the K-band data, which
is the most complete data set. They are plotted in Fig. A1 of the
Appendix, excluding IRAS\,17347--3139 and IRAS\,17411--3154. We added
K band photometry from the 2MASS and DENIS \citep{Epchtein97}
near-infrared surveys, and from the literature
\citep{Hyland72,Manchado89,Persi90,vanderVeen90,Blommaert93,Noguchi93,Xiong94,Jiang96}.
Whenever a model light curve was fitted to the observational data (see
the following section) this is also shown. The number of K-band
observations ($N_{o/b}$ = own and contributed) used is given in Table
\ref{table1}.

   \section{\label{resul-lc} Light curve analysis}


We characterize the variability of the sources on the basis of the
amplitudes and periods listed in Table \ref{table1}. If periods are
given, the light curves were successfully fitted by a periodic model
light curve (see next section) and the amplitudes are those derived
from the fit. In the other case, amplitudes are taken as the
difference between the observed maximum and minimum brightness.

Conclusions on variability can be made for 23 of the 24 sources of
Table \ref{table1}. We found variability with amplitudes
$\Delta$K\,$>$\,0.2 for all sources except IRAS\,02095$-$2355,
IRAS\,17347--3139, IRAS\,18454$-$1226, and IRAS\,23492$+$0846. The
latest sources only show some scattering and can be considered as
essentially non-variable. IRAS\,17347--3139 is a bipolar planetary
nebula \citep{PereaCalderon06}. The three other stars are rather
bright (K\,$<$\,2.1\Mag) and belong to the bluest sources in the
sample ([12]$-$[25]\,$<$\,$-0.65$). Two of them were classified as
non-Mira M-giants because of their near-infrared colours by
\cite{Whitelock95}. This classification is corroborated by their
non-variability. For the same reasons, we identified the third
source IRAS IRAS\,18454$-$1226 with similar infrared and variability
properties as M-giant as well. The classifications adopted and their
origin are listed in the last two columns of Table \ref{table1}.

The variable sources can be split into a group of regularly pulsating
large-amplitude long-period variables (LALPVs) and a group with
ambiguous variability properties.


In order to characterize the variability properties of the LALPVs, we
fitted model light curves of several types to the observational data,
including data from the literature. As a first approach, we adopted a
symmetric sinusoidal model light curve (Type\,1). The least mean
square (LMS) method was applied for a range of periods between 200 and
3\,000\Day. The quality of the fit was determined with the $\chi^2$
test, weighting each observation with the inverse of the square of the
observational error.

We checked our results using the period dispersion minimization (PDM)
method \citep{Stellingwerf78}. The errors $\Delta P$ for the period
estimations were derived according to the PDM method from
$\Delta$P\,$\simeq$\,$P^2 / 2\,T$, where P is the trial period at the
primary minimum, and T is the time baseline of the observations.

We found that in some cases assuming a symmetric sinusoidal curve and
a constant mean magnitude is not the best choice. Some objects, like
IRAS\,17103$-$0559 (See Fig. A1), show clear evidence of large secular
variations in their near-infrared mean brightness.  We therefore also
used symmetric sinusoidal model light curves with a mean magnitude
linearly variable with time (Type\,2).  This type of light curve was
adopted, if we obtained a value of $\chi^2$ that is at least 30\%
smaller than the one derived from fitting Type\,1 light curve.

The third type of model light curve assumed a constant mean magnitude
but an asymmetric sine (Type\,3) characterized by an asymmetry factor
$f$. This factor can assume values in the range 0 and 1, with
$f$=\,0.5 corresponding to a symmetric light curve. A value $<$\,0.5
($>$\,0.5) corresponds to a steeper (shallower) rising branch compared
to the descending branch \citep{vanLangevelde93}. A good case for an
asymmetric sinusoidal light curve is given by IRAS\,05131+4530 (See
Fig. A1). For this source a similar period was previously reported
by \cite{Groenewegen99}. As before, this type was only adopted if
$\chi^2$ was found to be at least 30\% smaller than its value for
Type\,1 or Type\,2 light curves.


\begin{figure}
	\resizebox{\hsize}{!}{\includegraphics{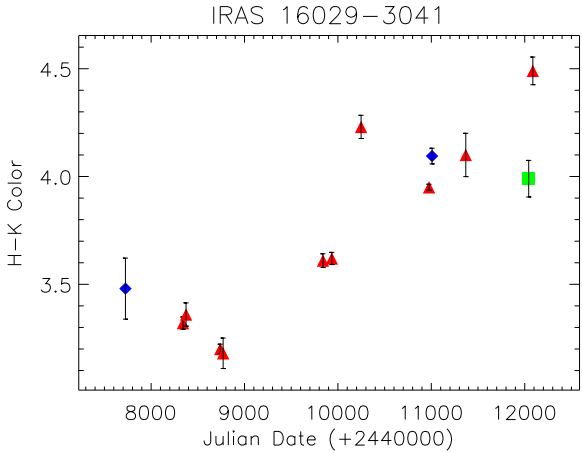}}
	\caption{\label{fig_monit:H-K_16029} Variation of the H--K
	colour with time for IRAS\,16029--3041. CVF data are plotted
	with triangles (red), MAGIC data with squares (green), and
	data from literature with diamonds (blue).}
\end{figure}


For nine sources we successfully fitted model light curves. They were
classified as LALPV-n in Table \ref{table1}, where n\,=\,1,2,3 depend
on the type of light curve adopted. The light curves are shown in
Fig. A1 and A2 of the Appendix, as functions of time and phase.  Their
periods, amplitudes, and corresponding errors are given in Table
\ref{table1}. In addition, the mean magnitudes $\overline{K}$ are
given for light curves of Type\,1, the slopes of the mean magnitude
$\dot{K}$ and the mean magnitudes $K_{o}$ at Julian Date 2\,450\,000
are listed for Type\,2 light curves, and $\overline{K}$ and the
asymmetry factors $f$ are given for Type\,3 light curves. Periods
between 360\Day, typical of Mira variables, and 1\,800\Day\, were
derived. The long periods of $>$1\,000\Day\, obtained for three stars
are in the range of OH/IR star periods found in earlier studies.


The objects with ambiguous variability properties consist of a group
of stars with irregular variability behaviour and others that probably
have an insufficient amount of observations to determine the
periodicity of their variations.

Irregular low-amplitude variability ($\Delta$K\,$\le$\,0.6\Mag) is
shown by IRAS\,18025$-$2113, IRAS\,18299$-$1705, and IRAS\,19147$+$5004
(TZ\,Cyg). IRAS\,18025$-$2113, a possible supergiant star
\citep{Josselin98}, and IRAS\,19147$+$5004, an Lb-type irregular
variable \citep{Kholopov98}, are bright ($\overline{K}$\,$<$\,3\Mag)
and blue in the mid-infrared ([12]$-$[25]\,$<$\,$-0.1$). Their
presence indicates that a fraction of the stars populating the blue end
of the ``O-rich AGB sequence'' are actually not genuine Mira
variables.

In contrast, the third low-amplitude variable IRAS\,18299$-$1705 is
fainter ($\overline{K}$\,=\,5.2) and has one of the reddest
[12]$-$[25] colours in the sample. The low amplitude indicates that
this star belongs to the group of ``non-variable'' OH/IR stars. This
group shows irregular low-amplitude variability (if any) in contrast
to the regular pulsating OH/IR stars with large amplitudes. Because of
the cessation of large-amplitude pulsations, they are considered as
stars that are leaving the AGB and in which the strong mass-loss has
finished \citep{Engels02}. In the case of IRAS\,18299$-$1705, the
circumstellar shell has been diluted so far that the central star has
reappeared again and can be detected in the near-infrared. Therefore,
we tentatively classify this star as a proto-planetary nebula. Note,
however, that in principle similar observational properties can also
be associated to AGB stars during the helium-burning cycle following a
thermal pulse, when a sharp decrease in the mass-loss rate is also
expected, leading to the development of a detached dust shell.

A peculiar variability behaviour is shown by IRAS\,16029--3041 and
IRAS\,18576$+$0341. IRAS\,16029$-$3041 shows strong variability with
one of the largest amplitudes ($\Delta$K\,$\approx$\,2.75\Mag) in our
sample. This object actually shows the reddest near-infrared colours,
but these colours have been changing during the years of
monitoring. In Fig. \ref{fig_monit:H-K_16029} the evolution of its
H--K colour as a function of the Julian Date is shown, which has
increased by $\approx$\,1\Mag\, in the past 10\,years. This change in
the near-infrared colour is accompanied by a simultaneous fading in
the K and H bands, as might be expected in case of high, increasing
extinction. This object must have a very thick CSE as derived from its
lack of an optical counterpart, which suggests a large mass-loss
experienced in the recent past. The progressive reddening is
interpreted as the result of a continuous increase in the thickness of
its CSE, which has yet not finished, probably due to a very recent
episode of enhanced mass loss.

A special case is also IRAS\,18576$+$0341. This star is not on the
AGB, but is a luminous blue variable star \citep{Clark03}. It has a
rather slowly varying light curve, which reached a maximum at
K\,=\,6.5\Mag\, in 1995 and descended since then by $\approx$1.5\Mag\,
until the last available measurement by the end of 2002.

The remaining five stars, classified as ``variable'' in Table
\ref{table1}, show a wide range of amplitudes
(0.8\Mag\,$<$\,$\Delta$K\,$<$\,3.3\Mag), but have a relatively small
number of observations (N\,$\la$\,15). For some of them we could fit a
periodic model light curve, however the significance was low. For
others, irregular variations look more likely. Additional photometric
data points taken with a higher observing frequency are required to
classify the light curves of these variables correctly.

   \section{\label{discu}Discussion}


\begin{figure*}
\begin{center}
   \includegraphics[width=13cm]{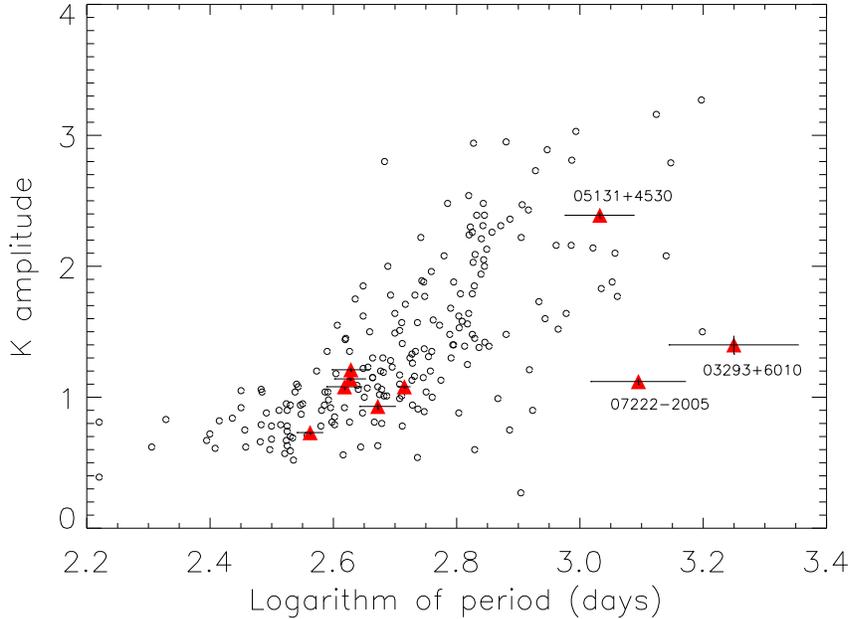}
	\caption{\label{fig_monit:Peri_Amp} K amplitude ($\Delta$K) as
	a function of the period. We have plotted our data (filled
	triangles), together with data extracted from the bibliography
	(open circles).}
\end{center}
\end{figure*}


The monitoring programme confirmed that the large majority of the IRAS
sources along the `O-rich AGB sequence' are indeed
variable. Non-variable M-giants are part of the population only at the
blue end of the sequence. One of them, IRAS\,18454--1226, was
detected as an OH maser \citep{teLintel-Hekkert91a}, implying that
non-pulsating M-giants may also possess a circumstellar gas and dust
shell able to support OH emission.

Variability has been found on all amplitude scales. Low-amplitude
variability ($\Delta$K\,$\leq$\,0.6 mag) has been found for two stars
among the blue sources that can be characterized as irregular
variables. Irregular, low-amplitude variations were also found for the
very red source IRAS\,18299--1705. Together with the presence of a
rather bright near-infrared counterpart, this star is probably already
in the post-AGB stage. IRAS\,18299-1705 shows that post-AGB stars can
be found relatively close to the red end of the `O-rich AGB sequence'
and that there is an overlap between the locations of heavily obscured
late AGB stars pulsating with large amplitudes and the
``non-variable'' post-AGB stars having, in general, redder [12]$-$[25]
colours. A similar conclusion was drawn by JE05, who discovered
several bright near-infrared counterparts to very red objects on the
`O-rich AGB sequence'.

Large-amplitude variations were found for 16 sources. Two are
peculiar. One, IRAS\,18576+0341, is not an AGB star and will not be
discussed further. IRAS\,16029--3041 has a strong double-peaked OH
maser and the IRAS Low Resolution Spectrum has an index 32,
characterizing oxygen-rich dust shells with the 10$\mu$m feature in
absorption. It is definitely an AGB star. The erratic light curve (see
Fig. A2) connected to a systematic reddening of the near-infrared
colours is suggested to be the result of a recent episode of enhanced
mass loss.

The service-time character of the monitoring programme had the
disadvantage that for several sources (labeled ``variable'' in Table
\ref{table1}) the observing frequency was too low to prove periodicity
in their light curves. For nine sources, however, regular pulsation
could be proven, and we found the expected wide range of periods
between 360 and 1\,800\Day, and amplitudes up to
$\Delta$K\,=\,2.5\Mag. Due to the long duration of the monitor
programme, observations were usually spread over several pulsation
cycles. We did not find significant cycle-to-cycle variations in 6
cases, but for three stars a systematic decrease of the mean
near-infrared brightness was found. Such secular variations in the
mean brightness have been reported in several other long-period
variables \citep{Bedding02,Whitelock03}. The secular variations might
be connected to changes in the mass-loss rate, which lead to a varying
obscuration of the star in the near-infrared.

For IRAS\,05131+4530 and IRAS\,19129+2803, a significant asymmetry of
the light curve was found. These asymmetries are commonly observed in
the optical light curves of Mira variables \citep{Vardya88} but are
less pronounced in the infrared.

In our sample we find stars with very similar periods, but very
different amplitudes. In Fig. \ref{fig_monit:Peri_Amp} we plotted our
results with data taken from the literature
\citep{Olivier01,Wood98b,Whitelock94,LeBertre93,Engels83}. Up to
log\,P\,=\,2.55, a small range of amplitudes
0.7\Mag$\leq$\,$\Delta$K\,$\leq$\,1.1\Mag\, is observed, while the
amplitudes increase steadily with longer periods. However, the scatter
is large. Two sources of our sample, IRAS\,03293+6010 and
IRAS\,07222--2005, have relatively low amplitudes for their periods.
IRAS\,03293+6010 has red mid-infrared colours and the longest period
(P\,=\,1\,800\Day) in the sample. IRAS\,07222--2005, in contrast, has
very blue mid-infrared colours and an unusually long period
(P\,=\,1\,200\Day) for this colour.

We speculate that IRAS\,03293+6010 is in an advanced evolutionary
stage, in which the large-amplitude pulsation is fading due to the
rapidly decreasing envelope mass, leading to very lengthened periods
and decreasing amplitudes. The star might be close to the transition
to the post-AGB stage and may develop currently into a
``non-variable'' OH/IR star.

For IRAS\,07222--2005, it is not the amplitude of its pulsation but
the length of the period that is peculiar. There are no other stars
known with such a long period located at the blue end of the `O-rich
AGB sequence'. The star has a strong silicate feature in emission
(IRAS LRS=29) and a low gas outflow velocity of 8.1 km\,s$^{-1}$
according to the OH maser \citep{teLintel-Hekkert91a}.
IRAS\,07222--2005 is therefore not an M-Supergiant. Similar to
IRAS\,03293+6010, the long period of IRAS\,07222--2005 might be due to
a small remaining envelope mass, while approaching the end of the
AGB. Contrary to IRAS\,03293+6010, however, IRAS\,07222--2005 may be
finishing the AGB with a modest mass-loss rate and, accordingly, with
a blue colour.  A reduced mass loss is expected for stars with their
C/O ratio being close to unity, thus hampering dust formation
\citep{Willems88, Chan88}. Alternatively, IRAS\,07222--2005 can be
similar to some MCs OH/IR stars with P\,$>$\,1000\Day\, and rather
blue colours, like e.g. IRAS\,04509--6922 with
$<$\,H--K\,$>$\,=\,0.85, $\Delta$K\,=\,1.5\Mag, and P\,=\,1290\Day\,
(see Table 5 of \citealt{Wood92}). They are rather luminous and
therefore probably massive. The lower metallicity of the LMC compared
to the Galaxy might have decreased their mass-loss
rate. Interestingly, IRAS\,07222--2005 is located in the anticentre
direction (l\,=\,234$^{\circ}$, b\,=\,$-$2$^{\circ}$) compared to
other monitored objects which are mostly at
(10$^{\circ}$\,$<$\,l\,$<$\,45$^{\circ}$), and, it shows a relatively
low expansion velocity\footnote{The expansion velocity is derived from
a questionable OH 1612\,MHz detection by
\cite{teLintel-Hekkert90}. The measurement is not listed in Table\,2
of \cite{teLintel-Hekkert91a} (printed version), but is contained in
the electronic version of the Table. The maser and derived velocities
therefore require confirmation (te Lintel Hekkert,
priv. communication).}, which is compatible with lower metallicity as
in the LMC.

   \section{\label{conclu}Conclusions}

The monitoring programme has shown that the variability properties
along the `O-rich AGB sequence' are diversified. In general, periods
and amplitudes increase with redder colours, but the scatter is large,
and AGB stars in evolutionary phases of non- or low-amplitude
variability are found at the borders of the sequence. The transition
to post-AGB stars in the IRAS colour-colour diagram appears to be
smooth. The variability properties certainly hold important clues to
the evolutionary stage of the stars on the AGB. To uncover the links
between these properties and particular phases of AGB evolution
requires, however, larger samples and therefore long-term monitoring
programmes to study the variability properties, especially at the red
end of the `O-rich AGB sequence', should be encouraged.

   \begin{acknowledgements}

Based on service-time observations made with the Carlos S\'anchez
telescope of the Observatorio del Teide (Tenerife, Spain), operated by
the Instituto de Astrof\'\i sica de Canarias, and on observations
collected at the German-Spanish Astronomical Centre, Calar Alto,
operated jointly by the Max-Planck-Institut f\"ur Astronomie and the
Instituto de Astrof\'\i sica de Andaluc\'\i a (CSIC). This work was
partially funded through grants PB94$-$1274 from the Spanish
Direcci\'on General de Investigaci\'on Cient\'\i fica y T\'ecnica
(DGICYT) and AYA2003$-$06473, AYA2003$-$09499, and AYA2004-03136 from
the Spanish Ministerio de Ciencia y Tecnolog\'\i a. This publication
makes use of data products from the Two Micron All Sky Survey, which
is a joint project of the University of Massachusetts and the Infrared
Processing and Analysis Center/California Institute of Technology,
funded by the National Aeronautics and Space Administration and the
National Science Foundation. We also thank to the referee for valuable
suggestions.

   \end{acknowledgements}

   \bibliographystyle{aa} 
   \bibliography{/home/fran/RESEARCH/bibliography/references} 

   \appendix

\begin{table*}
{\bf Table A.} Summary of the near infrared photometric measurements.
\begin{center}
\begin{tabular}{cccccl}

\hline\hline\noalign{\smallskip}
Julian Date &  \multicolumn{4}{c}{NIR magnitude [mag]}  & Notes\\
 (2400000+)  & J & H & K & L' &  \\
\noalign{\smallskip}\hline\noalign{\smallskip}

& & & & & \\
\multicolumn{6}{c}{\bf{IRAS 02095$-$2355}} \\
& & & & & \\
48488 &  3.32 $\pm$ 0.01 &  2.39 $\pm$ 0.01 &        ---       &        ---       &    \\
48509 &  3.26 $\pm$ 0.04 &  2.38 $\pm$ 0.02 &  2.04 $\pm$ 0.02 &        ---       &    \\
48865 &  3.41 $\pm$ 0.05 &  2.43 $\pm$ 0.02 &  2.10 $\pm$ 0.01 &        ---       &    \\
48880 &  3.34 $\pm$ 0.02 &  2.45 $\pm$ 0.02 &  2.12 $\pm$ 0.02 &        ---       &    \\
48922 &  3.29 $\pm$ 0.06 &  2.38 $\pm$ 0.02 &  2.06 $\pm$ 0.02 &        ---       &    \\
48944 &  3.24 $\pm$ 0.05 &  2.35 $\pm$ 0.02 &  2.02 $\pm$ 0.02 &  1.80 $\pm$ 0.03 &    \\
49229 &  3.35 $\pm$ 0.25 &  2.34 $\pm$ 0.06 &  2.03 $\pm$ 0.06 &  1.81 $\pm$ 0.06 &    \\
49320 &  3.24 $\pm$ 0.03 &  2.39 $\pm$ 0.09 &  2.08 $\pm$ 0.14 &        ---       &    \\
49326 &  3.36 $\pm$ 0.02 &  2.45 $\pm$ 0.02 &  2.16 $\pm$ 0.02 &        ---       &    \\
49385 &  3.42 $\pm$ 0.07 &  2.49 $\pm$ 0.05 &  2.14 $\pm$ 0.05 &        ---       &    \\   
49694 &  3.38 $\pm$ 0.03 &  2.50 $\pm$ 0.03 &  2.14 $\pm$ 0.02 &  2.01 $\pm$ 0.01 &    \\
49737 &  3.33 $\pm$ 0.03 &  2.46 $\pm$ 0.03 &  2.13 $\pm$ 0.03 &  1.85 $\pm$ 0.10 &    \\
49935 &  3.23 $\pm$ 0.03 &  2.36 $\pm$ 0.02 &  2.03 $\pm$ 0.02 &        ---       &    \\
49936 &  3.44 $\pm$ 0.05 &  2.47 $\pm$ 0.03 &  2.12 $\pm$ 0.03 &        ---       &    \\
49988 &  3.26 $\pm$ 0.03 &  2.38 $\pm$ 0.02 &  2.03 $\pm$ 0.01 &        ---       &    \\
49997 &  3.21 $\pm$ 0.02 &  2.31 $\pm$ 0.02 &  2.00 $\pm$ 0.02 &  1.74 $\pm$ 0.08 &    \\
50056 &  3.20 $\pm$ 0.02 &  2.32 $\pm$ 0.02 &  2.03 $\pm$ 0.02 &  1.74 $\pm$ 0.06 &    \\
50305 &  3.34 $\pm$ 0.04 &  2.42 $\pm$ 0.03 &  2.08 $\pm$ 0.03 &        ---       &    \\
50467 &  3.23 $\pm$ 0.02 &  2.36 $\pm$ 0.02 &  2.01 $\pm$ 0.02 &  1.72 $\pm$ 0.05 &    \\
51832 &  3.24 $\pm$ 0.01 &  2.35 $\pm$ 0.01 &  2.02 $\pm$ 0.01 &        ---       &    \\   
& & & & & \\

& & & & & \\
\multicolumn{6}{c}{\bf{IRAS 03293+6010}} \\
& & & & & \\
48362 & 13.41 $\pm$ 0.75 & 11.06 $\pm$ 0.06 &  8.09 $\pm$ 0.03 &        ---       &    \\
48686 & 13.59 $\pm$ 0.16 & 10.10 $\pm$ 0.04 &  7.20 $\pm$ 0.04 &        ---       &    \\
48944 & 13.17 $\pm$ 0.32 &  9.99 $\pm$ 0.02 &  7.12 $\pm$ 0.02 &        ---       &    \\
49320 & 13.16 $\pm$ 0.26 &  9.85 $\pm$ 0.12 &  7.14 $\pm$ 0.16 &        ---       &    \\
49325 &       ---        &  9.62 $\pm$ 0.02 &  6.93 $\pm$ 0.02 &        ---       &    \\
49328 & 12.83 $\pm$ 0.20 &  9.64 $\pm$ 0.03 &  6.93 $\pm$ 0.02 &  3.02 $\pm$ 0.06 &    \\
49385 & 12.96 $\pm$ 0.40 &  9.66 $\pm$ 0.06 &  7.02 $\pm$ 0.03 &        ---       &    \\
49415 & 12.64 $\pm$ 0.21 &  9.70 $\pm$ 0.07 &  7.02 $\pm$ 0.04 &        ---       &    \\
49737 & 13.95 $\pm$ 1.08 & 10.23 $\pm$ 0.04 &  7.56 $\pm$ 0.02 &  3.55 $\pm$ 0.10 &    \\
49936 &        ---       & 10.71 $\pm$ 0.04 &  7.92 $\pm$ 0.03 &        ---       &    \\
51366 &        ---       &        ---       &  7.34 $\pm$ 0.01 &        ---       & Imaging photometry \\
51832 & 14.49 $\pm$ 0.73 & 10.77 $\pm$ 0.07 &  7.95 $\pm$ 0.05 &        ---       &    \\
& & & & & \\

& & & & & \\
\multicolumn{6}{c}{\bf{IRAS 04130+3918}} \\
& & & & & \\
48362 &  7.15 $\pm$ 0.01 &  5.16 $\pm$ 0.02 &  3.54 $\pm$ 0.02 &        ---       &    \\
48509 &  6.47 $\pm$ 0.04 &  4.58 $\pm$ 0.02 &  3.04 $\pm$ 0.02 &        ---       &    \\
48686 &  7.69 $\pm$ 0.03 &  5.63 $\pm$ 0.03 &  3.90 $\pm$ 0.03 &        ---       &    \\
48865 &  7.58 $\pm$ 0.03 &  5.35 $\pm$ 0.02 &  3.72 $\pm$ 0.02 &        ---       &    \\
48880 &  7.37 $\pm$ 0.01 &  5.31 $\pm$ 0.01 &  3.65 $\pm$ 0.01 &        ---       &    \\
48922 &  6.82 $\pm$ 0.06 &  4.85 $\pm$ 0.02 &  3.29 $\pm$ 0.02 &        ---       &    \\
48944 &  6.59 $\pm$ 0.04 &  4.70 $\pm$ 0.01 &  3.15 $\pm$ 0.01 &  1.25 $\pm$ 0.02 &    \\
49228 &  8.08 $\pm$ 0.07 &  5.68 $\pm$ 0.03 &  4.03 $\pm$ 0.05 &  2.00 $\pm$ 0.03 &    \\
49324 &  7.41 $\pm$ 0.02 &  5.27 $\pm$ 0.02 &  3.64 $\pm$ 0.02 &        ---       &    \\
49386 &  6.91 $\pm$ 0.02 &  4.92 $\pm$ 0.02 &  3.37 $\pm$ 0.02 &  1.47 $\pm$ 0.11 &    \\
49415 &  6.54 $\pm$ 0.04 &  4.65 $\pm$ 0.03 &  3.13 $\pm$ 0.04 &  1.18 $\pm$ 0.05 &    \\
49426 &  6.57 $\pm$ 0.02 &  4.61 $\pm$ 0.02 &  3.10 $\pm$ 0.02 &  1.12 $\pm$ 0.05 &    \\
49737 &  7.76 $\pm$ 0.03 &  5.62 $\pm$ 0.02 &  3.88 $\pm$ 0.02 &  1.68 $\pm$ 0.07 &    \\
49936 &  6.75 $\pm$ 0.05 &  4.72 $\pm$ 0.03 &  3.12 $\pm$ 0.03 &        ---       &    \\
49988 &  6.87 $\pm$ 0.03 &  4.82 $\pm$ 0.02 &  3.21 $\pm$ 0.01 &        ---       &    \\
& & & & & \\

\hline
\end{tabular}
\end{center}
\hspace{12cm}{\footnotesize {\em Continued next page.}}
\end{table*}

\begin{table*}
{\bf Table A.} Summary of the near infrared photometric measurements ({\em continued})
\begin{center}
\begin{tabular}{cccccl}

\hline\hline\noalign{\smallskip}
Julian Date &  \multicolumn{4}{c}{NIR magnitude [mag]}  & Notes\\
 (2400000+)  & J & H & K & L' &  \\
\noalign{\smallskip}\hline\noalign{\smallskip}

& & & & & \\
\multicolumn{6}{c}{{\bf IRAS 04130+3918} ({\em continued})} \\
& & & & & \\
50013 &  7.06 $\pm$ 0.02 &  4.97 $\pm$ 0.02 &  3.34 $\pm$ 0.02 &        ---       &    \\
50043 &  7.30 $\pm$ 0.03 &  5.19 $\pm$ 0.02 &  3.51 $\pm$ 0.02 &        ---       &    \\
50075 &  7.64 $\pm$ 0.02 &  5.50 $\pm$ 0.02 &  3.78 $\pm$ 0.03 &        ---       &    \\
51832 &  6.34 $\pm$ 0.05 &  4.47 $\pm$ 0.03 &  3.00 $\pm$ 0.04 &        ---       &    \\
& & & & & \\

& & & & & \\
\multicolumn{6}{c}{\bf{IRAS 05131+4530}} \\
& & & & & \\
48327 & 14.13 $\pm$ 0.38 & 10.69 $\pm$ 0.05 &  8.00 $\pm$ 0.05 &        ---       &    \\
48341 & 14.59 $\pm$ 1.10 & 10.02 $\pm$ 0.30 &  7.26 $\pm$ 0.25 &        ---       &    \\
48362 & 14.71 $\pm$ 1.32 &  9.80 $\pm$ 0.03 &  7.05 $\pm$ 0.03 &        ---       &    \\
48686 & 15.01 $\pm$ 0.84 & 11.34 $\pm$ 0.07 &  8.28 $\pm$ 0.05 &        ---       &    \\
48728 & 13.05 $\pm$ 0.21 & 10.09 $\pm$ 0.05 &  7.32 $\pm$ 0.04 &        ---       &    \\
48922 & 11.46 $\pm$ 0.06 &  8.34 $\pm$ 0.02 &  5.82 $\pm$ 0.02 &        ---       &    \\
48944 & 11.38 $\pm$ 0.07 &  8.28 $\pm$ 0.02 &  5.73 $\pm$ 0.02 &        ---       &    \\
49004 & 11.20 $\pm$ 0.28 &  8.10 $\pm$ 0.25 &  5.65 $\pm$ 0.36 &        ---       &    \\
49229 & 12.15 $\pm$ 0.62 &  8.72 $\pm$ 0.07 &  6.28 $\pm$ 0.06 &        ---       &    \\
49324 &        ---       &  9.07 $\pm$ 0.04 &  6.53 $\pm$ 0.02 &        ---       &    \\
49328 & 12.50 $\pm$ 0.20 &  9.12 $\pm$ 0.03 &  6.58 $\pm$ 0.02 &  3.34 $\pm$ 0.06 &    \\
49386 & 13.07 $\pm$ 0.38 &  9.61 $\pm$ 0.03 &  6.94 $\pm$ 0.03 &        ---       &    \\
49400 & 13.53 $\pm$ 0.56 &  9.64 $\pm$ 0.04 &  6.92 $\pm$ 0.03 &        ---       &    \\
49415 & 13.12 $\pm$ 0.56 &  9.79 $\pm$ 0.05 &  7.02 $\pm$ 0.04 &        ---       &    \\
49427 & 14.37 $\pm$ 2.07 &  9.88 $\pm$ 0.05 &  7.07 $\pm$ 0.02 &        ---       &    \\
49456 & 14.44 $\pm$ 0.60 & 10.01 $\pm$ 0.03 &  7.22 $\pm$ 0.02 &        ---       &    \\
49621 &        ---       & 12.33 $\pm$ 0.10 &  8.95 $\pm$ 0.03 &        ---       &    \\
49694 & 14.09 $\pm$ 0.61 & 12.65 $\pm$ 0.68 &  8.07 $\pm$ 0.02 &  4.09 $\pm$ 0.04 &    \\
49736 &        ---       & 10.66 $\pm$ 0.03 &  7.51 $\pm$ 0.03 &        ---       &    \\
49936 & 11.28 $\pm$ 0.05 &  8.23 $\pm$ 0.03 &  5.72 $\pm$ 0.03 &        ---       &    \\
50043 &        ---       &  8.07 $\pm$ 0.04 &  5.66 $\pm$ 0.03 &        ---       &    \\
50057 & 11.09 $\pm$ 0.05 &  7.91 $\pm$ 0.03 &  5.50 $\pm$ 0.02 &        ---       &    \\
50075 &        ---       &  7.98 $\pm$ 0.03 &  5.63 $\pm$ 0.03 &        ---       &    \\
50172 &        ---       &  8.24 $\pm$ 0.02 &  5.74 $\pm$ 0.02 &        ---       &    \\
50467 & 12.69 $\pm$ 0.20 &  9.31 $\pm$ 0.03 &  6.71 $\pm$ 0.02 &        ---       &    \\
51832 & 13.86 $\pm$ 0.94 & 11.38 $\pm$ 0.12 &  8.26 $\pm$ 0.05 &        ---       &    \\
52041 & 12.38 $\pm$ 0.03 &  8.55 $\pm$ 0.02 &  6.11 $\pm$ 0.03 &        ---       & Imaging photometry \\
& & & & & \\ 

& & & & & \\
\multicolumn{6}{c}{\bf{IRAS 06297+4045}} \\
& & & & & \\
48327 &  5.51 $\pm$ 0.02 &  4.00 $\pm$ 0.01 &  2.79 $\pm$ 0.01 &        ---       &    \\
48341 &  5.44 $\pm$ 0.01 &  3.83 $\pm$ 0.01 &  2.62 $\pm$ 0.01 &        ---       &    \\
48355 &  5.41 $\pm$ 0.01 &  3.80 $\pm$ 0.01 &  2.63 $\pm$ 0.02 &        ---       &    \\
48362 &  5.38 $\pm$ 0.01 &  3.77 $\pm$ 0.01 &  2.58 $\pm$ 0.02 &        ---       &    \\
48374 &  5.37 $\pm$ 0.01 &  3.78 $\pm$ 0.02 &  2.61 $\pm$ 0.02 &        ---       &    \\
48664 &  7.13 $\pm$ 0.02 &  5.18 $\pm$ 0.02 &  3.71 $\pm$ 0.02 &        ---       &    \\
48686 &  7.35 $\pm$ 0.02 &  5.33 $\pm$ 0.02 &  3.80 $\pm$ 0.03 &        ---       &    \\
48728 &  7.59 $\pm$ 0.04 &  5.50 $\pm$ 0.03 &  3.90 $\pm$ 0.03 &        ---       &    \\
48865 &  5.65 $\pm$ 0.04 &  3.82 $\pm$ 0.02 &  2.68 $\pm$ 0.03 &        ---       &    \\
48922 &  5.22 $\pm$ 0.06 &  3.59 $\pm$ 0.02 &  2.48 $\pm$ 0.02 &        ---       &    \\
48944 &  5.20 $\pm$ 0.04 &  3.64 $\pm$ 0.01 &  2.50 $\pm$ 0.01 &  1.01 $\pm$ 0.02 &    \\
49320 &  6.30 $\pm$ 0.03 &  4.48 $\pm$ 0.08 &  3.12 $\pm$ 0.13 &        ---       &    \\
49324 &  6.18 $\pm$ 0.02 &  4.36 $\pm$ 0.02 &  3.03 $\pm$ 0.02 &        ---       &    \\
49385 &  5.41 $\pm$ 0.03 &  3.75 $\pm$ 0.02 &  2.60 $\pm$ 0.02 &  1.06 $\pm$ 0.04 &    \\
49415 &  5.36 $\pm$ 0.03 &  3.75 $\pm$ 0.03 &  2.58 $\pm$ 0.03 &  1.00 $\pm$ 0.06 &    \\
49426 &  5.30 $\pm$ 0.03 &  3.64 $\pm$ 0.02 &  2.50 $\pm$ 0.02 &  0.94 $\pm$ 0.05 &    \\
49456 &  5.16 $\pm$ 0.02 &  3.54 $\pm$ 0.02 &  2.44 $\pm$ 0.01 &  0.95 $\pm$ 0.07 &    \\
49597 &  5.76 $\pm$ 0.02 &  4.10 $\pm$ 0.01 &  2.92 $\pm$ 0.02 &        ---       &    \\
49621 &  5.89 $\pm$ 0.01 &  4.21 $\pm$ 0.01 &  3.05 $\pm$ 0.01 &        ---       &    \\
& & & & & \\

\hline
\end{tabular}
\end{center}
\hspace{12cm}{\footnotesize {\em Continued next page.}}
\end{table*}

\begin{table*}
{\bf Table A.} Summary of the near infrared photometric measurements ({\em continued})
\begin{center}
\begin{tabular}{cccccl}

\hline\hline\noalign{\smallskip}
Julian Date &  \multicolumn{4}{c}{NIR magnitude [mag]}  & Notes\\
 (2400000+)  & J & H & K & L' &  \\
\noalign{\smallskip}\hline\noalign{\smallskip}

& & & & & \\
\multicolumn{6}{c}{{\bf IRAS 06297+4045} ({\em continued})} \\
& & & & & \\
49694 &  6.61 $\pm$ 0.02 &  4.76 $\pm$ 0.02 &  3.44 $\pm$ 0.02 &  1.83 $\pm$ 0.01 &    \\
49736 &  6.94 $\pm$ 0.02 &  5.00 $\pm$ 0.01 &  3.57 $\pm$ 0.02 &        ---       &    \\
49840 &  6.56 $\pm$ 0.01 &  4.75 $\pm$ 0.01 &  3.31 $\pm$ 0.03 &        ---       &    \\
51548 &  5.19 $\pm$ 0.02 &  3.51 $\pm$ 0.06 &  2.37 $\pm$ 0.04 &        ---       &    \\
51860 &  6.26 $\pm$ 0.01 &  4.56 $\pm$ 0.01 &  3.30 $\pm$ 0.01 &        ---       &    \\
52041 &  5.79 $\pm$ 0.01 &  4.42 $\pm$ 0.03 &  2.43 $\pm$ 0.05 &        ---       & Imaging photometry \\
& & & & & \\

& & & & & \\
\multicolumn{6}{c}{\bf{IRAS 07222$-$2005} } \\
& & & & & \\
48327 &  5.78 $\pm$ 0.02 &  4.86 $\pm$ 0.02 &  4.10 $\pm$ 0.02 &        ---       &    \\
48341 &  5.98 $\pm$ 0.02 &  4.69 $\pm$ 0.01 &  3.98 $\pm$ 0.02 &        ---       &    \\
48354 &  5.99 $\pm$ 0.01 &  4.67 $\pm$ 0.02 &  3.97 $\pm$ 0.02 &        ---       &    \\
48664 &  6.56 $\pm$ 0.02 &  5.37 $\pm$ 0.03 &  4.54 $\pm$ 0.02 &        ---       &    \\
48686 &  6.64 $\pm$ 0.03 &  5.42 $\pm$ 0.03 &  4.61 $\pm$ 0.03 &        ---       &    \\
48728 &  6.78 $\pm$ 0.05 &  5.59 $\pm$ 0.03 &  4.76 $\pm$ 0.04 &        ---       &    \\
48740 &  6.87 $\pm$ 0.01 &  5.65 $\pm$ 0.01 &  4.78 $\pm$ 0.01 &        ---       &    \\
48944 &  7.27 $\pm$ 0.05 &  6.07 $\pm$ 0.02 &  5.12 $\pm$ 0.02 &        ---       &    \\
49004 &  7.19 $\pm$ 0.22 &  5.95 $\pm$ 0.25 &  5.23 $\pm$ 0.36 &        ---       &    \\
49385 &  6.49 $\pm$ 0.05 &  5.40 $\pm$ 0.04 &  4.62 $\pm$ 0.03 &  3.25 $\pm$ 0.06 &    \\
49415 &  6.40 $\pm$ 0.04 &  5.24 $\pm$ 0.03 &  4.45 $\pm$ 0.04 &  3.41 $\pm$ 0.06 &    \\
49426 &  6.31 $\pm$ 0.03 &  5.13 $\pm$ 0.03 &  4.38 $\pm$ 0.02 &  3.14 $\pm$ 0.06 &    \\
49456 &  6.30 $\pm$ 0.02 &  5.11 $\pm$ 0.02 &  4.36 $\pm$ 0.02 &  3.37 $\pm$ 0.08 &    \\
49694 &  6.25 $\pm$ 0.03 &  4.92 $\pm$ 0.03 &  4.18 $\pm$ 0.02 &  3.26 $\pm$ 0.01 &    \\
49736 &  6.35 $\pm$ 0.02 &  5.04 $\pm$ 0.01 &  4.32 $\pm$ 0.02 &        ---       &    \\
50044 &  7.30 $\pm$ 0.04 &  6.05 $\pm$ 0.04 &  5.15 $\pm$ 0.03 &        ---       &    \\
51832 &  7.03 $\pm$ 0.06 &  5.94 $\pm$ 0.05 &  5.00 $\pm$ 0.06 &        ---       &    \\
& & & & & \\

& & & & & \\
\multicolumn{6}{c}{\bf{IRAS 16029$-$3041}} \\
& & & & & \\
48341 & 13.32 $\pm$ 0.32 &  9.86 $\pm$ 0.02 &  6.54 $\pm$ 0.02 &        ---       &    \\
48374 & 13.57 $\pm$ 0.41 & 10.15 $\pm$ 0.05 &  6.79 $\pm$ 0.02 &        ---       &    \\
48740 & 12.10 $\pm$ 0.03 &  8.91 $\pm$ 0.02 &  5.71 $\pm$ 0.01 &        ---       &    \\
48771 & 12.88 $\pm$ 0.08 &  9.60 $\pm$ 0.05 &  6.42 $\pm$ 0.05 &        ---       &    \\
49840 & 12.55 $\pm$ 0.03 &  9.27 $\pm$ 0.01 &  5.66 $\pm$ 0.03 &        ---       &    \\
49935 & 12.84 $\pm$ 0.11 &  9.59 $\pm$ 0.02 &  5.97 $\pm$ 0.02 &        ---       &    \\
50174 &        ---       &        ---       &  8.41 $\pm$ 0.02 &        ---       &    \\
50247 &        ---       &        ---       &  7.84 $\pm$ 0.02 &        ---       &    \\
50248 &        ---       & 12.06 $\pm$ 0.05 &  7.83 $\pm$ 0.02 &  2.33 $\pm$ 0.05 &    \\
50976 & 14.21 $\pm$ 0.40 & 10.44 $\pm$ 0.01 &  6.49 $\pm$ 0.01 &        ---       &    \\
51369 & 13.99 $\pm$ 0.02 & 11.20 $\pm$ 0.10 &  7.10 $\pm$ 0.01 &        ---       &    \\
51752 &        ---       &        ---       &  7.94 $\pm$ 0.04 &        ---       & Imaging photometry \\
52043 &        ---       & 10.46 $\pm$ 0.06 &  6.47 $\pm$ 0.06 &        ---       & Imaging photometry* \\
52087 &        ---       & 11.38 $\pm$ 0.05 &  6.89 $\pm$ 0.04 &        ---       &    \\
& & & & & \\

& & & & & \\
\multicolumn{6}{c}{\bf{IRAS 16037$+$4218}} \\
& & & & & \\
48327 &  4.62 $\pm$ 0.01 &  3.80 $\pm$ 0.01 &  3.36 $\pm$ 0.01 &        ---       &    \\
48341 &  4.42 $\pm$ 0.02 &  3.64 $\pm$ 0.01 &  3.18 $\pm$ 0.01 &        ---       &    \\
48355 &  4.44 $\pm$ 0.01 &  3.65 $\pm$ 0.02 &  3.19 $\pm$ 0.01 &        ---       &    \\
48374 &  4.42 $\pm$ 0.01 &  3.63 $\pm$ 0.02 &  3.17 $\pm$ 0.01 &        ---       &    \\
48484 &  5.15 $\pm$ 0.03 &  4.37 $\pm$ 0.10 &  4.17 $\pm$ 0.16 &        ---       &    \\
48485 &  4.97 $\pm$ 0.02 &  4.22 $\pm$ 0.03 &  3.66 $\pm$ 0.04 &        ---       &    \\
48499 &  5.10 $\pm$ 0.05 &  4.31 $\pm$ 0.07 &  3.74 $\pm$ 0.07 &        ---       &    \\
48509 &  5.21 $\pm$ 0.04 &  4.45 $\pm$ 0.02 &  3.83 $\pm$ 0.04 &        ---       &    \\
48664 &  4.76 $\pm$ 0.01 &  3.96 $\pm$ 0.01 &  3.45 $\pm$ 0.01 &        ---       &    \\
& & & & & \\

\hline
\end{tabular}
\end{center}
\hspace{12cm}{\footnotesize {\em Continued next page.}}
\end{table*}

\begin{table*}
{\bf Table A.} Summary of the near infrared photometric measurements ({\em continued})
\begin{center}
\begin{tabular}{cccccl}

\hline\hline\noalign{\smallskip}
Julian Date &  \multicolumn{4}{c}{NIR magnitude [mag]}  & Notes\\
 (2400000+)  & J & H & K & L' &  \\
\noalign{\smallskip}\hline\noalign{\smallskip}

& & & & & \\
\multicolumn{6}{c}{{\bf IRAS 16037$+$4218} ({\em continued})}  \\
& & & & & \\
48686 &  4.68 $\pm$ 0.02 &  3.87 $\pm$ 0.02 &  3.36 $\pm$ 0.02 &        ---       &    \\
48728 &  4.58 $\pm$ 0.04 &  3.76 $\pm$ 0.02 &  3.25 $\pm$ 0.03 &        ---       &    \\
48770 &  4.69 $\pm$ 0.02 &  3.89 $\pm$ 0.01 &  3.36 $\pm$ 0.01 &        ---       &    \\
48824 &  4.99 $\pm$ 0.01 &  4.21 $\pm$ 0.01 &  3.63 $\pm$ 0.01 &        ---       &    \\
48880 &  5.39 $\pm$ 0.02 &  4.56 $\pm$ 0.02 &  3.94 $\pm$ 0.02 &        ---       &    \\
48921 &  5.44 $\pm$ 0.07 &  4.60 $\pm$ 0.03 &  3.96 $\pm$ 0.02 &        ---       &    \\
49002 &  4.94 $\pm$ 0.03 &  4.17 $\pm$ 0.03 &  3.64 $\pm$ 0.01 &  2.78 $\pm$ 0.09 &    \\
49226 &  5.48 $\pm$ 0.02 &  4.58 $\pm$ 0.02 &  3.96 $\pm$ 0.01 &        ---       &    \\
49386 &  4.89 $\pm$ 0.01 &  4.17 $\pm$ 0.01 &  3.64 $\pm$ 0.01 &  2.75 $\pm$ 0.09 &    \\
49400 &  4.75 $\pm$ 0.02 &  4.01 $\pm$ 0.02 &  3.53 $\pm$ 0.01 &  2.51 $\pm$ 0.03 &    \\
49427 &  4.69 $\pm$ 0.02 &  3.89 $\pm$ 0.02 &  3.38 $\pm$ 0.01 &  2.60 $\pm$ 0.03 &    \\
49439 &  4.63 $\pm$ 0.03 &  3.83 $\pm$ 0.04 &  3.32 $\pm$ 0.02 &        ---       &    \\
49456 &  4.72 $\pm$ 0.01 &  3.82 $\pm$ 0.01 &  3.37 $\pm$ 0.01 &  2.64 $\pm$ 0.06 &    \\
49500 &  4.86 $\pm$ 0.02 &  4.03 $\pm$ 0.01 &  3.64 $\pm$ 0.02 &        ---       &    \\
49597 &  5.51 $\pm$ 0.02 &  4.68 $\pm$ 0.01 &  4.03 $\pm$ 0.02 &        ---       &    \\
49608 &  5.52 $\pm$ 0.02 &  4.70 $\pm$ 0.01 &  4.05 $\pm$ 0.01 &  3.04 $\pm$ 0.15 &    \\
49735 &  4.81 $\pm$ 0.02 &  4.03 $\pm$ 0.03 &  3.52 $\pm$ 0.06 &        ---       &    \\
49839 &  4.56 $\pm$ 0.01 &  3.78 $\pm$ 0.01 &  3.33 $\pm$ 0.03 &        ---       &    \\
50201 &  4.68 $\pm$ 0.02 &  3.90 $\pm$ 0.02 &  3.40 $\pm$ 0.02 &        ---       &    \\
50226 &  4.95 $\pm$ 0.02 &  4.15 $\pm$ 0.02 &  3.59 $\pm$ 0.02 &        ---       &    \\
50232 &  4.95 $\pm$ 0.02 &  4.23 $\pm$ 0.02 &  3.66 $\pm$ 0.02 &  2.84 $\pm$ 0.04 &    \\
50976 &  5.07 $\pm$ 0.01 &  4.33 $\pm$ 0.01 &  3.76 $\pm$ 0.01 &        ---       &    \\
51360 &  5.32 $\pm$ 0.04 &  4.64 $\pm$ 0.05 &        ---       &        ---       & Imaging photometry \\
51364 &  5.28 $\pm$ 0.02 &  4.50 $\pm$ 0.01 &  3.87 $\pm$ 0.01 &        ---       &    \\
51367 &        ---       &        ---       &  3.83 $\pm$ 0.13 &        ---       & Imaging photometry \\
51745 &  5.44 $\pm$ 0.01 &  4.57 $\pm$ 0.01 &  3.89 $\pm$ 0.09 &        ---       & Imaging photometry \\
52041 &  5.47 $\pm$ 0.04 &  4.69 $\pm$ 0.03 &  3.59 $\pm$ 0.04 &        ---       & Imaging photometry* \\
52117 &  5.60 $\pm$ 0.01 &  4.87 $\pm$ 0.01 &  4.21 $\pm$ 0.01 &        ---       &    \\
52118 &  5.58 $\pm$ 0.07 &  4.84 $\pm$ 0.04 &  4.18 $\pm$ 0.05 &        ---       &    \\
& & & & & \\

& & & & & \\
\multicolumn{6}{c}{\bf{IRAS 16437$-$3140}} \\
& & & & & \\
48355 &        ---       & 12.38 $\pm$ 0.13 &  8.91 $\pm$ 0.03 &        ---       &    \\
48374 &        ---       & 12.76 $\pm$ 0.49 &  9.06 $\pm$ 0.03 &        ---       &    \\
49860 &        ---       &  9.66 $\pm$ 0.03 &  6.52 $\pm$ 0.02 &  2.44 $\pm$ 0.75 &    \\
50226 &        ---       & 12.52 $\pm$ 0.14 &  9.08 $\pm$ 0.02 &        ---       &    \\
50976 &        ---       & 12.79 $\pm$ 0.09 &  9.73 $\pm$ 0.01 &        ---       &    \\
51369 &        ---       & 11.40 $\pm$ 0.15 &  8.38 $\pm$ 0.01 &        ---       &    \\
51745 &        ---       & 10.86 $\pm$ 0.06 &  7.56 $\pm$ 0.03 &        ---       & Imaging photometry \\
52043 &        ---       & 12.00 $\pm$ 0.07 &  8.56 $\pm$ 0.05 &        ---       & Imaging photometry* \\
& & & & & \\

& & & & & \\
\multicolumn{6}{c}{\bf{IRAS 17103$-$0559}} \\
& & & & & \\
48327 &  6.01 $\pm$ 0.01 &  4.74 $\pm$ 0.01 &  3.68 $\pm$ 0.01 &        ---       &    \\
48341 &  5.69 $\pm$ 0.02 &  4.37 $\pm$ 0.01 &  3.42 $\pm$ 0.01 &        ---       &    \\
48355 &  5.35 $\pm$ 0.01 &  4.11 $\pm$ 0.01 &  3.23 $\pm$ 0.02 &        ---       &    \\
48374 &  5.11 $\pm$ 0.01 &  3.89 $\pm$ 0.02 &  3.07 $\pm$ 0.02 &        ---       &    \\
48441 &  4.91 $\pm$ 0.04 &  3.62 $\pm$ 0.04 &  2.82 $\pm$ 0.04 &        ---       &    \\
48486 &  4.97 $\pm$ 0.03 &  3.63 $\pm$ 0.04 &  2.83 $\pm$ 0.03 &        ---       &    \\
48740 &  6.23 $\pm$ 0.01 &  4.70 $\pm$ 0.01 &  3.60 $\pm$ 0.01 &        ---       &    \\
48771 &  5.85 $\pm$ 0.11 &  4.53 $\pm$ 0.12 &  3.51 $\pm$ 0.11 &        ---       &    \\
48824 &  5.22 $\pm$ 0.02 &  3.96 $\pm$ 0.01 &  3.06 $\pm$ 0.02 &        ---       &    \\
48865 &  5.02 $\pm$ 0.06 &  3.61 $\pm$ 0.03 &  2.82 $\pm$ 0.03 &        ---       &    \\
48880 &  4.86 $\pm$ 0.01 &  3.59 $\pm$ 0.01 &  2.79 $\pm$ 0.01 &        ---       &    \\
& & & & & \\

\hline
\end{tabular}
\end{center}
\hspace{12cm}{\footnotesize {\em Continued next page.}}
\end{table*}

\begin{table*}
{\bf Table A.} Summary of the near infrared photometric measurements ({\em continued})
\begin{center}
\begin{tabular}{cccccl}

\hline\hline\noalign{\smallskip}
Julian Date &  \multicolumn{4}{c}{NIR magnitude [mag]}  & Notes\\
 (2400000+)  & J & H & K & L' &  \\
\noalign{\smallskip}\hline\noalign{\smallskip}

& & & & & \\
\multicolumn{6}{c}{{\bf IRAS 17103$-$0559} ({\em continued})}  \\
& & & & & \\
49228 &  5.96 $\pm$ 0.09 &  4.50 $\pm$ 0.03 &        ---       &        ---       &    \\
49405 &  5.81 $\pm$ 0.05 &  4.33 $\pm$ 0.04 &  3.39 $\pm$ 0.02 &  2.08 $\pm$ 0.08 &    \\
49840 &  5.82 $\pm$ 0.01 &  4.36 $\pm$ 0.01 &  3.34 $\pm$ 0.03 &        ---       &    \\
49935 &  7.17 $\pm$ 0.02 &  5.59 $\pm$ 0.02 &  4.35 $\pm$ 0.02 &        ---       &    \\
49952 &  7.54 $\pm$ 0.02 &  5.95 $\pm$ 0.02 &  4.68 $\pm$ 0.02 &        ---       &    \\
50174 &  6.28 $\pm$ 0.03 &  4.97 $\pm$ 0.03 &  3.86 $\pm$ 0.03 &        ---       &    \\
50201 &  6.23 $\pm$ 0.02 &  4.89 $\pm$ 0.02 &  3.81 $\pm$ 0.02 &        ---       &    \\
50232 &  6.31 $\pm$ 0.02 &  4.99 $\pm$ 0.02 &  3.90 $\pm$ 0.02 &  2.22 $\pm$ 0.04 &    \\
50247 &  7.08 $\pm$ 0.10 &  5.39 $\pm$ 0.02 &  4.20 $\pm$ 0.03 &        ---       &    \\
50976 &  7.26 $\pm$ 0.01 &  5.92 $\pm$ 0.01 &  4.55 $\pm$ 0.01 &        ---       &    \\
51365 &        ---       &        ---       &  5.66 $\pm$ 0.10 &        ---       & Imaging photometry* \\
51751 &  9.39 $\pm$ 0.15 &  7.59 $\pm$ 0.14 &        ---       &        ---       & Imaging photometry* \\
52041 &  7.89 $\pm$ 0.03 &  6.03 $\pm$ 0.05 &        ---       &        ---       & Imaging photometry* \\

& & & & & \\

& & & & & \\
\multicolumn{6}{c}{\bf{IRAS 17105$-$2804}} \\
& & & & & \\
48327 &  6.35 $\pm$ 0.01 &  5.61 $\pm$ 0.01 &  4.89 $\pm$ 0.01 &        ---       &    \\
48341 &  6.68 $\pm$ 0.02 &  5.56 $\pm$ 0.01 &  4.80 $\pm$ 0.01 &        ---       &    \\
48355 &  6.79 $\pm$ 0.01 &  5.64 $\pm$ 0.02 &  4.87 $\pm$ 0.02 &        ---       &    \\
48374 &  6.98 $\pm$ 0.01 &  5.81 $\pm$ 0.02 &  5.01 $\pm$ 0.02 &        ---       &    \\
48441 &  7.57 $\pm$ 0.04 &  6.34 $\pm$ 0.04 &  5.36 $\pm$ 0.04 &        ---       &    \\
48838 &  7.09 $\pm$ 0.03 &  5.97 $\pm$ 0.04 &  5.07 $\pm$ 0.05 &        ---       &    \\
49860 &  5.79 $\pm$ 0.02 &  4.73 $\pm$ 0.03 &  3.99 $\pm$ 0.02 &        ---       &    \\
50247 &  6.26 $\pm$ 0.02 &  5.27 $\pm$ 0.02 &  4.52 $\pm$ 0.02 &        ---       &    \\
50301 &  5.67 $\pm$ 0.02 &  4.71 $\pm$ 0.04 &  4.07 $\pm$ 0.05 &        ---       &    \\
50976 &  6.69 $\pm$ 0.01 &  5.71 $\pm$ 0.01 &  4.93 $\pm$ 0.01 &        ---       &    \\
51310 &  7.07 $\pm$ 0.01 &  6.06 $\pm$ 0.01 &  5.24 $\pm$ 0.02 &        ---       &    \\
51364 &  7.19 $\pm$ 0.02 &  6.14 $\pm$ 0.01 &  5.28 $\pm$ 0.01 &        ---       &    \\
51751 &        ---       &        ---       &  4.90 $\pm$ 0.04 &        ---       & Imaging photometry* \\
52041 &  6.55 $\pm$ 0.02 &  5.42 $\pm$ 0.02 &  4.23 $\pm$ 0.02 &        ---       & Imaging photometry* \\
& & & & & \\

& & & & & \\
\multicolumn{6}{c}{\bf{IRAS 17313$-$1531}} \\
& & & & & \\
48327 &  7.60 $\pm$ 0.01 &  5.66 $\pm$ 0.01 &  4.27 $\pm$ 0.01 &        ---       &    \\
48341 &  7.74 $\pm$ 0.02 &  5.70 $\pm$ 0.01 &  4.29 $\pm$ 0.01 &        ---       &    \\
48374 &  8.01 $\pm$ 0.01 &  5.96 $\pm$ 0.02 &  4.54 $\pm$ 0.02 &        ---       &    \\
48441 &  8.78 $\pm$ 0.04 &  6.60 $\pm$ 0.04 &  4.99 $\pm$ 0.04 &        ---       &    \\
48838 &  7.91 $\pm$ 0.03 &  5.75 $\pm$ 0.04 &  4.20 $\pm$ 0.05 &        ---       &    \\
49935 &  9.40 $\pm$ 0.02 &  7.16 $\pm$ 0.02 &  5.35 $\pm$ 0.02 &        ---       &    \\
50269 &  8.02 $\pm$ 0.02 &  6.06 $\pm$ 0.02 &  4.58 $\pm$ 0.02 &        ---       &    \\
50301 &  8.24 $\pm$ 0.02 &  6.31 $\pm$ 0.04 &  4.79 $\pm$ 0.05 &        ---       &    \\
50976 &  8.85 $\pm$ 0.01 &  6.89 $\pm$ 0.01 &  5.21 $\pm$ 0.01 &        ---       &    \\
51364 &  9.60 $\pm$ 0.03 &  7.41 $\pm$ 0.01 &  5.71 $\pm$ 0.01 &        ---       &    \\
51365 &       ---        &       ---        &  5.70 $\pm$ 0.09 &        ---       & Imaging photometry*\\
52494 &  8.06 $\pm$ 0.02 &  5.91 $\pm$ 0.10 &  4.54 $\pm$ 0.01 &        ---       & Imaging photometry*\\
& & & & & \\

& & & & & \\
\multicolumn{6}{c}{\bf{IRAS 17347$-$3139}} \\
& & & & & \\
51753 &        ---       & 12.66 $\pm$ 0.08 & 10.12 $\pm$ 0.07 &        ---       & Imaging photometry \\
52041 &        ---       & 13.12 $\pm$ 0.07 & 10.23 $\pm$ 0.06 &        ---       & Imaging photometry*\\
52494 &        ---       & 12.78 $\pm$ 0.04 & 10.13 $\pm$ 0.04 &        ---       & Imaging photometry*\\
& & & & & \\

\hline
\end{tabular}
\end{center}
\hspace{12cm}{\footnotesize {\em Continued next page.}}
\end{table*}

\begin{table*}
{\bf Table A.} Summary of the near infrared photometric measurements ({\em continued})
\begin{center}
\begin{tabular}{cccccl}

\hline\hline\noalign{\smallskip}
Julian Date &  \multicolumn{4}{c}{NIR magnitude [mag]}  & Notes\\
 (2400000+)  & J & H & K & L' &  \\
\noalign{\smallskip}\hline\noalign{\smallskip}

& & & & & \\
\multicolumn{6}{c}{\bf{IRAS 17411$-$3154}} \\
& & & & & \\
49860 &        ---       &        ---       &        ---       &  3.82 $\pm$ 0.11 &    \\
& & & & & \\

& & & & & \\
\multicolumn{6}{c}{\bf{IRAS 18025$-$2113}} \\
& & & & & \\
48824 &  4.28 $\pm$ 0.02 &  3.16 $\pm$ 0.01 &  2.51 $\pm$ 0.02 &        ---       &    \\
48838 &  4.32 $\pm$ 0.03 &  3.17 $\pm$ 0.04 &  2.45 $\pm$ 0.04 &        ---       &    \\
48865 &  4.43 $\pm$ 0.05 &  3.18 $\pm$ 0.03 &  2.53 $\pm$ 0.03 &        ---       &    \\
50204 &  4.22 $\pm$ 0.05 &  3.14 $\pm$ 0.05 &  2.50 $\pm$ 0.05 &        ---       &    \\
50300 &  3.99 $\pm$ 0.04 &  2.92 $\pm$ 0.05 &  2.27 $\pm$ 0.04 &        ---       &    \\
50976 &  4.34 $\pm$ 0.01 &  3.22 $\pm$ 0.01 &  2.55 $\pm$ 0.01 &        ---       &    \\
51364 &  4.37 $\pm$ 0.02 &  3.32 $\pm$ 0.01 &  2.67 $\pm$ 0.01 &        ---       &    \\
51365 &        ---       &        ---       &  2.92 $\pm$ 0.21 &        ---       & Imaging photometry \\
51384 &  4.42 $\pm$ 0.01 &  3.36 $\pm$ 0.01 &  2.73 $\pm$ 0.01 &        ---       &    \\
& & & & & \\

& & & & & \\
\multicolumn{6}{c}{\bf{IRAS 18299$-$1705}} \\
& & & & & \\
49229 &  8.13 $\pm$ 0.05 &  6.45 $\pm$ 0.03 &  5.00 $\pm$ 0.03 &        ---       &    \\
49936 &  8.09 $\pm$ 0.05 &  6.55 $\pm$ 0.03 &  5.23 $\pm$ 0.03 &        ---       &    \\
50232 &  8.22 $\pm$ 0.02 &  6.65 $\pm$ 0.02 &  5.31 $\pm$ 0.02 &        ---       &    \\
50327 &  8.29 $\pm$ 0.02 &  6.69 $\pm$ 0.02 &  5.33 $\pm$ 0.02 &        ---       &    \\
50976 &  8.00 $\pm$ 0.01 &  6.38 $\pm$ 0.01 &  5.02 $\pm$ 0.01 &        ---       &    \\
51364 &  8.18 $\pm$ 0.02 &  6.62 $\pm$ 0.01 &  5.30 $\pm$ 0.01 &        ---       &    \\
51365 &        ---       &        ---       &  5.26 $\pm$ 0.02 &        ---       & Imaging photometry* \\
51753 &  8.18 $\pm$ 0.03 &  6.39 $\pm$ 0.02 &  5.12 $\pm$ 0.01 &        ---       & Imaging photometry \\
52494 &  8.14 $\pm$ 0.02 &  6.44 $\pm$ 0.02 &  5.18 $\pm$ 0.01 &        ---       & Imaging photometry*\\
& & & & & \\

& & & & & \\
\multicolumn{6}{c}{\bf{IRAS 18314$-$1131}} \\
& & & & & \\
49500 &  6.60 $\pm$ 0.03 &  4.61 $\pm$ 0.02 &  3.28 $\pm$ 0.03 &        ---       &    \\
49597 &  5.69 $\pm$ 0.05 &  4.10 $\pm$ 0.01 &  3.36 $\pm$ 0.02 &        ---       &    \\
49936 &  5.41 $\pm$ 0.05 &  3.82 $\pm$ 0.03 &  2.86 $\pm$ 0.03 &        ---       &    \\
49953 &  5.67 $\pm$ 0.03 &  3.99 $\pm$ 0.03 &  3.02 $\pm$ 0.03 &        ---       &    \\
50234 &  6.98 $\pm$ 0.02 &  5.28 $\pm$ 0.02 &  3.98 $\pm$ 0.02 &        ---       &    \\
50306 &  5.48 $\pm$ 0.03 &  4.09 $\pm$ 0.02 &  3.12 $\pm$ 0.02 &  1.39 $\pm$ 0.06 &    \\
50976 &  6.21 $\pm$ 0.01 &  4.65 $\pm$ 0.01 &  3.48 $\pm$ 0.01 &        ---       &    \\
51364 &  6.26 $\pm$ 0.02 &  4.70 $\pm$ 0.01 &  3.41 $\pm$ 0.01 &        ---       &    \\
51365 &        ---       &        ---       &  3.31 $\pm$ 0.04 &        ---       & Imaging photometry* \\
51753 &  8.67 $\pm$ 0.02 &  6.09 $\pm$ 0.02 &  4.56 $\pm$ 0.06 &        ---       & Imaging photometry \\
& & & & & \\

& & & & & \\
\multicolumn{6}{c}{\bf{IRAS 18327$-$0715}} \\
& & & & & \\
51360 &        ---       &        ---       &  8.06 $\pm$ 0.04 &        ---       & Imaging photometry* \\
51753 &        ---       &        ---       &  8.85 $\pm$ 0.03 &        ---       & Imaging photometry \\
& & & & & \\

& & & & & \\
\multicolumn{6}{c}{\bf{IRAS 18429$-$1721}} \\
& & & & & \\
49597 &  3.29 $\pm$ 0.02 &  2.33 $\pm$ 0.01 &  1.75 $\pm$ 0.01 &        ---       &    \\
49936 &  4.26 $\pm$ 0.05 &  3.41 $\pm$ 0.03 &  2.61 $\pm$ 0.03 &        ---       &    \\
50234 &  4.56 $\pm$ 0.02 &  3.54 $\pm$ 0.02 &  2.77 $\pm$ 0.02 &  1.50 $\pm$ 0.06 &    \\
50305 &  4.53 $\pm$ 0.04 &  3.52 $\pm$ 0.03 &  2.72 $\pm$ 0.03 &        ---       &    \\
50976 &  3.25 $\pm$ 0.01 &  2.30 $\pm$ 0.01 &  1.69 $\pm$ 0.01 &        ---       &    \\
51364 &  4.38 $\pm$ 0.02 &  3.45 $\pm$ 0.01 &  2.61 $\pm$ 0.01 &        ---       &    \\
51753 &  4.93 $\pm$ 0.06 &  3.62 $\pm$ 0.04 &  2.93 $\pm$ 0.04 &        ---       & Imaging photometry \\
& & & & & \\ 

\hline

\end{tabular}
\end{center}
\hspace{12cm}{\footnotesize {\em Continued next page.}}
\end{table*}

\begin{table*}
{\bf Table A.} Summary of the near infrared photometric measurements ({\em continued})
\begin{center}
\begin{tabular}{cccccl}

\hline\hline\noalign{\smallskip}
Julian Date &  \multicolumn{4}{c}{NIR magnitude [mag]}  & Notes\\
 (2400000+)  & J & H & K & L' &  \\
\noalign{\smallskip}\hline\noalign{\smallskip}

& & & & & \\
\multicolumn{6}{c}{\bf{IRAS 18454$-$1226}} \\
& & & & & \\
48481 &  3.39 $\pm$ 0.04 &  2.37 $\pm$ 0.04 &  1.97 $\pm$ 0.03 &        ---       &    \\
48499 &  3.38 $\pm$ 0.05 &  2.35 $\pm$ 0.07 &  1.93 $\pm$ 0.07 &        ---       &    \\
49942 &  3.47 $\pm$ 0.04 &  2.43 $\pm$ 0.03 &  2.01 $\pm$ 0.03 &        ---       &    \\
49952 &  3.40 $\pm$ 0.02 &  2.37 $\pm$ 0.01 &  1.95 $\pm$ 0.02 &        ---       &    \\
50193 &  3.29 $\pm$ 0.10 &  2.31 $\pm$ 0.11 &  1.99 $\pm$ 0.10 &        ---       &    \\
50234 &  3.45 $\pm$ 0.02 &  2.40 $\pm$ 0.02 &  2.01 $\pm$ 0.02 &  1.55 $\pm$ 0.06 &    \\
50269 &  3.46 $\pm$ 0.02 &  2.43 $\pm$ 0.02 &  2.00 $\pm$ 0.02 &  1.62 $\pm$ 0.05 &    \\
50301 &  3.46 $\pm$ 0.02 &  2.41 $\pm$ 0.04 &  2.00 $\pm$ 0.05 &        ---       &    \\
& & & & & \\

& & & & & \\
\multicolumn{6}{c}{\bf{IRAS 18576$+$0341}} \\
& & & & & \\
48486 & 12.03 $\pm$ 0.11 &  8.83 $\pm$ 0.04 &  6.86 $\pm$ 0.04 &        ---       &    \\
48830 & 12.78 $\pm$ 0.05 &  9.35 $\pm$ 0.03 &  7.37 $\pm$ 0.03 &        ---       &    \\
48945 & 12.82 $\pm$ 0.03 &  9.26 $\pm$ 0.02 &  7.27 $\pm$ 0.02 &        ---       &    \\
49228 &        ---       &  9.39 $\pm$ 0.05 &  7.55 $\pm$ 0.08 &        ---       &    \\
49936 & 11.92 $\pm$ 0.07 &  8.56 $\pm$ 0.03 &  6.62 $\pm$ 0.03 &        ---       &    \\
49953 & 12.49 $\pm$ 0.15 &  8.67 $\pm$ 0.03 &  6.69 $\pm$ 0.03 &        ---       &    \\
49989 &        ---       &  8.41 $\pm$ 0.03 &  6.49 $\pm$ 0.02 &        ---       &    \\
50201 & 11.36 $\pm$ 0.15 &  8.38 $\pm$ 0.02 &  6.50 $\pm$ 0.02 &        ---       &    \\
50300 &        ---       &  8.55 $\pm$ 0.05 &  6.63 $\pm$ 0.04 &        ---       &    \\
50976 & 11.61 $\pm$ 0.05 &  8.53 $\pm$ 0.01 &  6.59 $\pm$ 0.01 &        ---       &    \\
51365 &        ---       &        ---       &  6.99 $\pm$ 0.10 &        ---       & Imaging photometry*\\
51369 & 12.55 $\pm$ 0.01 &  8.93 $\pm$ 0.02 &  6.96 $\pm$ 0.01 &        ---       &    \\
51384 & 12.00 $\pm$ 0.10 &  8.86 $\pm$ 0.01 &  6.93 $\pm$ 0.01 &        ---       &    \\
51752 &        ---       &        ---       &  7.38 $\pm$ 0.01 &        ---       & Imaging photometry \\
52494 & 13.78 $\pm$ 0.05 & 10.00 $\pm$ 0.02 &  8.04 $\pm$ 0.01 &        ---       & Imaging photometry*\\
& & & & & \\

& & & & & \\
\multicolumn{6}{c}{\bf{IRAS 19129$+$2803}} \\
& & & & & \\
48487 &  6.61 $\pm$ 0.02 &  5.50 $\pm$ 0.01 &  4.74 $\pm$ 0.01 &        ---       &    \\
48509 &  6.65 $\pm$ 0.05 &  5.57 $\pm$ 0.02 &  4.80 $\pm$ 0.04 &        ---       &    \\
48830 &  6.73 $\pm$ 0.03 &  5.70 $\pm$ 0.02 &  4.85 $\pm$ 0.03 &        ---       &    \\
48865 &  6.62 $\pm$ 0.04 &  5.42 $\pm$ 0.02 &  4.69 $\pm$ 0.02 &        ---       &    \\
48880 &  6.49 $\pm$ 0.01 &  5.41 $\pm$ 0.01 &  4.64 $\pm$ 0.01 &        ---       &    \\
48922 &  6.52 $\pm$ 0.06 &  5.41 $\pm$ 0.02 &  4.65 $\pm$ 0.02 &        ---       &    \\
48945 &  6.60 $\pm$ 0.01 &  5.45 $\pm$ 0.01 &  4.67 $\pm$ 0.02 &        ---       &    \\
49227 &  7.37 $\pm$ 0.05 &  6.27 $\pm$ 0.06 &  5.34 $\pm$ 0.02 &        ---       &    \\
49405 &  6.86 $\pm$ 0.05 &  5.73 $\pm$ 0.04 &  4.93 $\pm$ 0.02 &        ---       &    \\
49597 &  8.37 $\pm$ 0.02 &  6.98 $\pm$ 0.01 &  5.81 $\pm$ 0.01 &        ---       &    \\
49840 &  7.02 $\pm$ 0.01 &  5.90 $\pm$ 0.01 &  5.08 $\pm$ 0.03 &        ---       &    \\
49936 &  8.00 $\pm$ 0.05 &  6.62 $\pm$ 0.03 &  5.60 $\pm$ 0.03 &        ---       &    \\
49953 &  8.17 $\pm$ 0.03 &  6.76 $\pm$ 0.03 &  5.68 $\pm$ 0.03 &        ---       &    \\
49997 &  8.42 $\pm$ 0.02 &  7.01 $\pm$ 0.02 &  5.80 $\pm$ 0.02 &        ---       &    \\
50247 &  6.65 $\pm$ 0.02 &  5.49 $\pm$ 0.02 &  4.75 $\pm$ 0.02 &        ---       &    \\
50248 &  6.70 $\pm$ 0.02 &  5.55 $\pm$ 0.02 &  4.78 $\pm$ 0.02 &  3.59 $\pm$ 0.05 &    \\
50249 &  6.71 $\pm$ 0.02 &  5.55 $\pm$ 0.02 &  4.78 $\pm$ 0.02 &  3.73 $\pm$ 0.05 &    \\
50300 &  6.96 $\pm$ 0.04 &  5.79 $\pm$ 0.05 &  4.98 $\pm$ 0.04 &        ---       &    \\
51365 &  7.02 $\pm$ 0.03 &  5.73 $\pm$ 0.02 &  4.98 $\pm$ 0.03 &        ---       & Imaging photometry* \\
51748 &  8.16 $\pm$ 0.02 &  6.59 $\pm$ 0.01 &  5.51 $\pm$ 0.01 &        ---       & Imaging photometry \\
52041 &  7.91 $\pm$ 0.02 &  6.33 $\pm$ 0.02 &  5.49 $\pm$ 0.03 &        ---       & Imaging photometry* \\
52117 &  8.30 $\pm$ 0.02 &  6.87 $\pm$ 0.01 &  5.75 $\pm$ 0.01 &        ---       &    \\
52118 &  8.38 $\pm$ 0.08 &  6.89 $\pm$ 0.05 &  5.75 $\pm$ 0.05 &        ---       &    \\
& & & & & \\

\hline

\end{tabular}
\end{center}
\hspace{12cm}{\footnotesize {\em Continued next page.}}
\end{table*}

\begin{table*}
{\bf Table A.} Summary of the near infrared photometric measurements ({\em continued})
\begin{center}
\begin{tabular}{cccccl}

\hline\hline\noalign{\smallskip}
Julian Date &  \multicolumn{4}{c}{NIR magnitude [mag]}  & Notes\\
 (2400000+)  & J & H & K & L' &  \\
\noalign{\smallskip}\hline\noalign{\smallskip}

& & & & & \\
\multicolumn{6}{c}{\bf{IRAS 19147$+$5004}} \\
& & & & & \\
48486 &  3.73 $\pm$ 0.03 &  2.88 $\pm$ 0.04 &  2.54 $\pm$ 0.03 &        ---       &    \\
48509 &  3.75 $\pm$ 0.05 &  2.91 $\pm$ 0.03 &  2.58 $\pm$ 0.04 &        ---       &    \\
48664 &  3.78 $\pm$ 0.02 &  2.92 $\pm$ 0.02 &  2.57 $\pm$ 0.02 &        ---       &    \\
48838 &  3.74 $\pm$ 0.02 &  2.83 $\pm$ 0.03 &  2.44 $\pm$ 0.04 &        ---       &    \\
48845 &  3.72 $\pm$ 0.03 &  2.84 $\pm$ 0.03 &  2.52 $\pm$ 0.02 &        ---       &    \\
48880 &  3.63 $\pm$ 0.01 &  2.77 $\pm$ 0.01 &  2.45 $\pm$ 0.01 &        ---       &    \\
48922 &  3.74 $\pm$ 0.07 &  2.85 $\pm$ 0.03 &  2.52 $\pm$ 0.02 &        ---       &    \\
48945 &  3.78 $\pm$ 0.02 &  2.93 $\pm$ 0.02 &  2.64 $\pm$ 0.04 &  2.27 $\pm$ 0.05 &    \\
49227 &  3.85 $\pm$ 0.02 &  2.88 $\pm$ 0.03 &  2.56 $\pm$ 0.02 &  2.26 $\pm$ 0.06 &    \\
49322 &  3.50 $\pm$ 0.07 &  2.58 $\pm$ 0.12 &  2.21 $\pm$ 0.11 &  2.07 $\pm$ 0.09 &    \\
49437 &  3.69 $\pm$ 0.02 &  2.85 $\pm$ 0.02 &  2.53 $\pm$ 0.02 &  2.30 $\pm$ 0.05 &    \\
49456 &  3.80 $\pm$ 0.02 &  2.84 $\pm$ 0.02 &  2.56 $\pm$ 0.01 &  2.26 $\pm$ 0.06 &    \\
49926 &  3.56 $\pm$ 0.07 &  2.73 $\pm$ 0.04 &  2.38 $\pm$ 0.05 &        ---       &    \\
49935 &  3.62 $\pm$ 0.02 &  2.75 $\pm$ 0.02 &  2.43 $\pm$ 0.02 &        ---       &    \\
49997 &  3.68 $\pm$ 0.02 &  2.88 $\pm$ 0.02 &  2.53 $\pm$ 0.02 &  2.36 $\pm$ 0.08 &    \\
50247 &  3.70 $\pm$ 0.02 &  2.84 $\pm$ 0.02 &  2.52 $\pm$ 0.02 &        ---       &    \\
50248 &  3.71 $\pm$ 0.02 &  2.85 $\pm$ 0.02 &  2.52 $\pm$ 0.02 &  2.16 $\pm$ 0.05 &    \\
50269 &  3.71 $\pm$ 0.02 &  2.90 $\pm$ 0.02 &  2.47 $\pm$ 0.02 &  2.27 $\pm$ 0.05 &    \\
50298 &  3.78 $\pm$ 0.02 &  2.93 $\pm$ 0.02 &  2.57 $\pm$ 0.02 &  2.23 $\pm$ 0.04 &    \\
& & & & & \\

& & & & & \\
\multicolumn{6}{c}{\bf{IRAS 23492$+$0846}} \\
& & & & & \\
48487 &  2.60 $\pm$ 0.01 &  1.70 $\pm$ 0.01 &  1.34 $\pm$ 0.01 &        ---       &    \\
49325 &  2.62 $\pm$ 0.02 &  1.70 $\pm$ 0.02 &  1.39 $\pm$ 0.02 &        ---       &    \\
49952 &  2.67 $\pm$ 0.02 &  1.76 $\pm$ 0.01 &  1.39 $\pm$ 0.02 &  1.13 $\pm$ 0.10 &    \\
49988 &  2.64 $\pm$ 0.03 &  1.71 $\pm$ 0.02 &  1.35 $\pm$ 0.01 &        ---       &    \\
50300 &  2.63 $\pm$ 0.04 &  1.74 $\pm$ 0.02 &  1.39 $\pm$ 0.04 &        ---       &    \\
50306 &  2.64 $\pm$ 0.03 &  1.73 $\pm$ 0.02 &  1.37 $\pm$ 0.02 &  1.07 $\pm$ 0.06 &    \\
& & & & & \\

\hline

\multicolumn{6}{l}{Notes:~* Data collected under non-photometric conditions.}\\

& & & & & \\
& & & & & \\
& & & & & \\
& & & & & \\
& & & & & \\
& & & & & \\
& & & & & \\
& & & & & \\
& & & & & \\
& & & & & \\
& & & & & \\
& & & & & \\
& & & & & \\
& & & & & \\
& & & & & \\
& & & & & \\
& & & & & \\
& & & & & \\
& & & & & \\
& & & & & \\
& & & & & \\
& & & & & \\
& & & & & \\
& & & & & \\
& & & & & \\
& & & & & \\
& & & & & \\

\end{tabular}
\end{center}
\end{table*}

\newpage

\begin{figure*}
\begin{center}
\vspace{0.5cm}
\epsfxsize=7.4cm
        \epsfbox{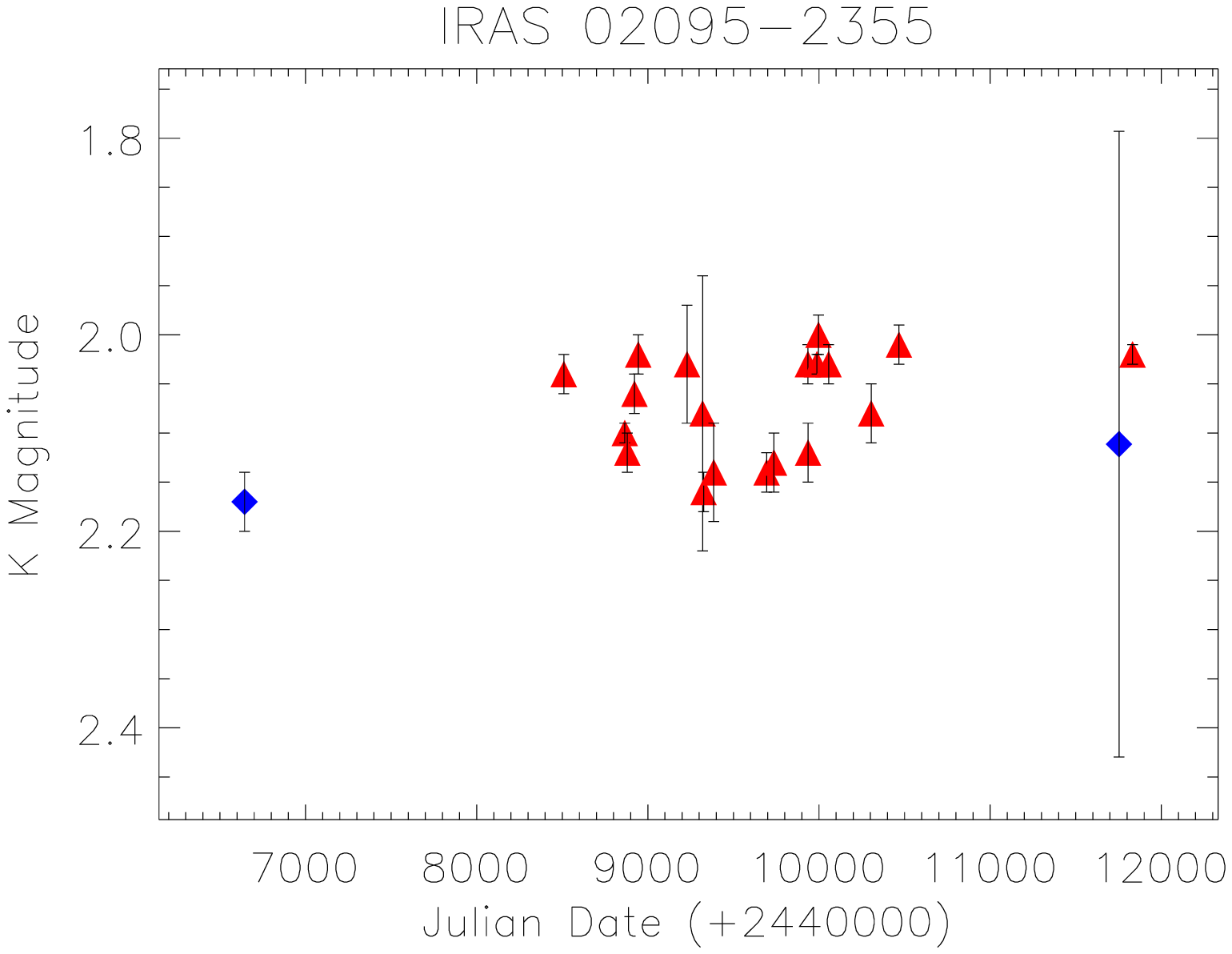}
\epsfxsize=7.4cm
	\epsfbox{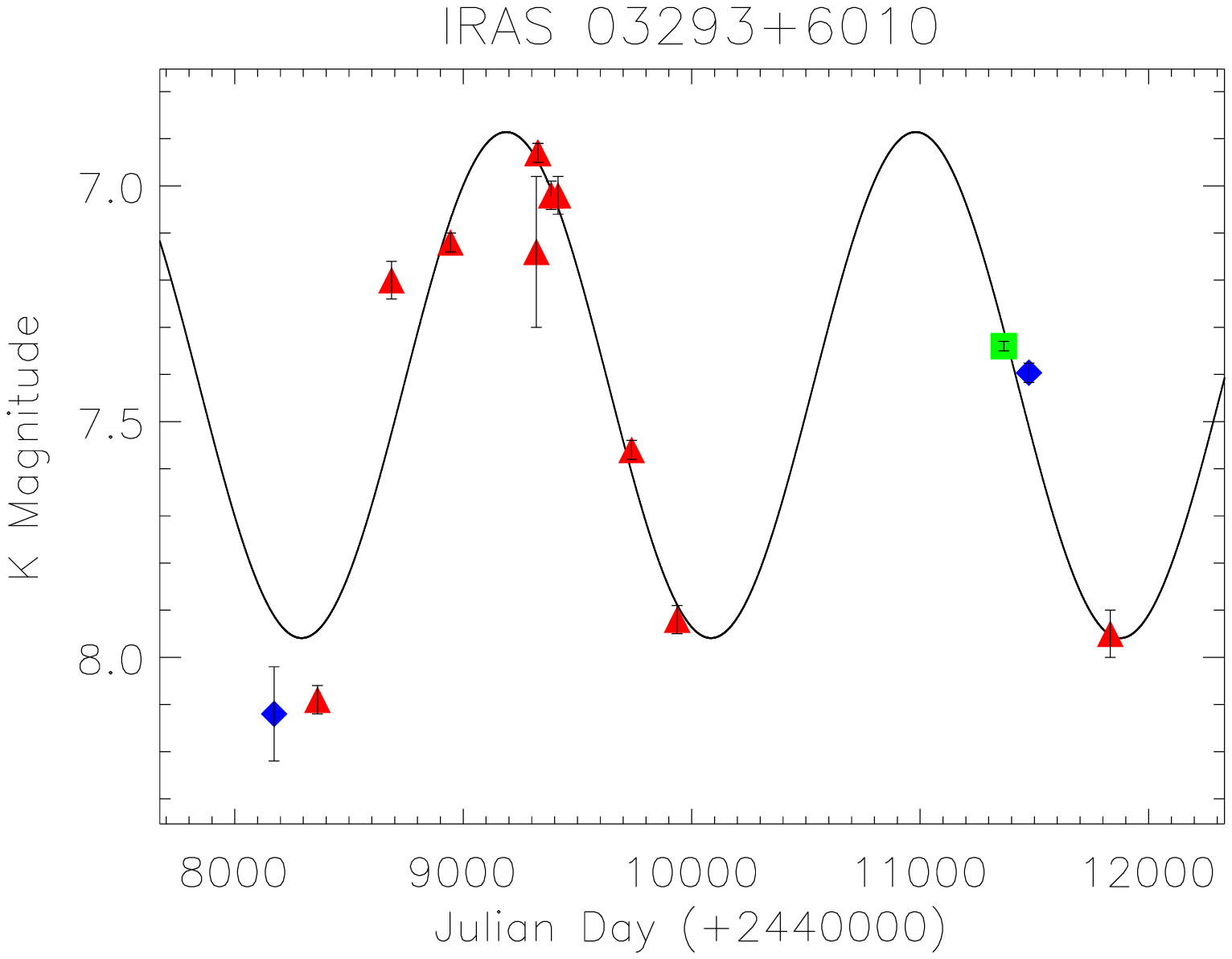}
\epsfxsize=7.4cm
	\epsfbox{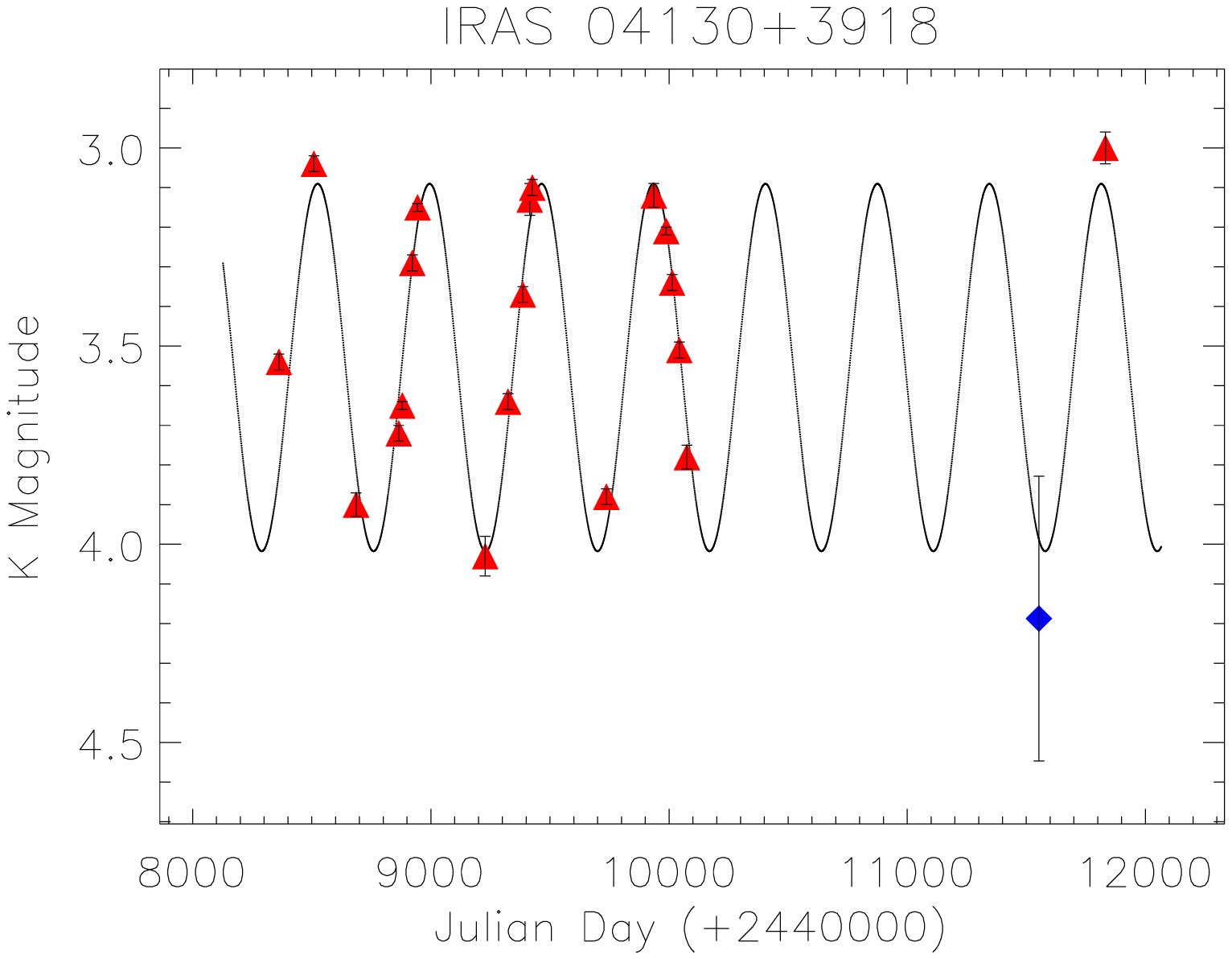}
\epsfxsize=7.4cm
	\epsfbox{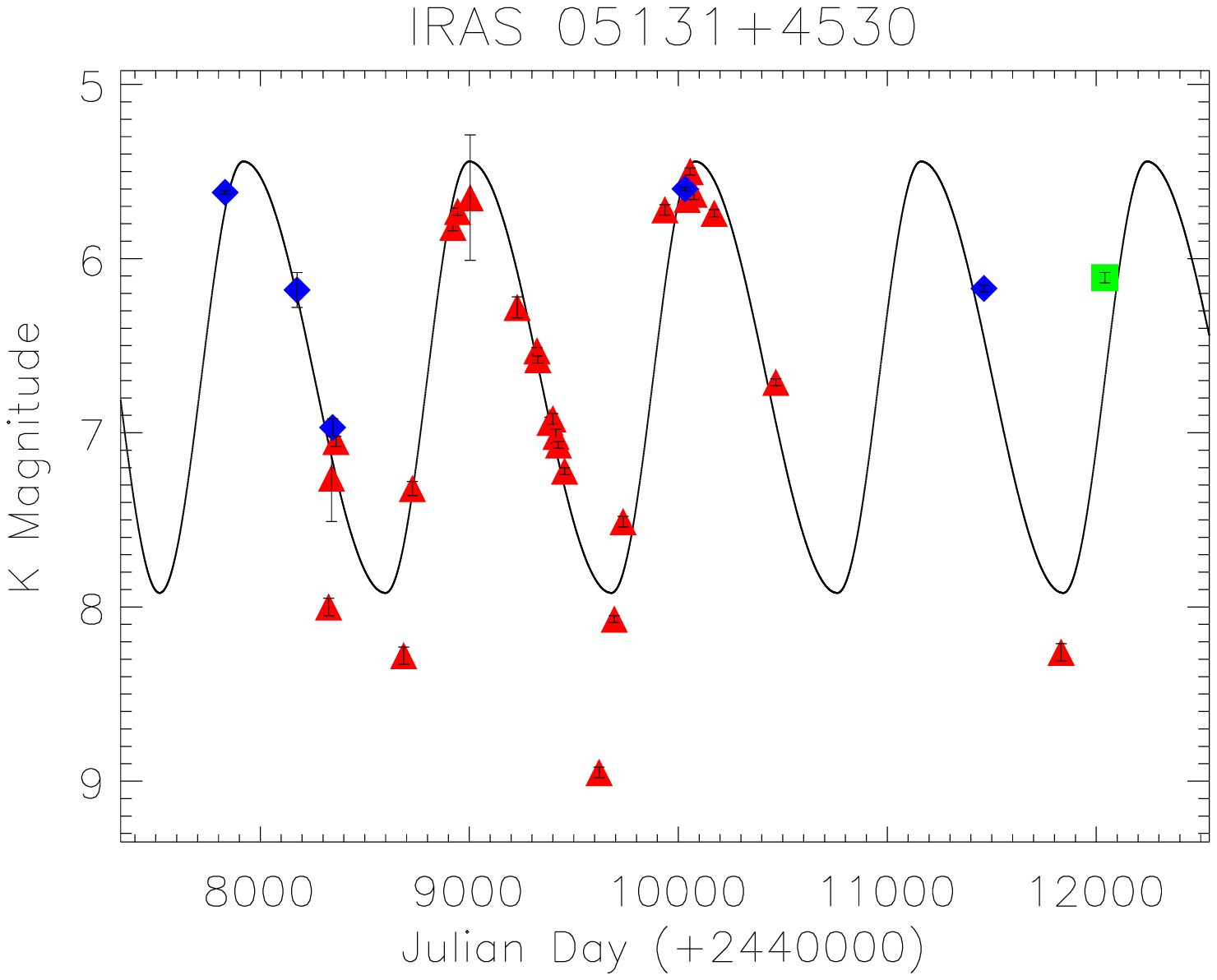}
\epsfxsize=7.4cm
	\epsfbox{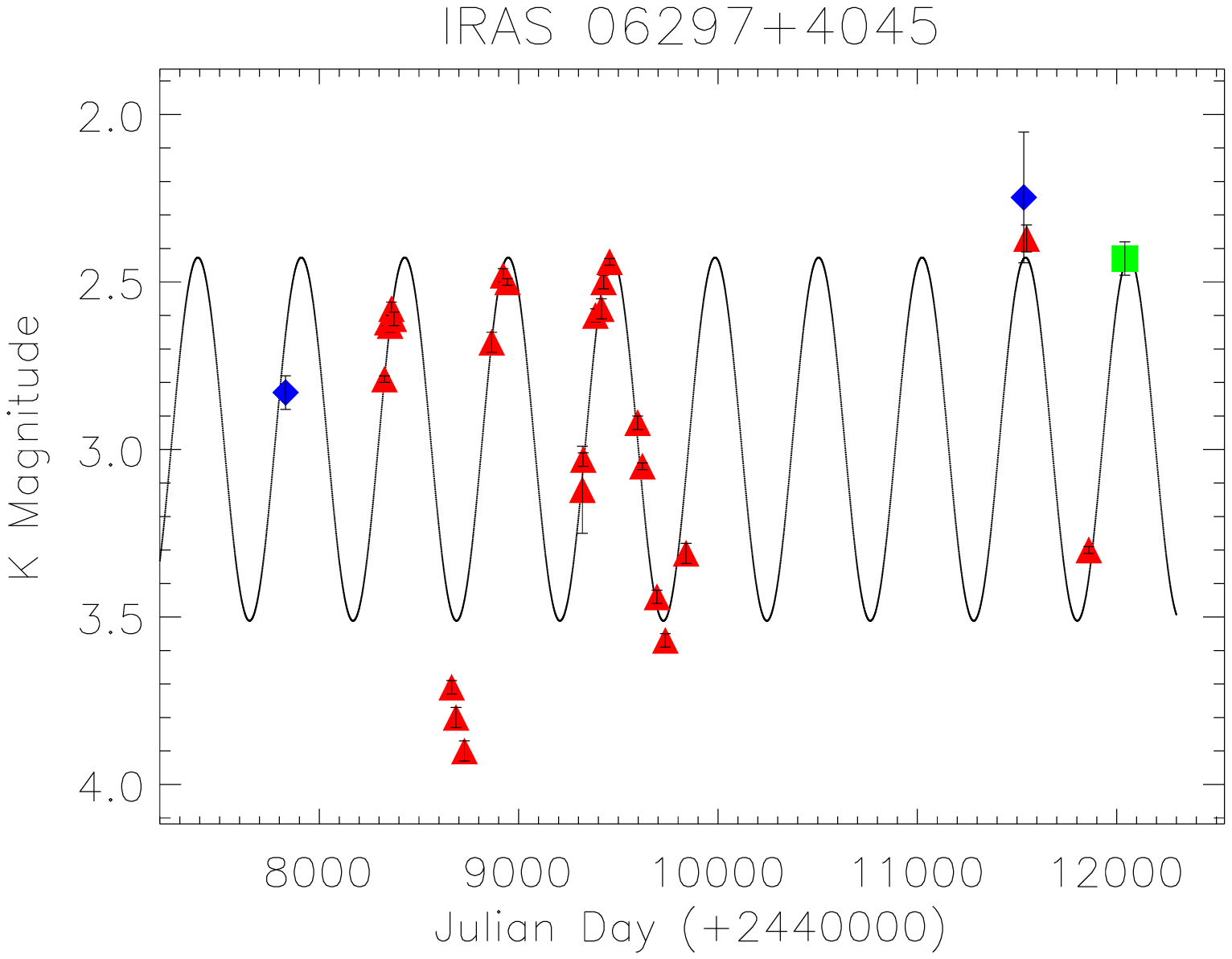}
\epsfxsize=7.4cm
	\epsfbox{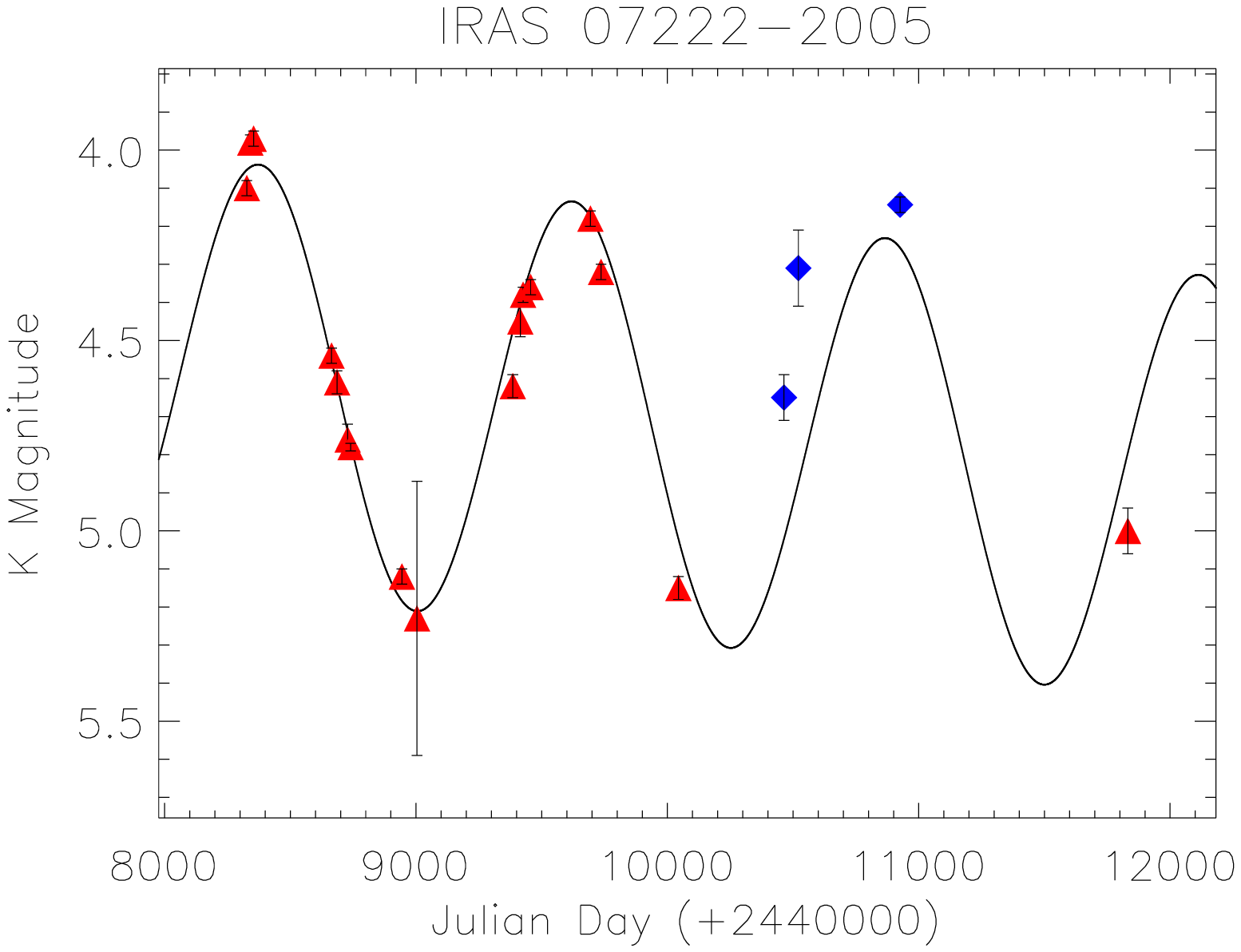}
\epsfxsize=7.4cm
	\epsfbox{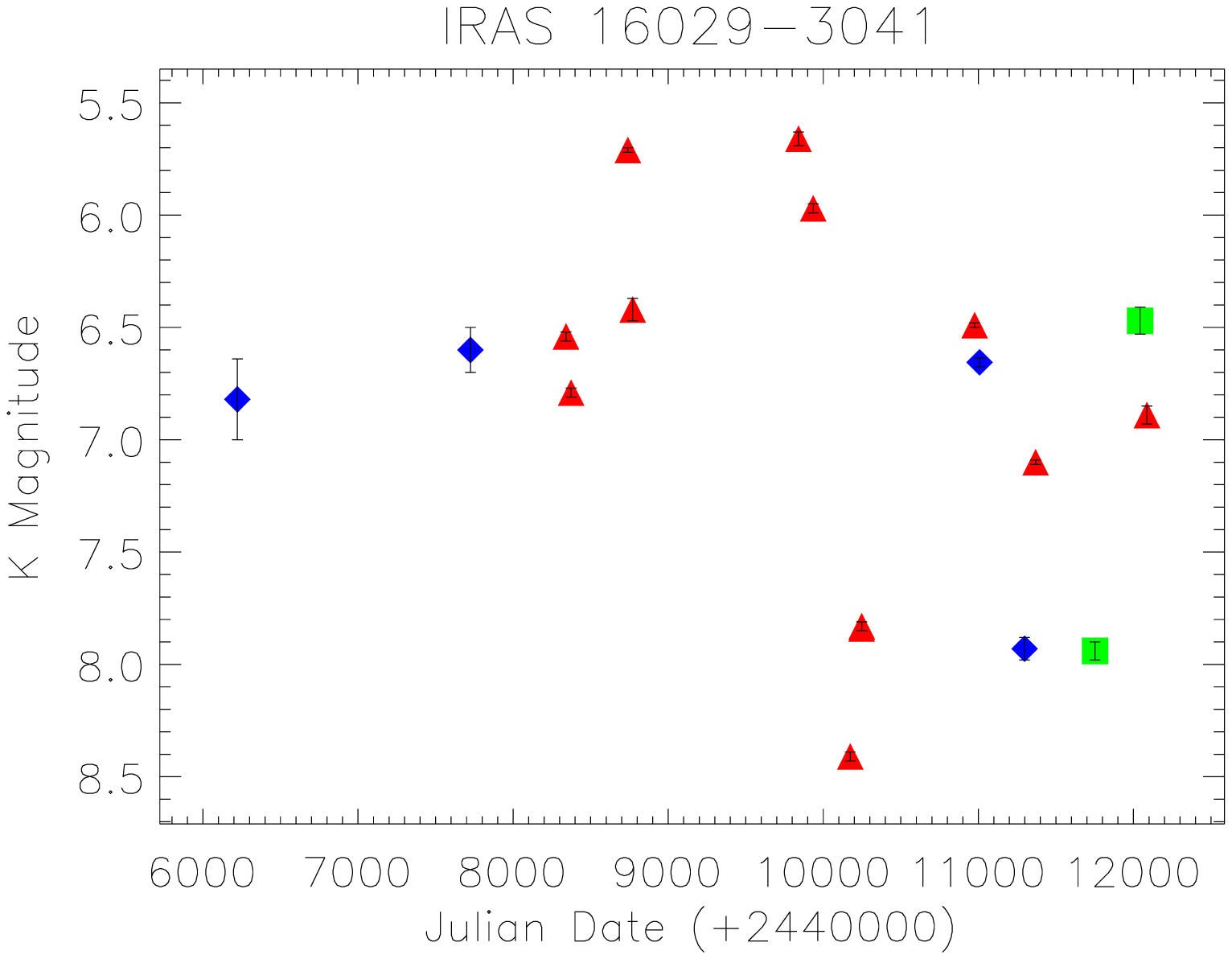}
\epsfxsize=7.4cm
	\epsfbox{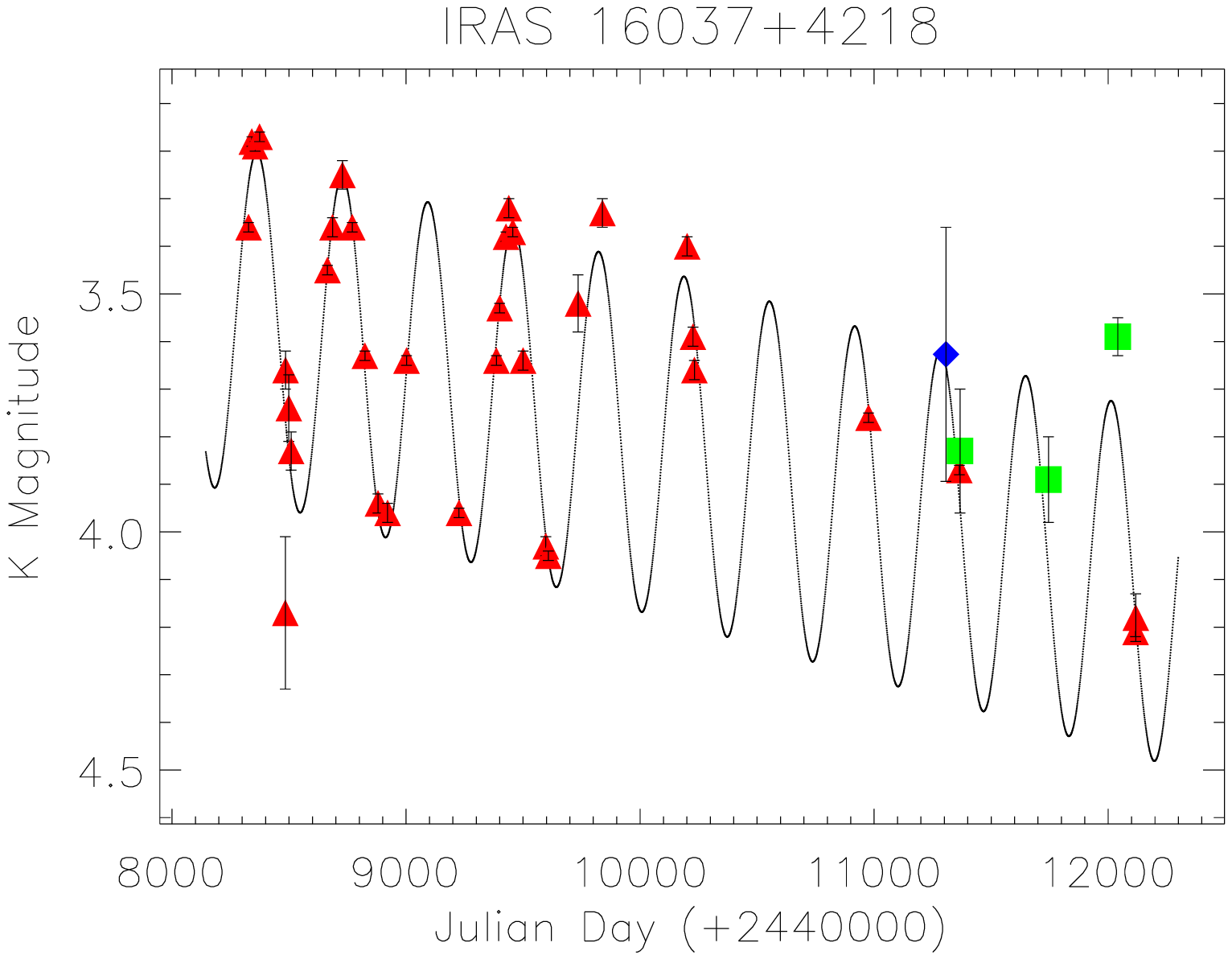}
\end{center}
{\bf Figure A1.} K-band measurements used in our variability
                 analysis. CVF data are plotted with triangles (red),
                 MAGIC data with squares (green), and data from
                 literature with diamonds (blue). For
                 IRAS\,06297+4045, one data point at JD\,2\,440\,510
                 from the literature has not been plotted. ({\em
                 continued next page})
\end{figure*}

\begin{figure*}
\begin{center}
\vspace{0.75cm}
\epsfxsize=7.4cm
	\epsfbox{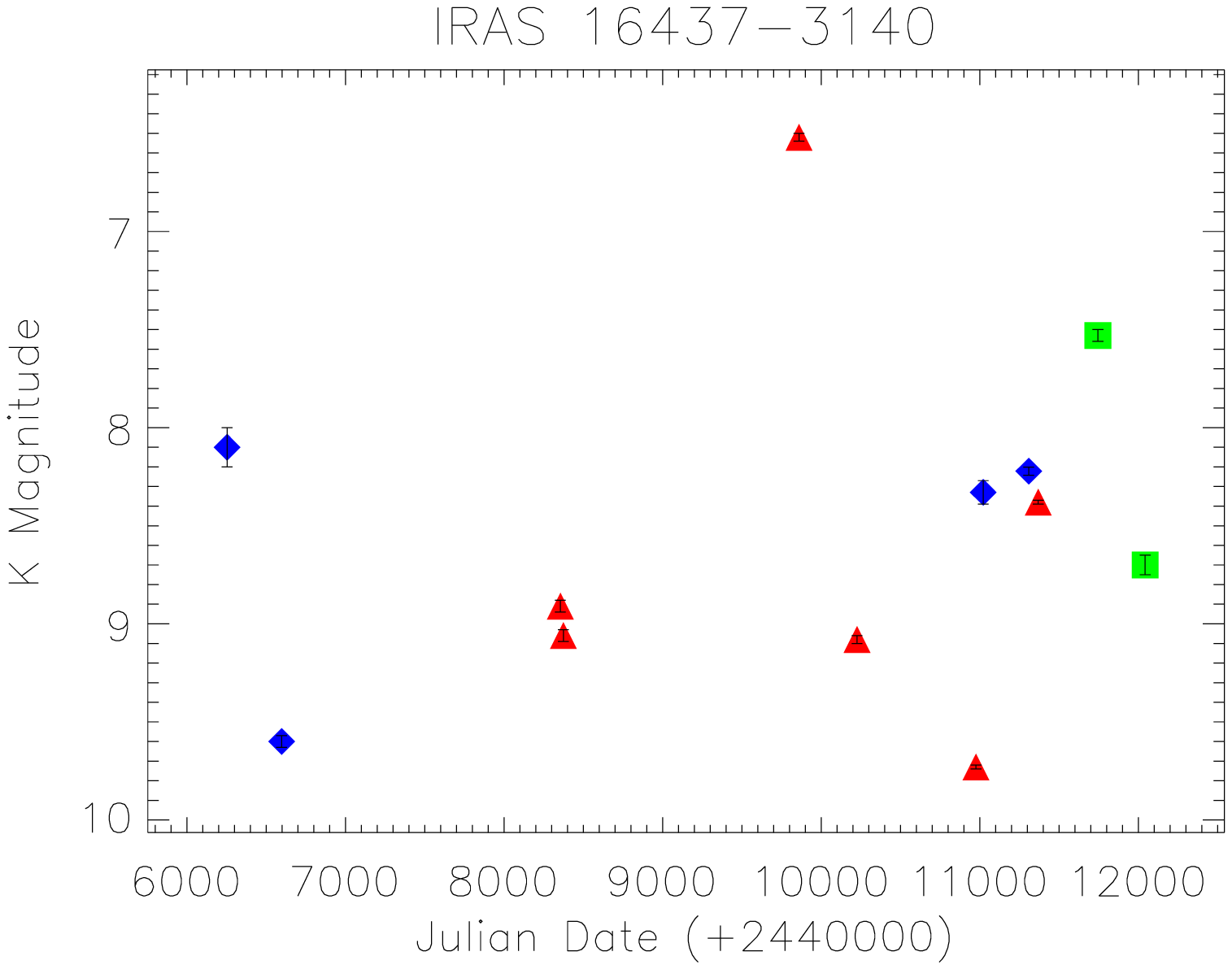}
\epsfxsize=7.4cm
        \epsfbox{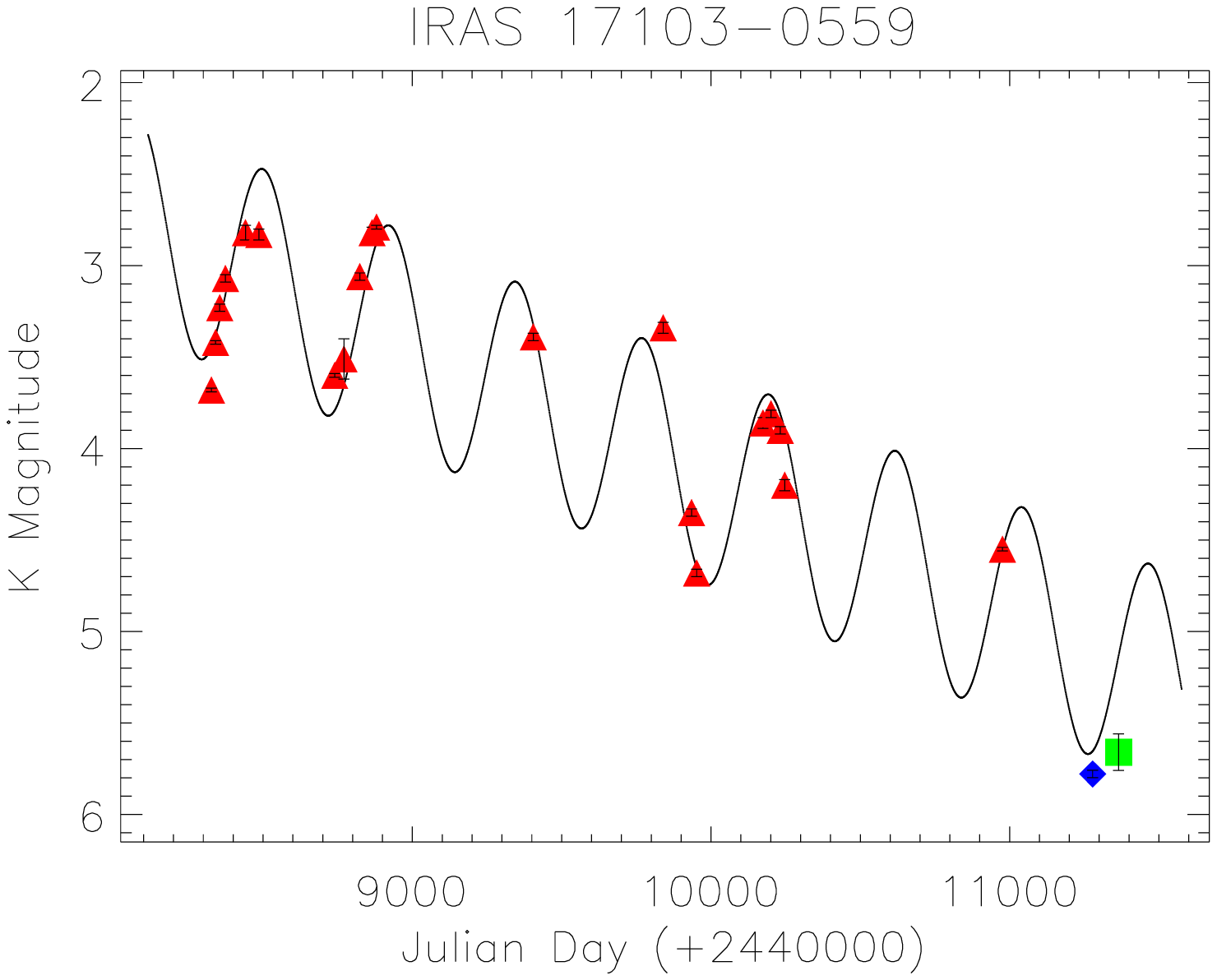}
\epsfxsize=7.4cm
	\epsfbox{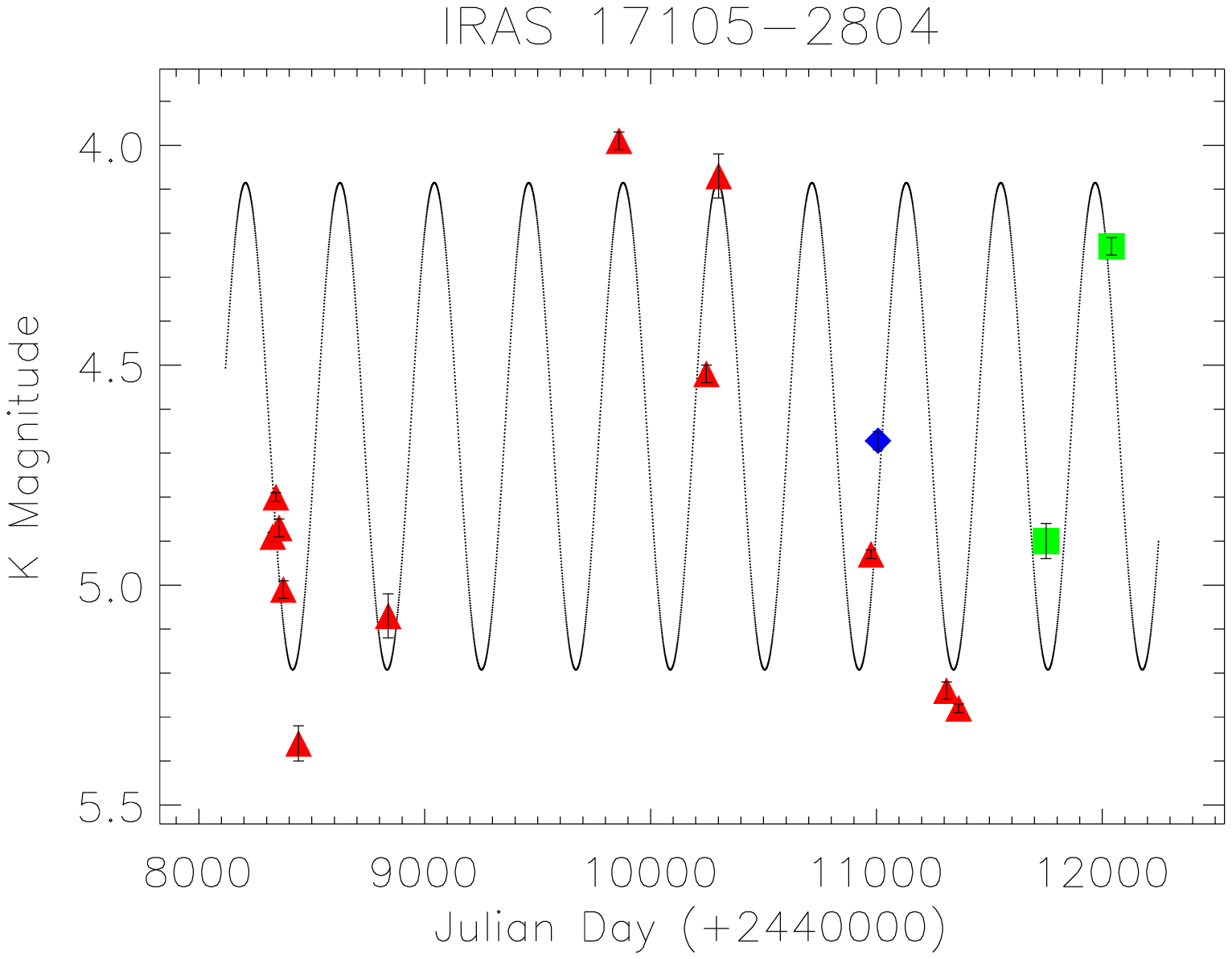}
\epsfxsize=7.4cm
	\epsfbox{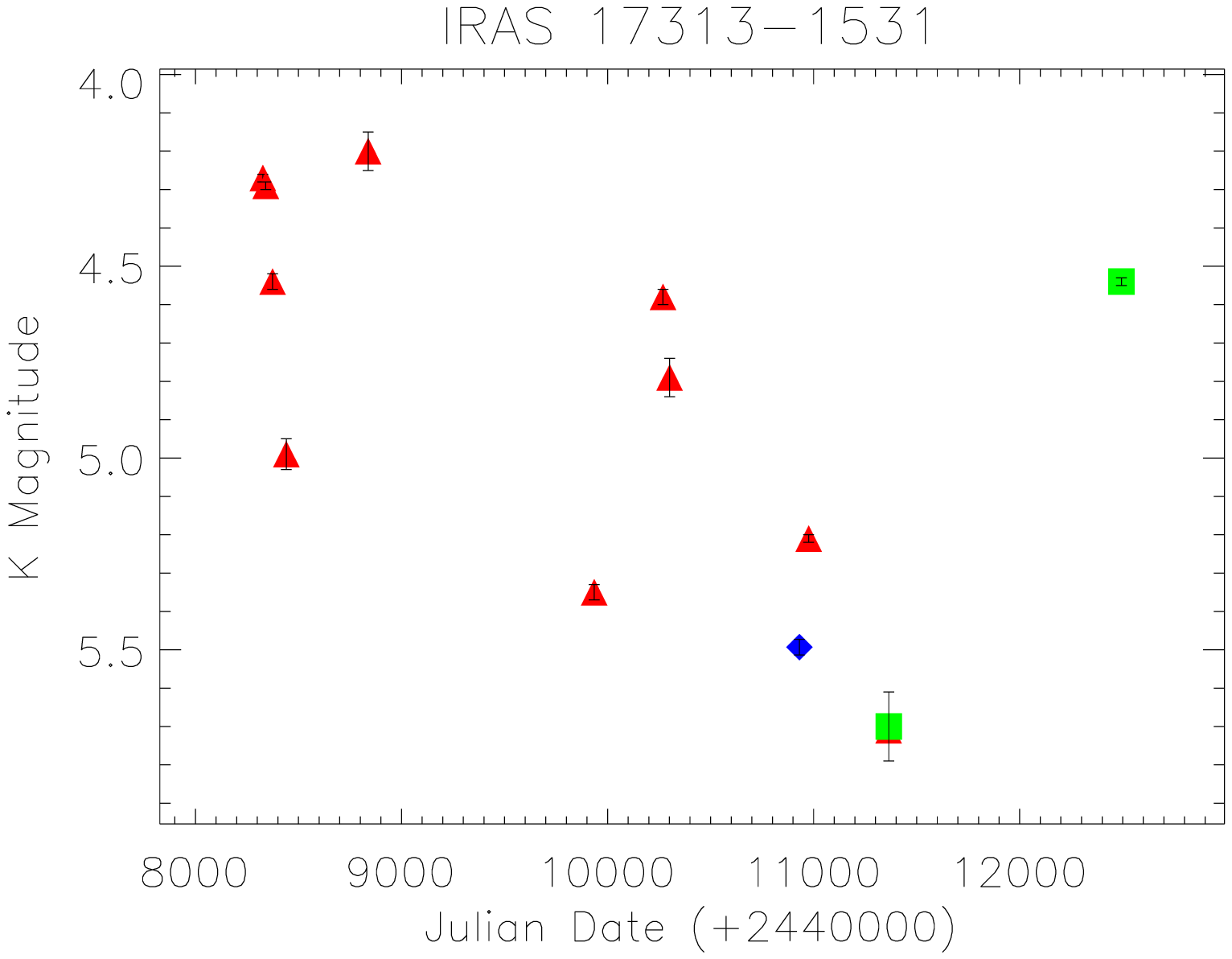}
\epsfxsize=7.4cm
	\epsfbox{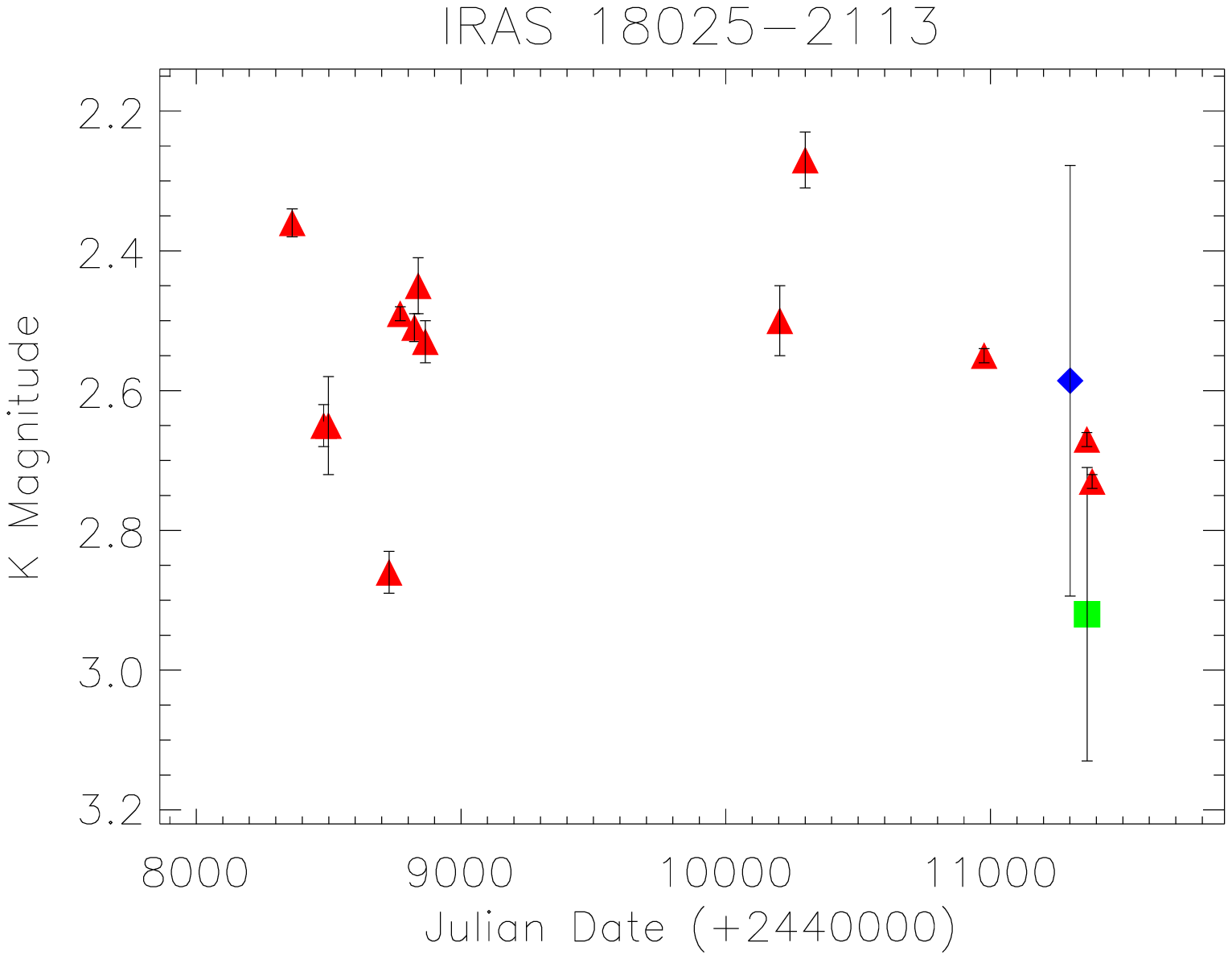}
\epsfxsize=7.4cm
	\epsfbox{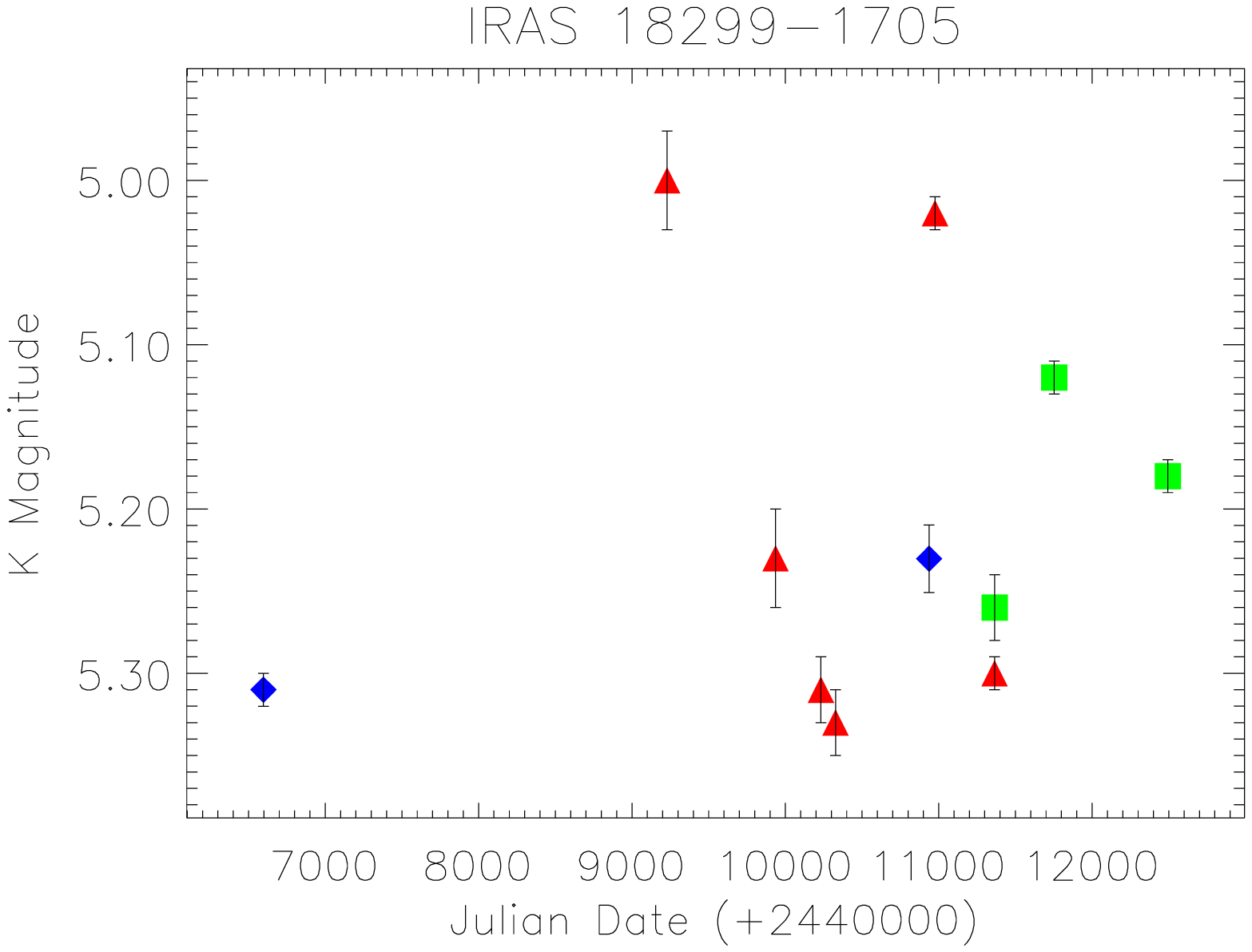}
\epsfxsize=7.4cm
	\epsfbox{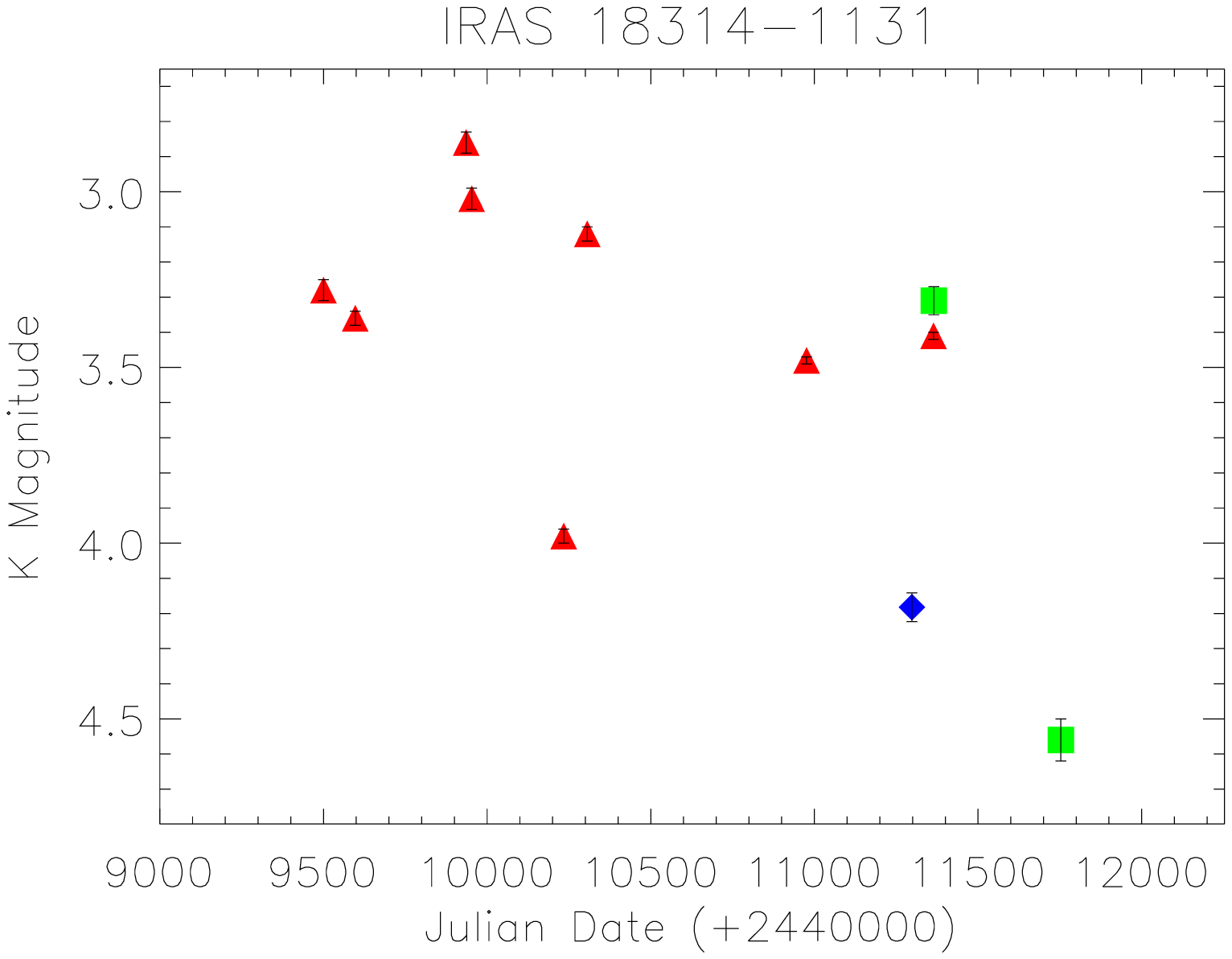}
\epsfxsize=7.4cm
	\epsfbox{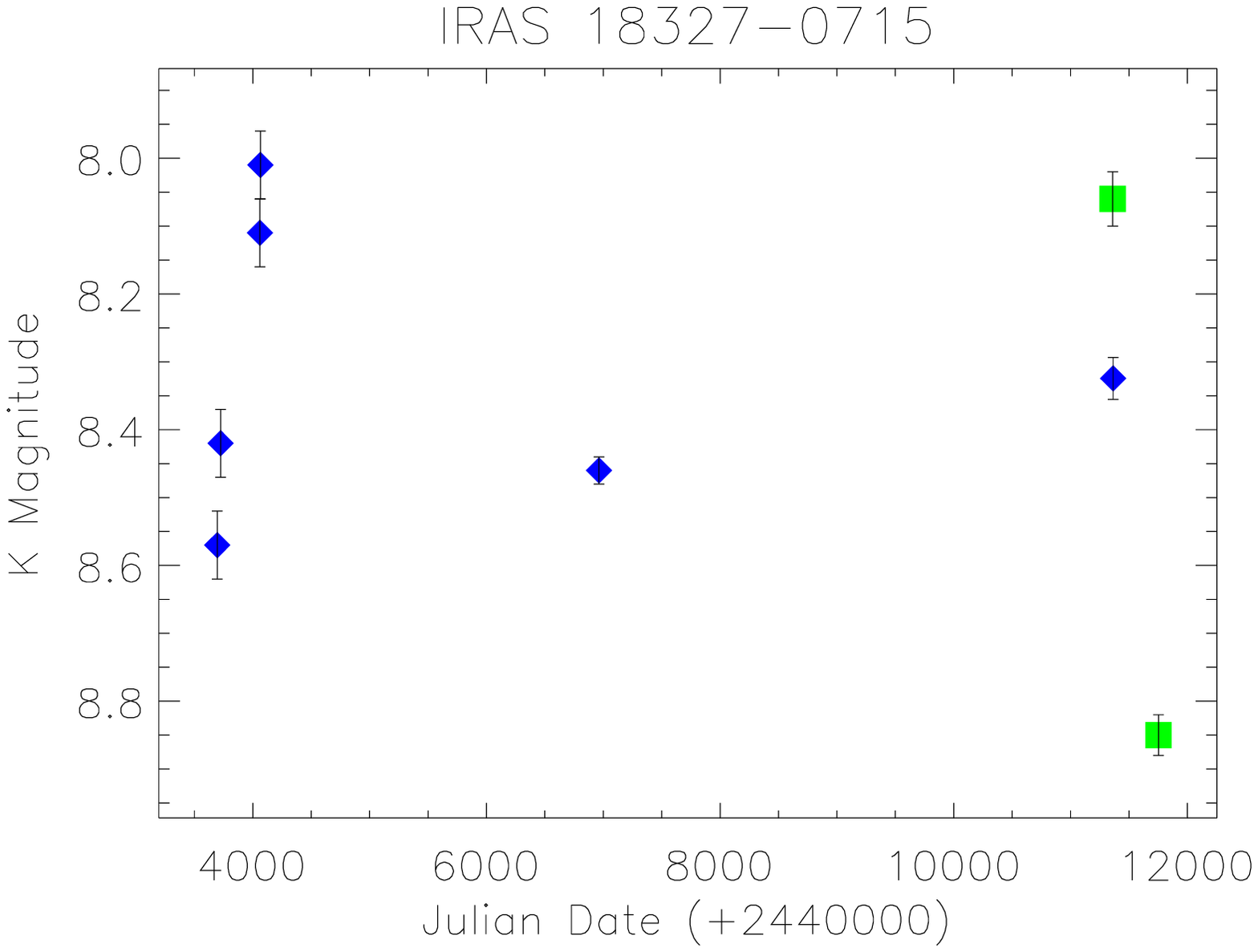}
\end{center}

{\bf Figure A1.} ({\em continued}) K-band measurements used in our
	         variability analysis. CVF data are plotted with
	         triangles (red), MAGIC data with squares (green), and
	         data from literature with diamonds (blue). For
	         IRAS\,06297+4045, one data point at JD\,2\,440\,510
	         from the literature has not been plotted. ({\em
	         continued next page})
\end{figure*}

\begin{figure*}
\begin{center}
\vspace{0.75cm}
\epsfxsize=7.4cm
	\epsfbox{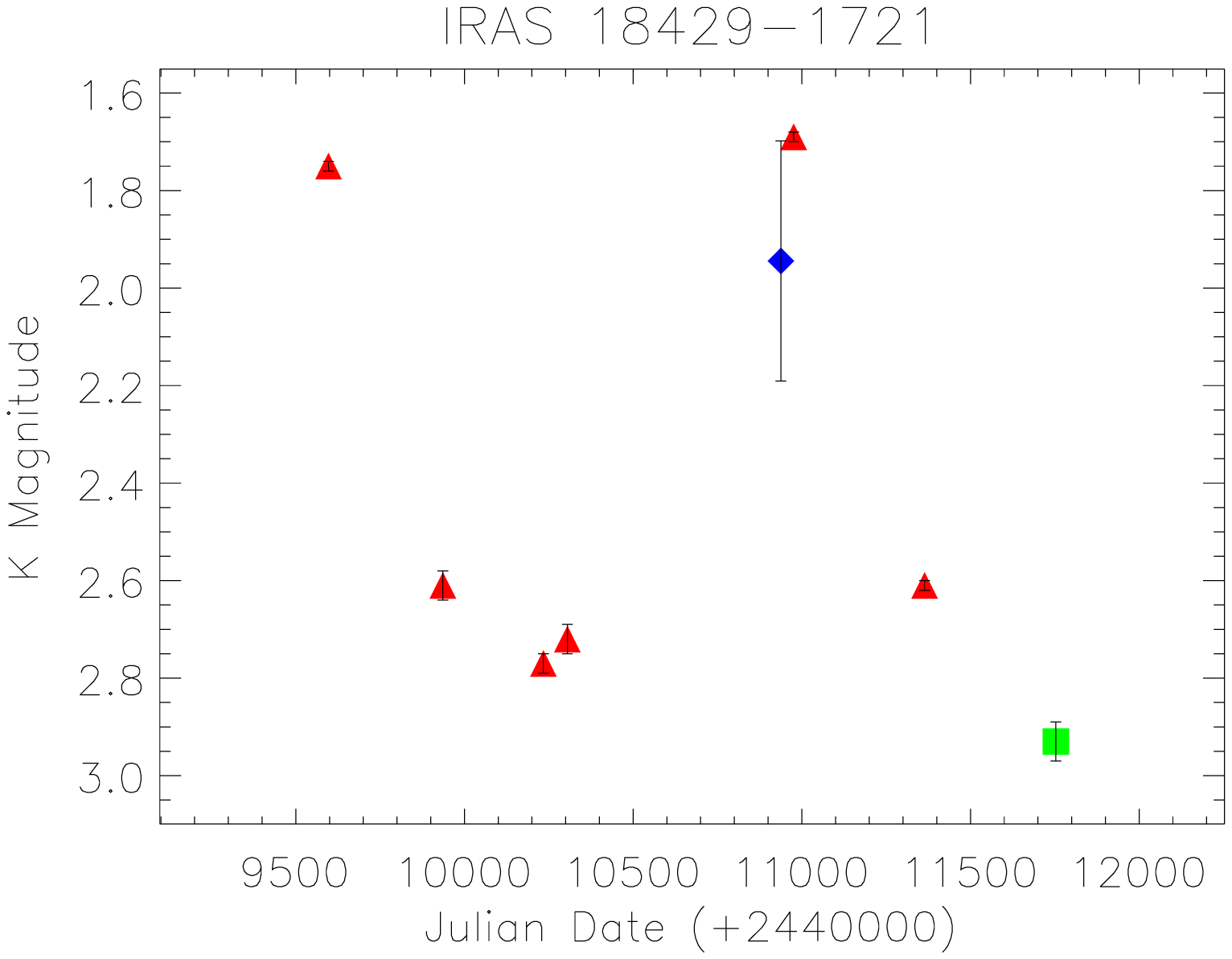}
\epsfxsize=7.4cm
	\epsfbox{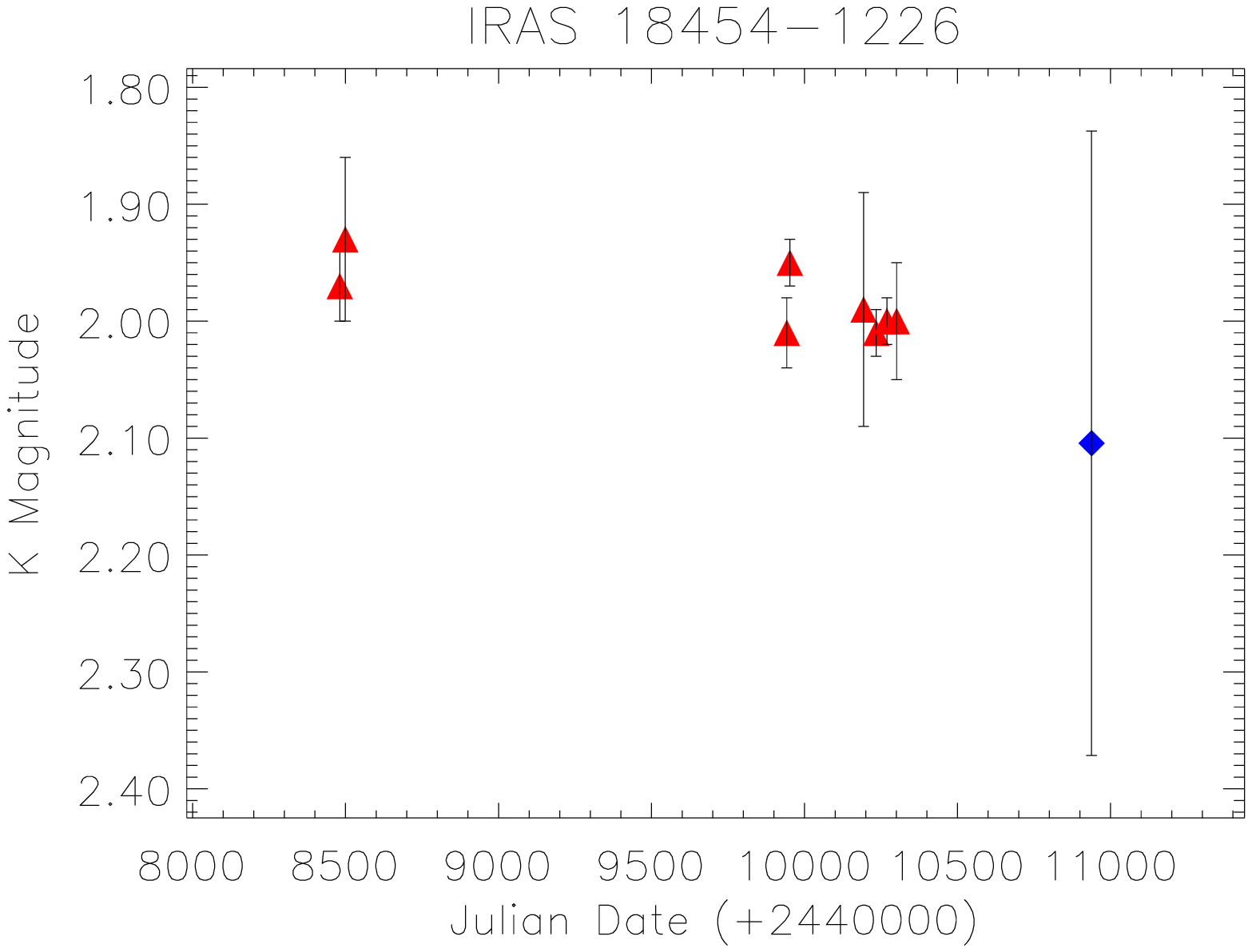}
\epsfxsize=7.4cm
	\epsfbox{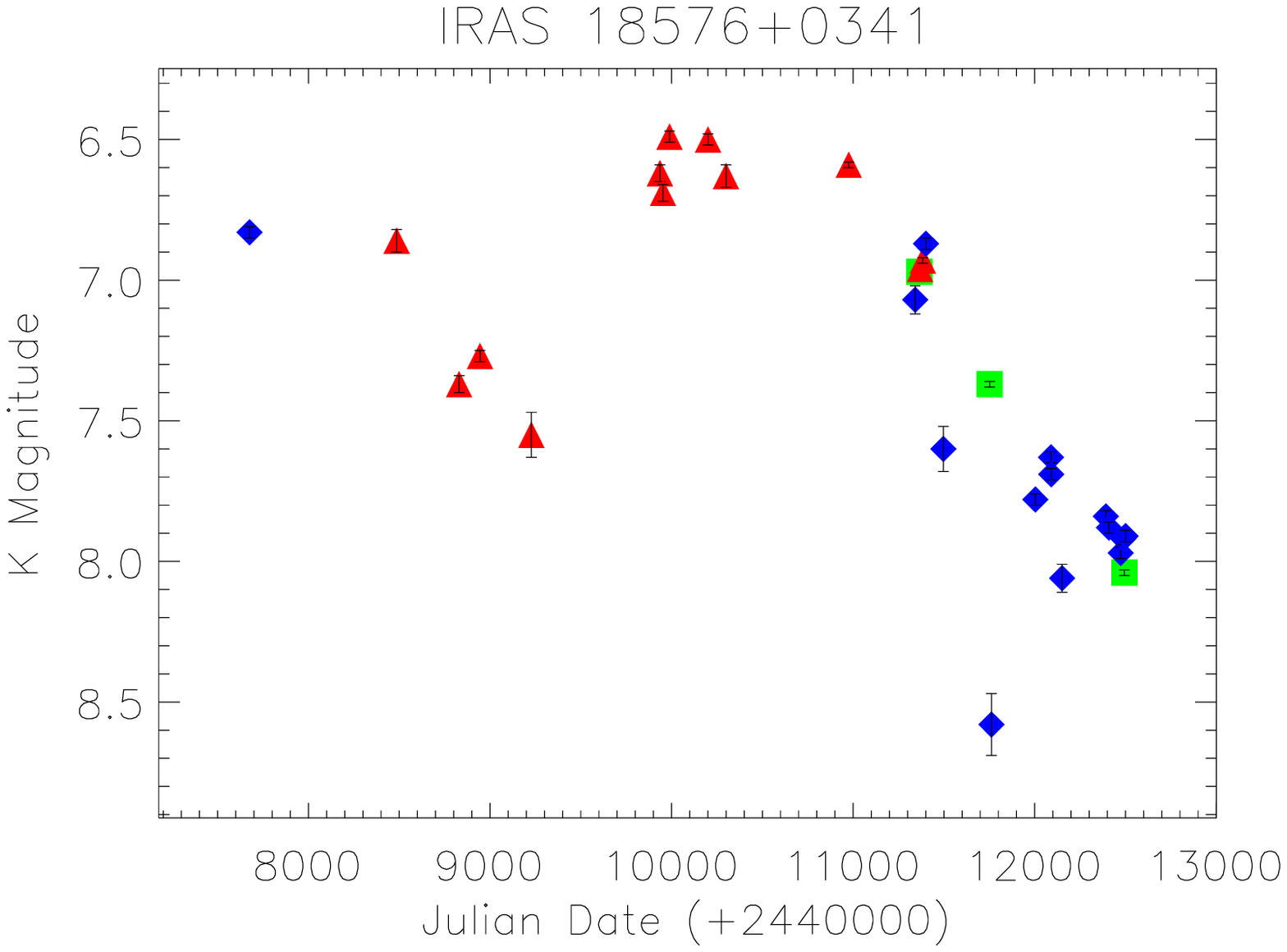}
\epsfxsize=7.4cm
	\epsfbox{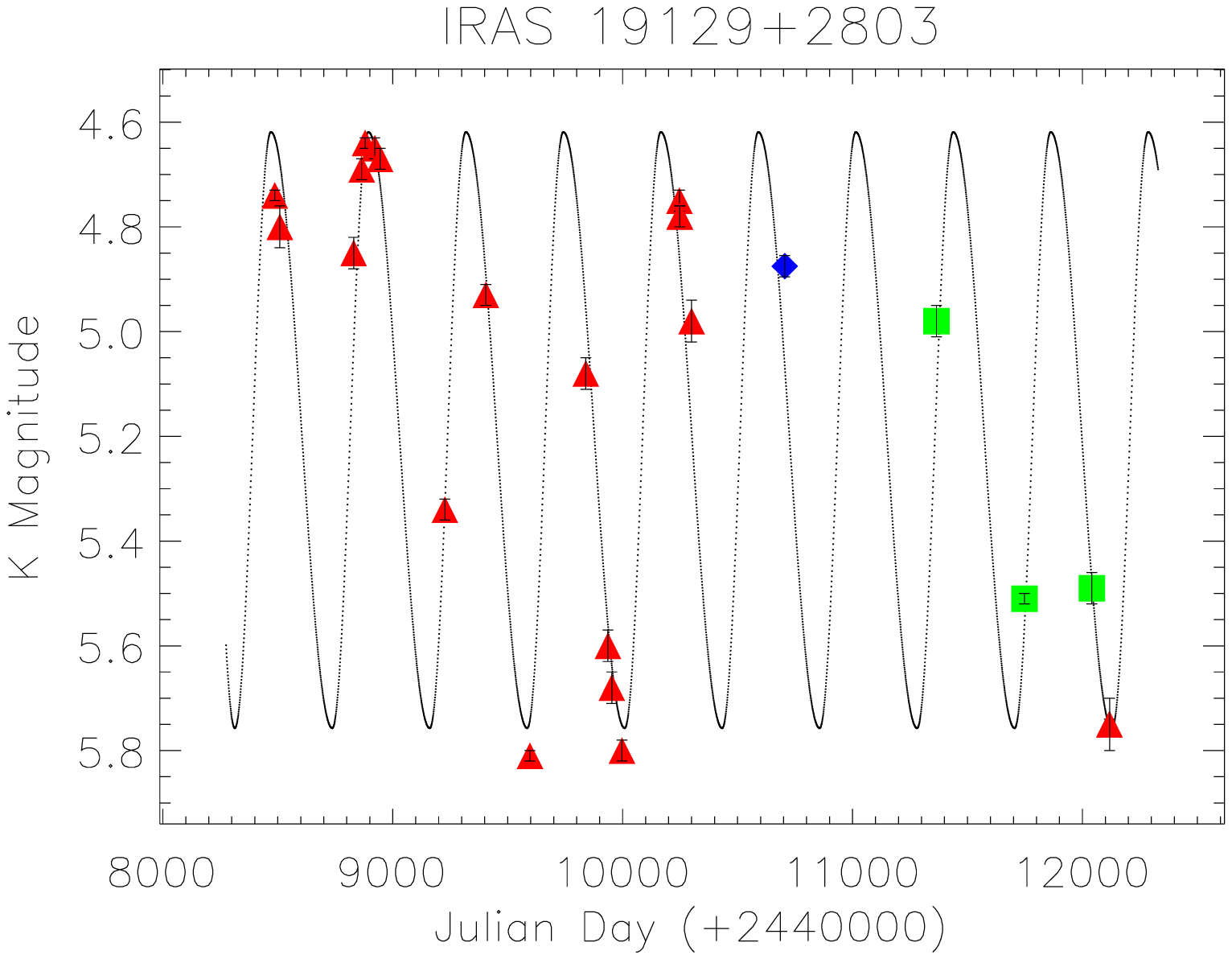}
\epsfxsize=7.4cm
	\epsfbox{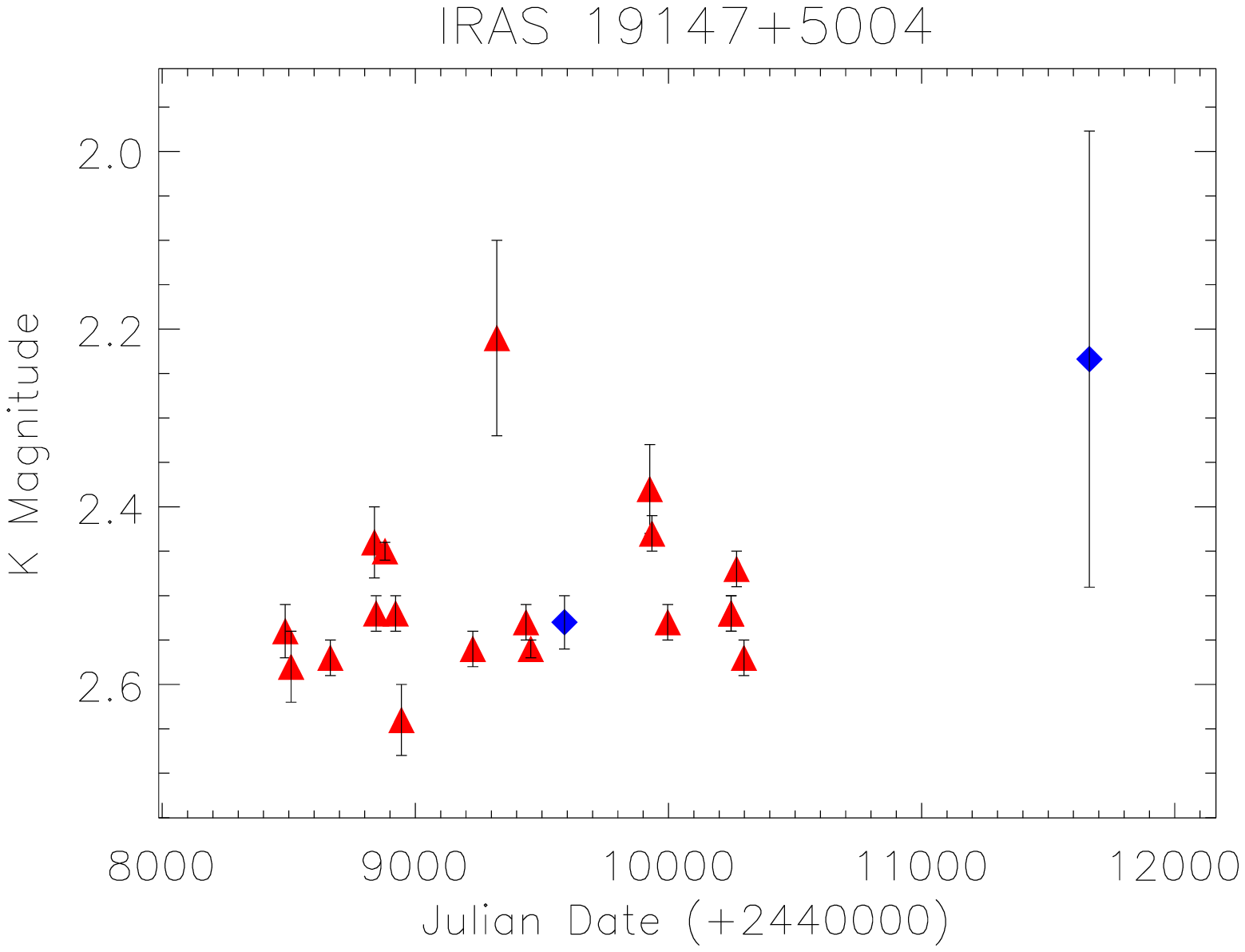}
\epsfxsize=7.4cm
	\epsfbox{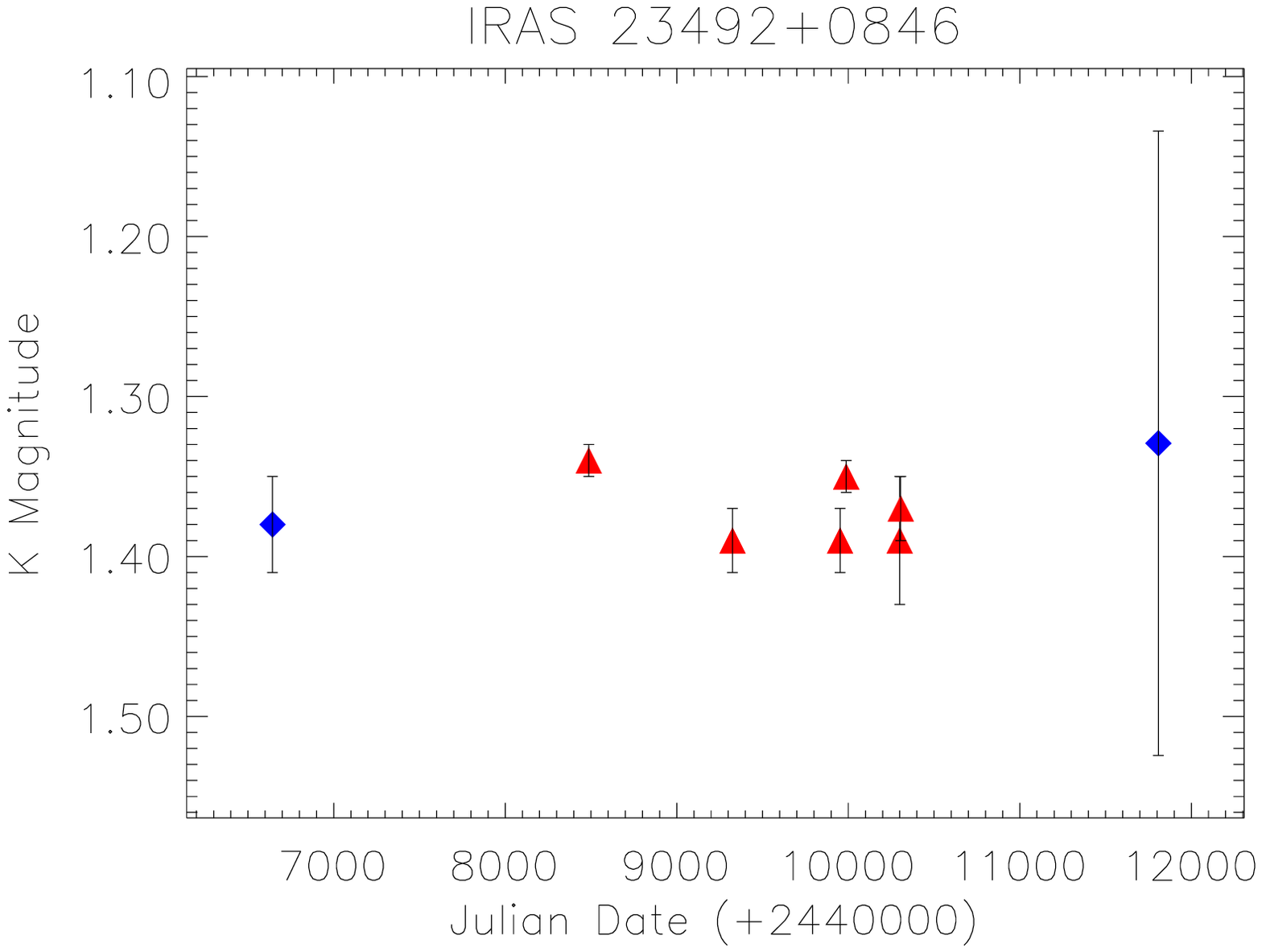}
\end{center}

{\bf Figure A1.} ({\em continued}) K-band measurements used in our
	         variability analysis. CVF data are plotted with
	         triangles (red), MAGIC data with squares (green), and
	         data from literature with diamonds (blue). For
	         IRAS\,06297+4045, one data point at JD\,2\,440\,510
	         from the literature has not been plotted.
\end{figure*}

\begin{figure*}
\begin{center}
\vspace{0.5cm}
\epsfxsize=7.4cm	
	\epsfbox{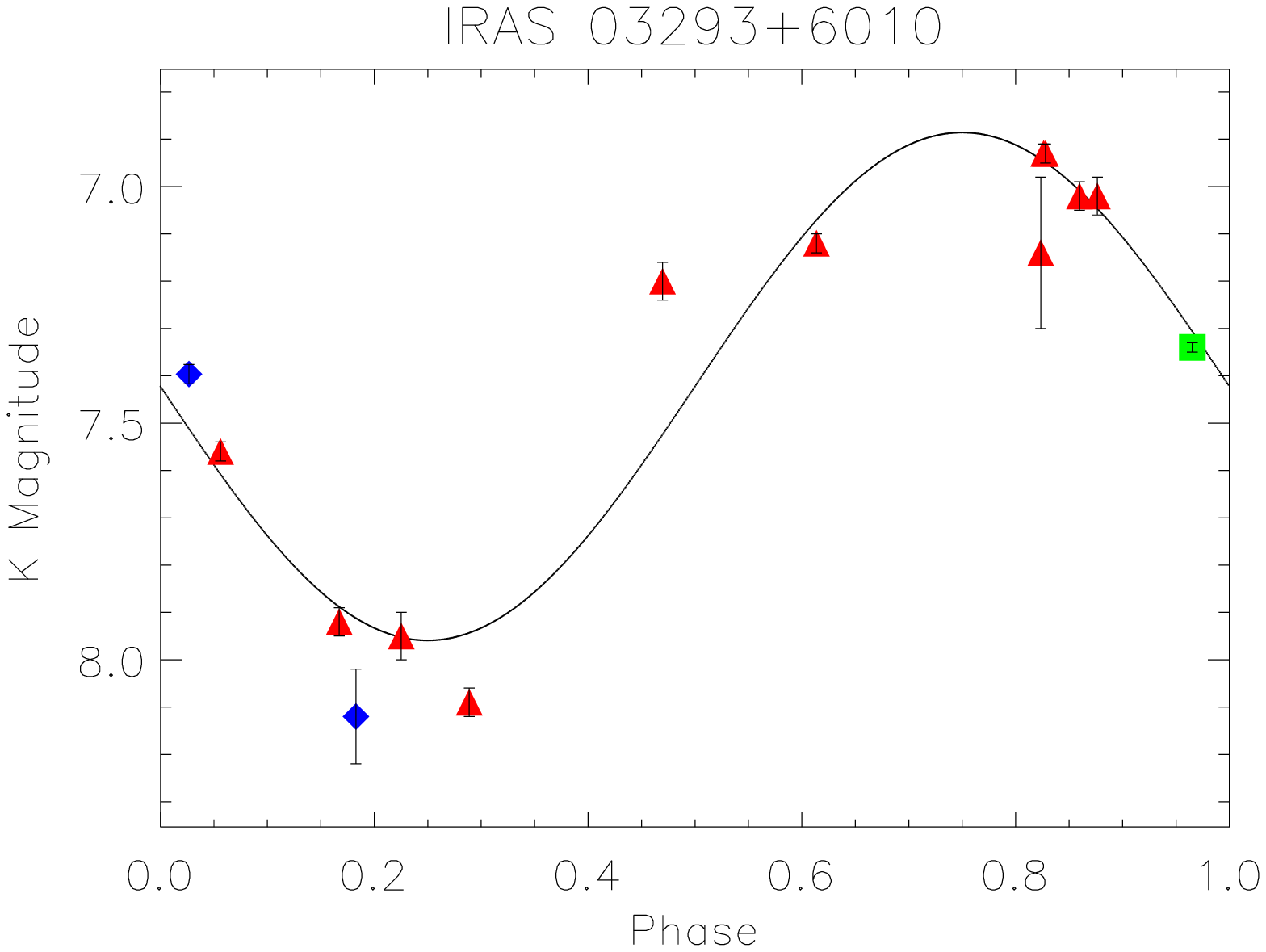}
\epsfxsize=7.4cm	
	\epsfbox{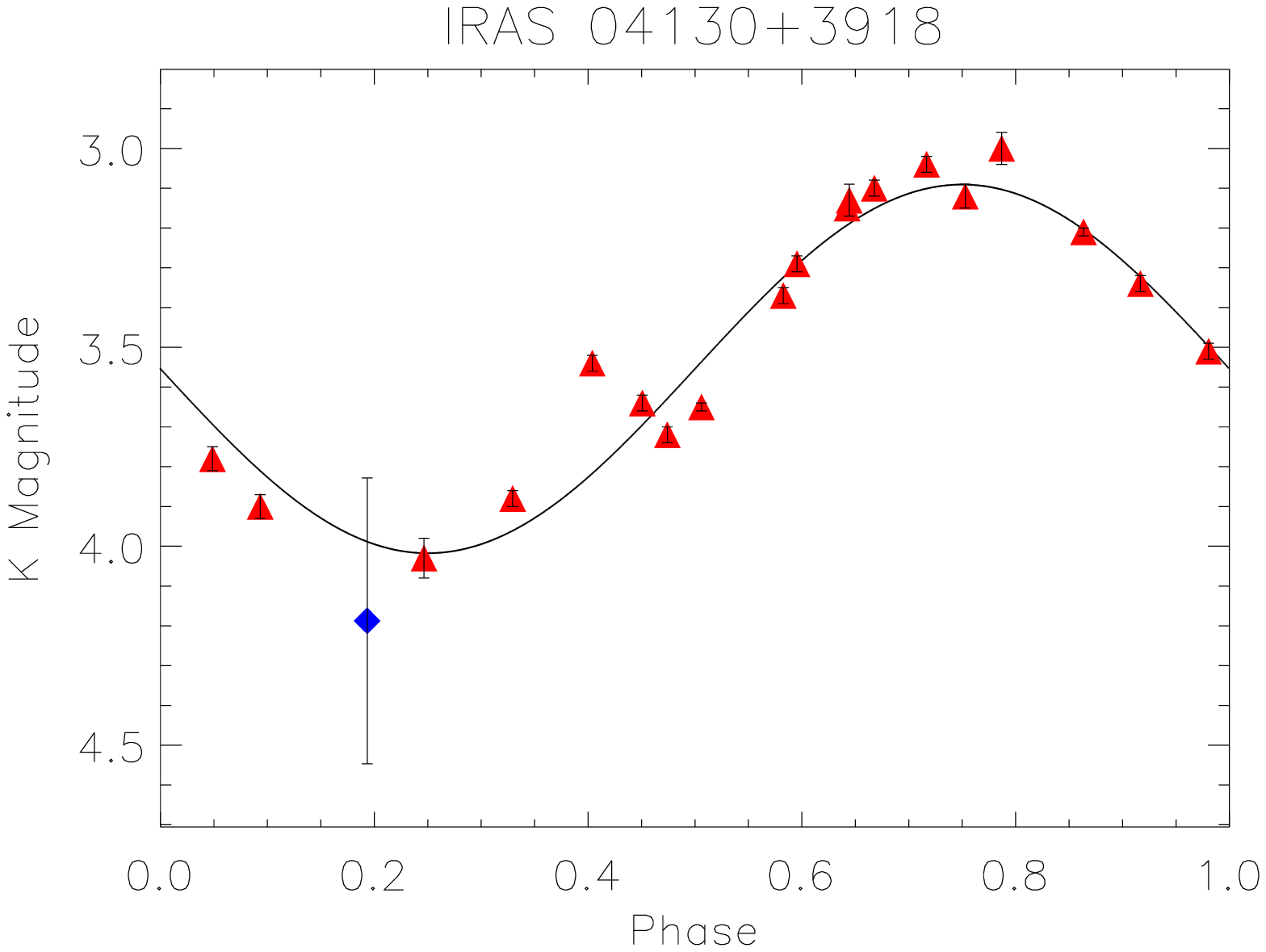}
\epsfxsize=7.4cm	
	\epsfbox{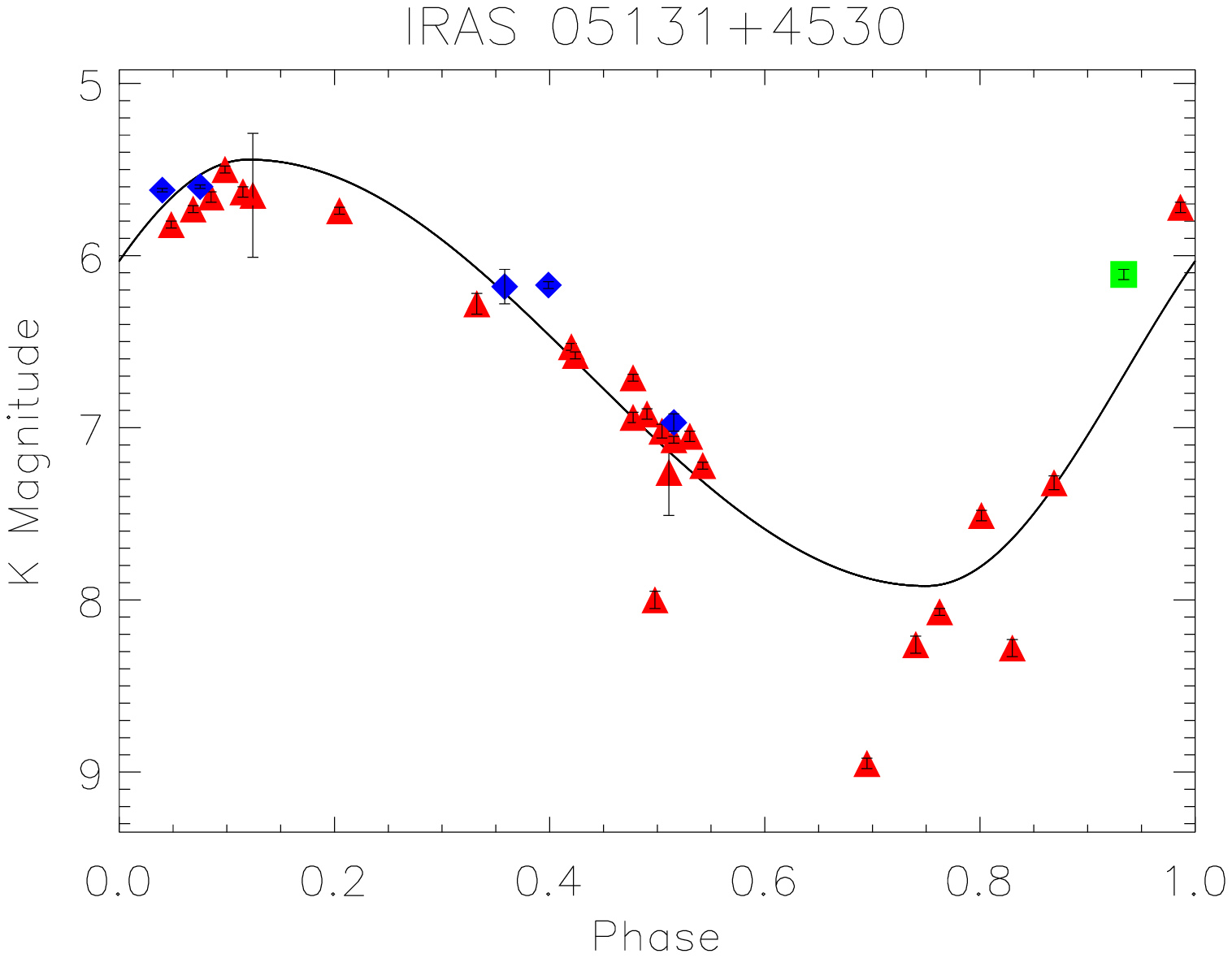}
\epsfxsize=7.4cm	
	\epsfbox{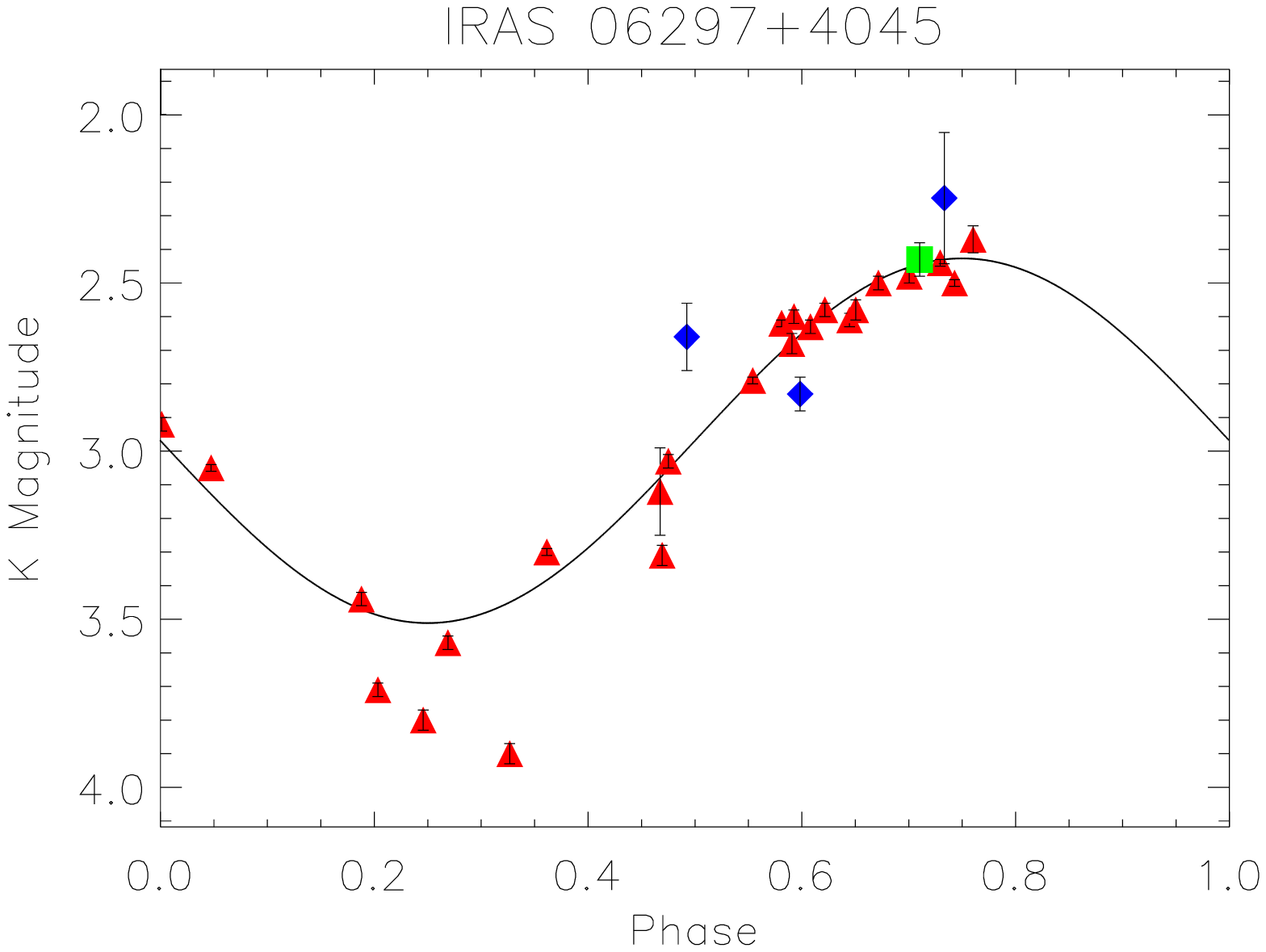}
\epsfxsize=7.4cm	
	\epsfbox{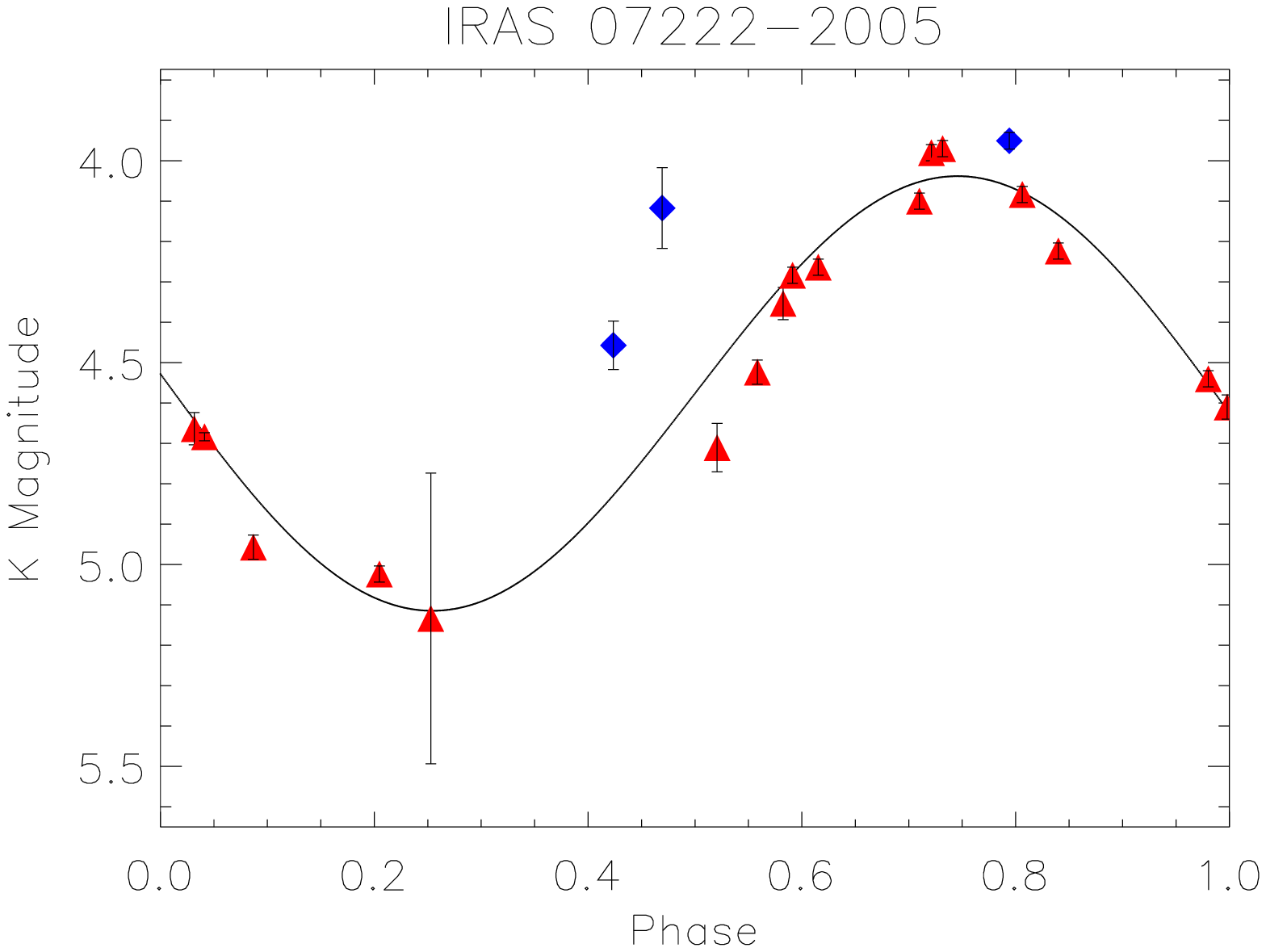}
\epsfxsize=7.4cm	
	\epsfbox{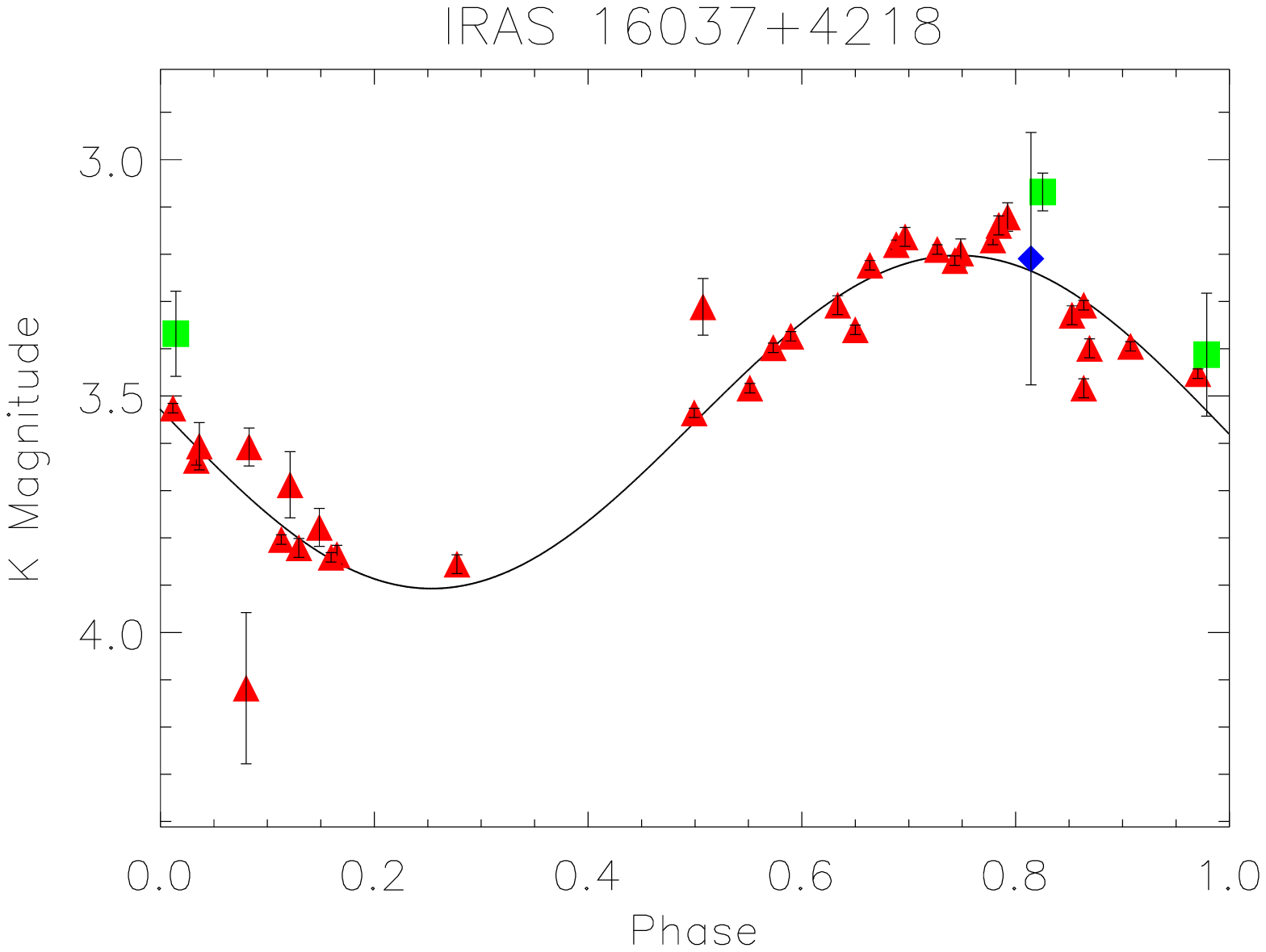}
\epsfxsize=7.4cm	
	\epsfbox{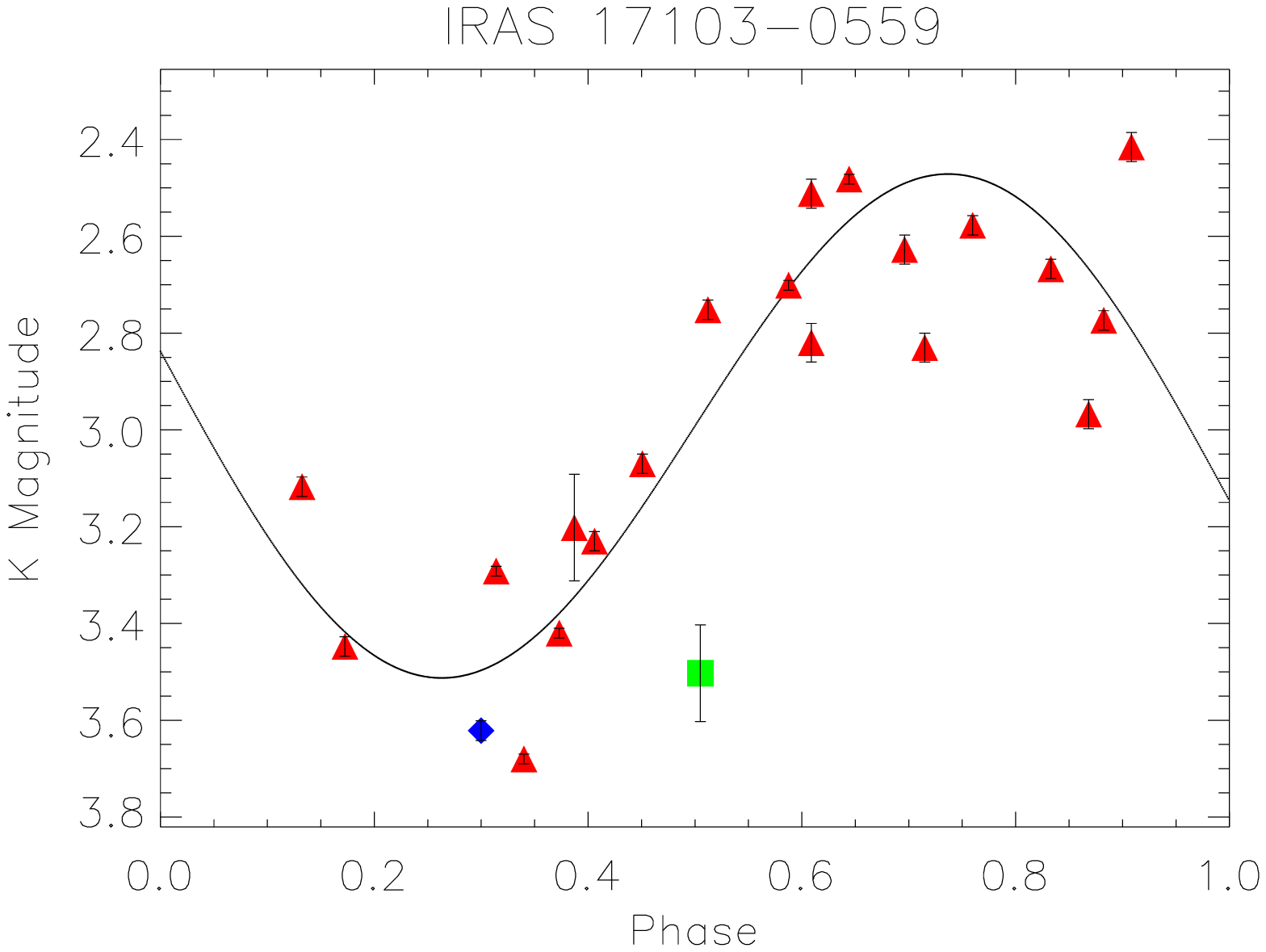}
\epsfxsize=7.4cm	
	\epsfbox{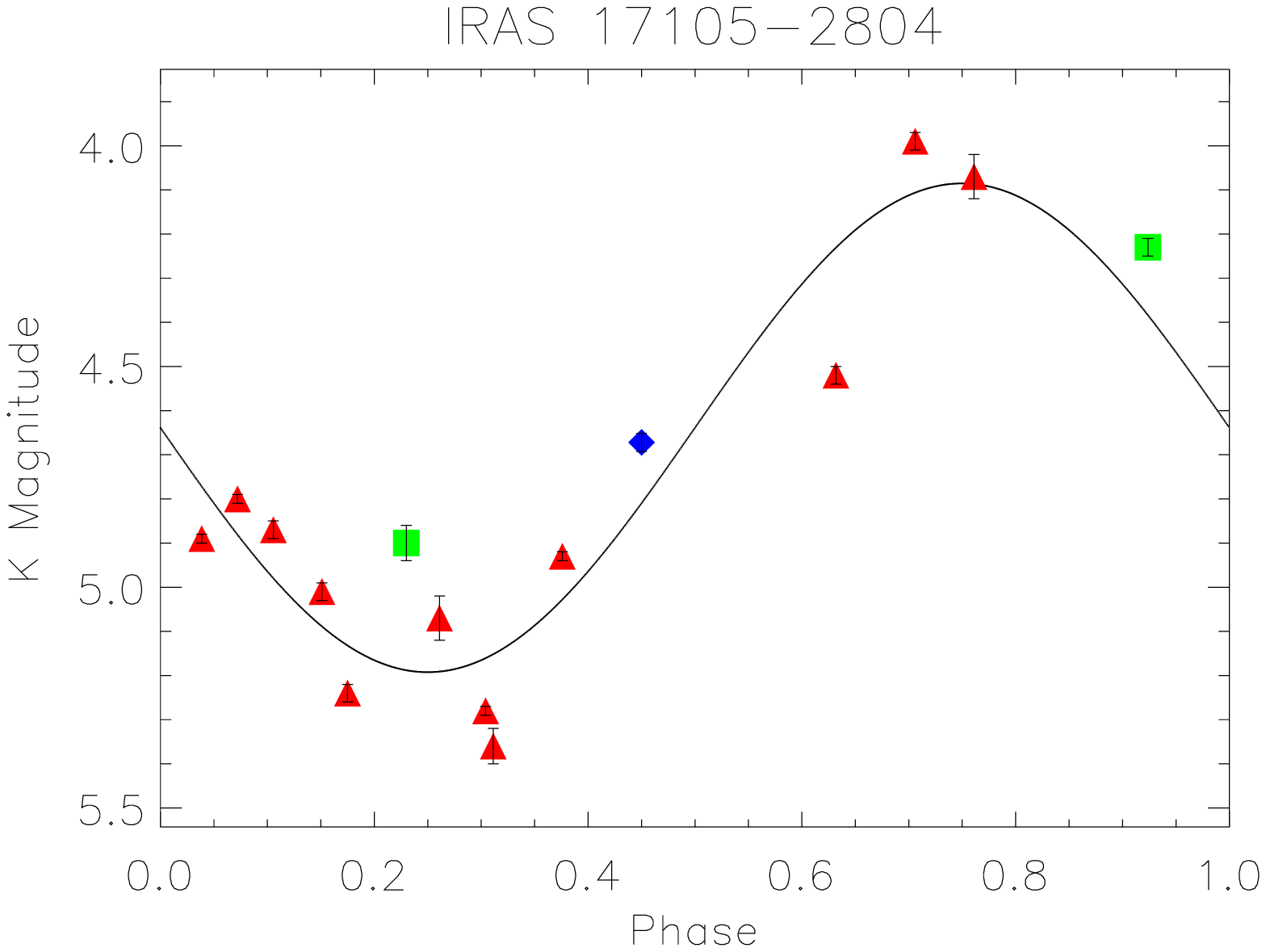}
	\label{fig_monit_app:light_curves}
\end{center}
{\bf Figure A2.} Light curves derived from the photometric data. CVF
		 data are plotted with triangles (red), MAGIC data
		 with squares (green), and data from the literature
		 with diamonds (blue). ({\em continued next page})
\end{figure*}

\begin{figure*}

\begin{center}
\vspace{0.75cm}
\epsfxsize=7.4cm	
	\epsfbox{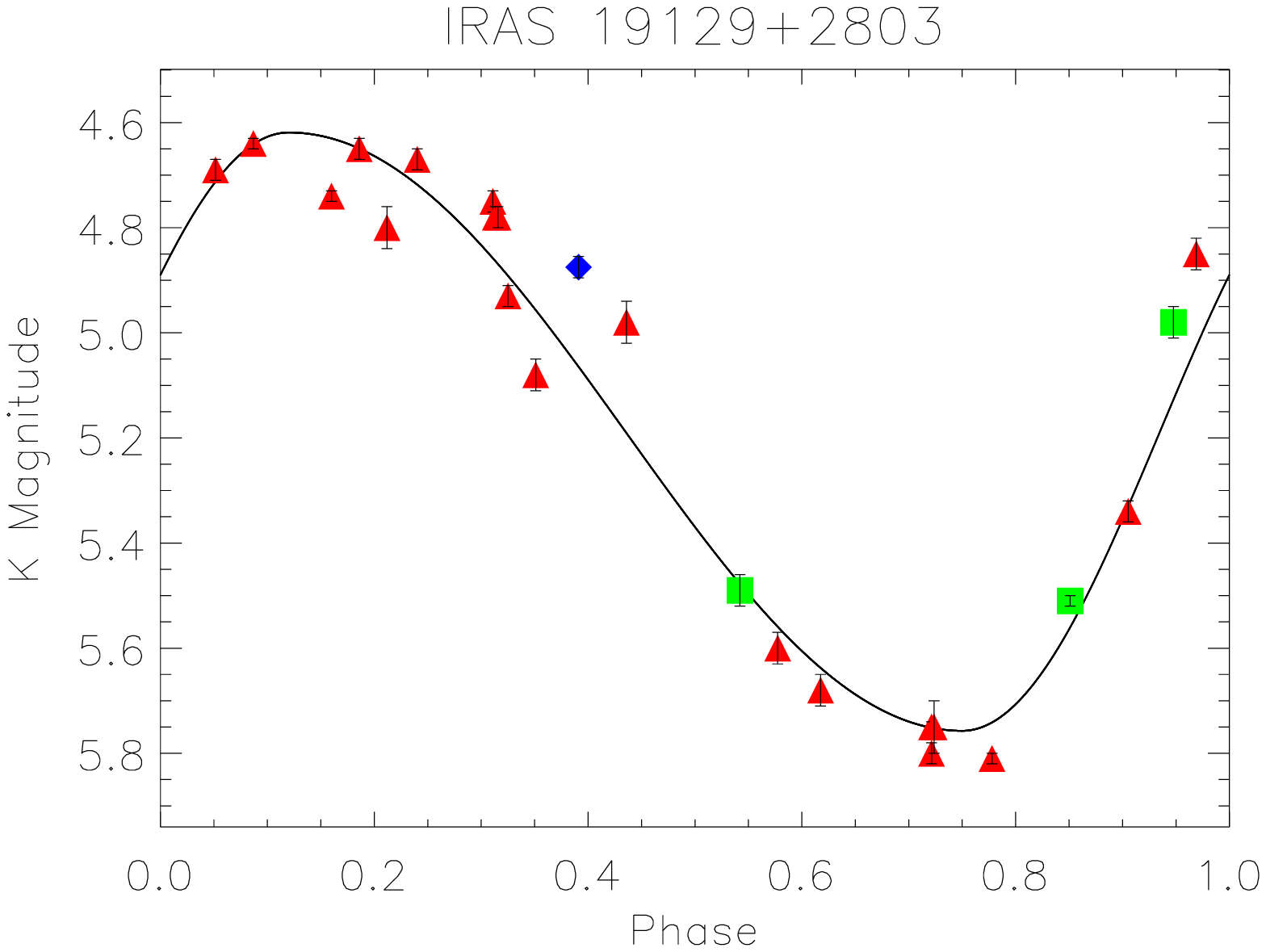}
\end{center}
{\bf Figure A2.} ({\em continued}) Light curves derived from the
		 photometric data. CVF data are plotted with triangles
		 (red), MAGIC data with squares (green), and data from
		 the literature with diamonds (blue).
\end{figure*}

\end{document}